\documentclass[a4paper,11pt]{article}
\usepackage[utf8]{inputenc}
\usepackage{enumitem}
\usepackage{amsmath}
\usepackage{tgpagella}
\usepackage[english]{babel}    
 
\usepackage[a4paper,margin=1in]{geometry}  
\usepackage{setspace}        
\usepackage{amsmath, amssymb, amsfonts}  
\usepackage{bm}                          

\usepackage{rotating}
\usepackage[rightcaption]{sidecap}
\usepackage{graphicx}    
\usepackage{caption}     
\usepackage{subcaption}  
\usepackage{booktabs}    
\usepackage{array}       
\usepackage{placeins}
\usepackage{makecell}
\usepackage{longtable}
\usepackage{float}
\usepackage{threeparttable} 
\usepackage{tabularx}
\usepackage{siunitx}    
\newcolumntype{L}{>{\raggedright\arraybackslash}X}
\usepackage{algorithm}    
\usepackage{algpseudocode} 
\usepackage{listings}      
\usepackage{xcolor}        

\usepackage[numbers,sort&compress]{natbib}  
\usepackage{hyperref}        
\hypersetup{
    colorlinks=true,
    linkcolor=blue,
    citecolor=blue,
    urlcolor=blue
}

\usepackage{filecontents}

\title{Lit2Vec: A Reproducible Workflow for Building a Legally Screened Chemistry Corpus from S2ORC for Downstream Retrieval and Text Mining}

\date{}

\author{
    Mahmoud Amiri$^{a,b}$,
    Jamile Mohammad Jafari$^{a,b}$, 
    Sara Mostafapour$^{a,b}$, \\
    Thomas Bocklitz (corresponding author)$^{a,b}$
}

\begin{document}

\maketitle

\noindent
{\footnotesize
$^a$ Leibniz Institute of Photonic Technology, Member of Leibniz Health Technologies, Member of the Leibniz Centre for Photonics in Infection Research (LPI), Albert-Einstein-Strasse 9, 07745 Jena, Germany. \\
$^b$ Institute of Physical Chemistry (IPC) and Abbe Center of Photonics (ACP), Friedrich Schiller University Jena, Member of the Leibniz Centre for Photonics in Infection Research (LPI), Helmholtzweg 4, 07743 Jena, Germany \\
}

{\small
We present Lit2Vec, a reproducible workflow for constructing and validating a chemistry corpus from the Semantic Scholar Open Research Corpus using conservative, metadata-based license screening. Using this workflow, we assembled an internal study corpus of 582,683 chemistry-specific full-text research articles with structured full text, token-aware paragraph chunks, paragraph-level embeddings generated with the intfloat/e5-large-v2 model, and record-level metadata including abstracts and licensing information. To support downstream retrieval and text-mining use cases, an eligible subset of the corpus was additionally enriched with machine-generated brief summaries and multi-label subfield annotations spanning 18 chemistry domains. Licensing was screened using metadata from Unpaywall, OpenAlex, and Crossref, and the resulting corpus was technically validated for schema compliance, embedding reproducibility, text quality, and metadata completeness. The primary contribution of this work is a reproducible workflow for corpus construction and validation, together with its associated schema and reproducibility resources. The released materials include the code, reconstruction workflow, schema, metadata/provenance artifacts, and validation outputs needed to reproduce the corpus from pinned public upstream resources. Public redistribution of source-derived text and broad text-derived representations is outside the scope of the general release. Researchers can reproduce the workflow by using the released pipeline with publicly available upstream datasets and metadata services.

\textbf{Scientific Contribution:} This work advances cheminformatics by introducing a reproducible workflow for constructing and validating a chemistry full-text corpus from public upstream resources using conservative, metadata-based license screening. In contrast to prior chemistry text resources that are often abstract-only, insufficiently documented, or not ready for semantic retrieval, Lit2Vec integrates full-text reconstruction, multi-source license screening, token-aware paragraph chunking, dense embeddings, and structured workflow outputs in a single transparent framework. The resulting workflow, schema, metadata/provenance artifacts, and validation resources provide a practical foundation for workflow-level reproducibility in chemistry literature retrieval and text mining under clearly documented access and licensing constraints.

}

\section*{List of abbreviations}
{\footnotesize
\begin{longtable}{p{0.10\textwidth} p{0.36\textwidth} p{0.10\textwidth} p{0.36\textwidth}}
\toprule
\textbf{Abbr.} & \textbf{Meaning} & \textbf{Abbr.} & \textbf{Meaning} \\
\midrule
\endfirsthead
\toprule
\textbf{Abbreviation} & \textbf{Meaning} & \textbf{Abbreviation} & \textbf{Meaning} \\
\midrule
\endhead
ANN & Approximate nearest neighbor & MAG & Microsoft Academic Graph \\
BART & Bidirectional and Auto-Regressive Transformers & MLP & Multilayer perceptron \\
BERTScore & Bidirectional Encoder Representations from Transformers Score & MMR & Maximal Marginal Relevance \\
CC & Creative Commons & NLP & Natural language processing \\
CC0 & Creative Commons Zero & OA & Open access \\
CC-BY & Creative Commons Attribution & RAG & Retrieval-augmented generation \\
CC-BY-NC & Creative Commons Attribution-NonCommercial & ReLU & Rectified linear unit \\
CC-BY-NC-ND & Creative Commons Attribution-NonCommercial-NoDerivatives & ROUGE & Recall-Oriented Understudy for Gisting Evaluation \\
CC-BY-NC-SA & Creative Commons Attribution-NonCommercial-ShareAlike & RTX & Ray tracing texel eXtreme \\
CC-BY-ND & Creative Commons Attribution-NoDerivatives & S2AG & Semantic Scholar Academic Graph \\
CC-BY-SA & Creative Commons Attribution-ShareAlike & S2ORC & Semantic Scholar Open Research Corpus \\
CUDA & Compute Unified Device Architecture & SMILES & Simplified Molecular Input Line Entry System \\
DOI & Digital Object Identifier & SPDX & Software Package Data Exchange \\
FAIR & Findable, Accessible, Interoperable, and Reusable & TL;DR & Too long; didn't read \\
FAISS & Facebook AI Similarity Search & JSONL & JSON Lines \\
F1 & F1 score & L2 & Euclidean norm \\
fp16 & 16-bit floating point & LLM & Large language model \\
GPT-4 & Generative Pre-trained Transformer 4 & InChI & International Chemical Identifier \\
GPT-4o & Generative Pre-trained Transformer 4 Omni & IQR & Interquartile range \\
\bottomrule
\end{longtable}
}

\begin{figure}[ht]
    \centering
    \includegraphics[width=\textwidth]{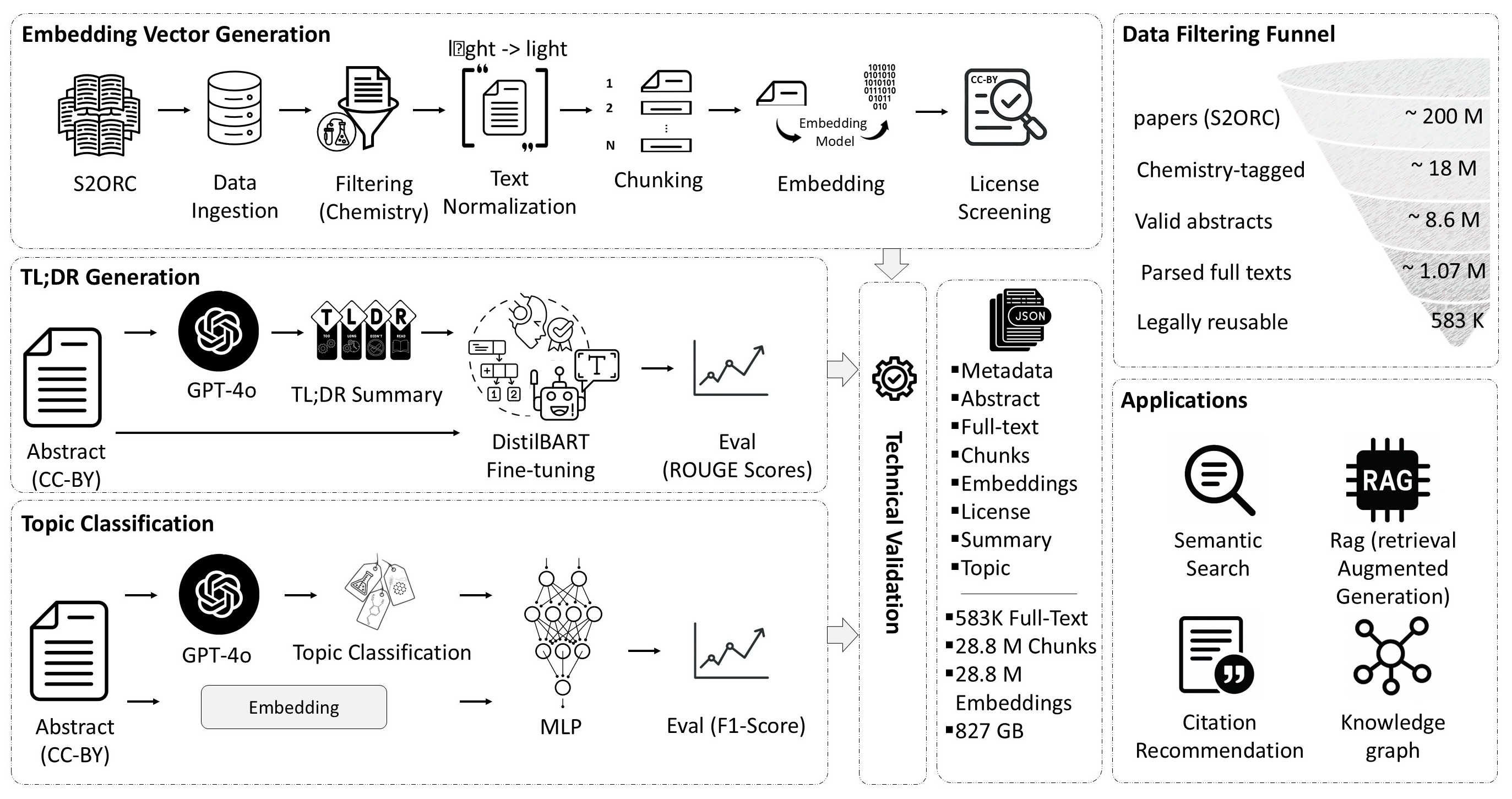}
    \caption{Overview of the Lit2Vec workflow. The primary pipeline constructs a legally screened chemistry corpus from S2ORC through domain filtering, text normalization, chunking, embedding, and license screening. Optional enrichment modules are applied to records with eligible abstracts to generate TL;DR summaries and subfield annotations.}
    \label{fig:lit2vec-workflow}
\end{figure}

\section{Introduction}

Recent progress in computational chemistry increasingly depends on the ability to analyze large bodies of scientific text with large language models (LLMs) and modern natural language processing (NLP) methods. In this setting, chemistry is not limited by a lack of published knowledge, but by the difficulty of converting a vast and heterogeneous literature into resources that are machine-readable, semantically structured, and legally reusable for downstream analysis.

This challenge is particularly acute because the chemistry literature is both large and operationally fragmented. Although automated techniques such as information extraction, literature mining, and retrieval-augmented generation (RAG) offer a path beyond manual review, their scientific usefulness depends on reproducible full-text corpora with standardized structure, rich metadata, and reviewable licensing provenance. Existing open collections only partially satisfy these requirements: much of the literature remains inaccessible, and even available full text is often not organized at the level of semantic segmentation and legal clarity needed for robust downstream workflows.

Prior chemistry corpora have supported tasks such as named entity recognition, question answering, and pretraining, but they do not fully satisfy the requirements of retrieval-oriented and reproducible chemistry NLP workflows. Common limitations include incomplete full-text coverage, lack of paragraph-level segmentation and dense embeddings for retrieval, heterogeneous licensing constraints, and inconsistent metadata normalization across chemistry subdomains. In addition, broad scientific corpora are not chemistry-structured, while large chemistry datasets primarily optimized for pretraining rather than real-time retrieval or question answering. A detailed review appears in Supplementary Section~\ref{sec:related-work}.

To address this gap, we introduce Literature to Vector (Lit2Vec) as a reproducible workflow for constructing a chemistry corpus for downstream retrieval and text-mining applications using conservative, metadata-based license screening. Using Semantic Scholar Open Research Corpus (S2ORC) as the upstream source, we assembled a study corpus of 582{,}683 full-text research articles through domain-based selection and rigorous screening checks. Licensing was screened by cross-referencing metadata from OpenAlex~\cite{openalex2025}, Unpaywall~\cite{unpaywall2025}, and Crossref~\cite{crossref2025}, with inclusion in the study corpus requiring metadata agreement under the study rule between at least two sources. The resulting screened corpus underpins the analyses reported here, and the released code, schema, manifests, and reconstruction workflow enable third parties to reproduce it from pinned public upstream sources. This screening provides a conservative, auditable basis for corpus construction by harmonizing license metadata from three independent services and excluding conflicting or insufficiently supported cases. It therefore improves transparency and reproducibility for workflow construction and auditing, while downstream redistribution decisions still depend on the accuracy and version-specific relevance of upstream records.

To our knowledge, Lit2Vec is the first chemistry-focused, reproducible workflow to reconstruct a legally screened full-text corpus from public upstream resources and organize it into a retrieval-ready framework with token-aware paragraph segmentation, dense embeddings, structured metadata, validation artifacts, and chemistry-specific enrichment.

Within this framework, articles are normalized into structured Markdown, segmented into approximately 28.8 million semantically coherent paragraph-level chunks, and embedded with the intfloat/e5-large-v2~\cite{wang2022text} model to produce 1024-dimensional dense vector representations. For a workflow overview, see Figure~\ref{fig:lit2vec-workflow}. For downstream artificial intelligence (AI) tasks such as RAG, document classification, scientific recommendation, and fact-grounded summarization, a large subset of records with eligible abstracts is further enriched with abstracts, abstract-level embeddings, LLM-generated Too Long; Didn't Read (TL;DR) summaries, and multi-label topic annotations. A detailed comparison of Lit2Vec with existing chemistry-related and general-purpose corpora is provided in Table~\ref{tab:chem_corpora_comparison}.

Lit2Vec supports a wide range of use cases across scientific research and infrastructure:

\begin{itemize}
    \item \textbf{RAG}: High-recall paragraph-level retrieval enables grounding LLMs in trusted scientific evidence, supporting applications such as chemical safety analysis and reasoning systems (demonstrated in Section~\ref{sec:RAG-Pipeline}).
    \item \textbf{Semantic search and recommendation}: Vector-based similarity enables discovery of related papers, techniques, or materials. We showcase a Facebook AI Similarity Search (FAISS)-powered recommender system in Section~\ref{sec:Recommendation-Pipeline}.
    \item \textbf{Model training and fine-tuning}: The released workflow, together with locally reconstructed records, can support training custom language and embedding models.
    \item \textbf{Knowledge graph construction}: Locally reconstructed Lit2Vec embeddings and metadata can be integrated with external resources (e.g., PubChem~\cite{pubchem2025}, ChEMBL~\cite{chembl2025}, and patent databases) to enhance chemical ontologies and build structured knowledge graphs.
    \item \textbf{Benchmarking and evaluation}: The corpus supports evaluation of LLM tasks including document retrieval, summarization, classification, question answering, hallucination detection, and citation fidelity.
    \item \textbf{Trends analysis}: With rich metadata and time-stamped documents, Lit2Vec enables trend analysis of materials, methods, or topics, for example the evolving use of gold versus silver in surface-enhanced Raman spectroscopy (SERS) (see Section~\ref{sec:trend-analysis}).
\end{itemize}

We therefore present Lit2Vec not simply as a static resource release, but as a reproducible foundation for chemistry-focused text infrastructure from public upstream sources. The released code, schema, manifests, metadata/provenance artifacts, and validation resources enable transparent reporting, reproducible analysis, and practical downstream use, while broad public redistribution of source-derived text and text-derived representations remains outside the scope of the general release.


\renewcommand{\arraystretch}{1}
\begin{table}[htbp]
    \centering
    \footnotesize
    \caption{Overview of the Lit2Vec pipeline, showing inputs, key processing steps, retained records, and retention rates relative to explicitly defined reference stages. Steps 9 and 10 (optional enrichment) were applied only to the subset of records with eligible abstracts.}
    \label{tab:pipeline_overview}
    \begin{tabular}{p{4cm} p{7cm} p{2cm} p{2cm}}
    \toprule
    \textbf{Step} & \textbf{Key Processing} & \textbf{Records Retained} & \textbf{\% of Reference Stage} \\
    \midrule
    1. Data Acquisition & Download JSONL dump & 200{,}000{,}000 & --- \\
    2. Domain Filtering & Restrict to fields\_of\_study = Chemistry & 18{,}621{,}044 & 9.3\% of S1 \\
    3. Abstract Alignment & Join with abstract corpus via corpus\_id & 8{,}660{,}979 & 46.5\% of S2 \\
    4. Full-Text Alignment & Join with full-text corpus via corpus\_id & 1{,}074{,}638 & 5.8\% of S2 \\
    5. Markdown Structuring & Parse annotations; normalize section headers & 966{,}946 & 89.9\% of S4 \\
    6. Chunking & Token-aware text segmentation & 963{,}697 & 99.7\% of S5 \\
    7. Embedding Generation & Encode with intfloat/e5-large-v2 & 963{,}697 & 100\% of S6 \\
    8. License Screening & Merge and verify license metadata across sources & 582{,}683 & 60.4\% of S7 \\    
    9. Topic Classification & Assign to 18 chemistry subfields & 460{,}397 & 79.0\% of S8 \\
    10. TL;DR Summarization & Generate two-sentence summary & 460{,}397 & 79.0\% of S8 \\
    \bottomrule
    \end{tabular}
\end{table}

\section{Methods}

In this study, reproducibility refers to deterministic regeneration of the screened study corpus and reported pipeline outputs from the same pinned upstream release using the released code, schema, archived manifests, frozen license metadata, and documented procedures. Thus, reproducibility is defined at the workflow level through regeneration from a fixed upstream snapshot, rather than through unrestricted redistribution of all source-derived content. A stepwise summary of the pipeline is given in Table~\ref{tab:pipeline_overview}.

\subsection{Data Collection}

We used the S2ORC dataset~\cite{lo2019s2orc} as the foundational source for constructing our corpus, specifically accessing the Semantic Scholar Academic Graph (S2AG) 2024-12-31 release via the official datasets API. This release includes structured metadata for approximately 200 million publications, abstracts for 100 million, and full-text parsed content for nearly 10 million open-access papers. We selected S2ORC for its scale, disciplinary breadth, and rich structural annotations that support downstream natural language processing (NLP) tasks.

To automate data acquisition, we developed a custom Python API client that interacts with the Semantic Scholar dataset API to retrieve metadata for the latest corpus release and generate download links for the papers, abstracts, and s2orc datasets. All files are distributed in compressed .json.gz format. We decompressed the papers dataset into JSONL (JSON Lines) format and ingested it into a MongoDB~\cite{mongodb2025} database for efficient querying.

Chemistry-related papers were identified directly from the fields\_of\_study metadata by selecting records labeled as Chemistry by Semantic Scholar's classification model, providing a clear and consistent basis for corpus construction.

We aligned abstracts and full-text records to the filtered chemistry subset using the shared corpus\_id field. Abstracts were decompressed, ingested via the same pipeline as the main metadata, and matched to chemistry-labeled papers. The s2orc full-text dataset was processed in parallel using the same method. Full-text entries were matched against the chemistry subset using corpus\_id. Only valid matches were retained for downstream processing. Corpus yields and coverage outcomes are reported in Section~\ref{sec:data-records}.

\subsection{Pre-processing}

Each JavaScript Object Notation (JSON) file in the full-text dataset contains raw text (content.text) and character-level annotations (content.annotations) that identify titles, abstracts, section headers, and paragraphs using start/end positions. Annotations stored as strings are parsed into dictionaries. We extract relevant spans, apply light Markdown formatting, and sort them by document order.

The converted output follows a standardized Markdown structure: the first title becomes a level-1 header (\#), the abstract is added under a level-2 header (\#\#), section headers are retained in order, and paragraphs are grouped with the nearest section. Content appearing before the first header is attached to it; content after the last header is added at the end.

Post-processing includes two steps. First, we normalize headers and remove low-content sections to reduce boilerplate. Headers matching a curated list of common scientific section names (e.g., \textit{Introduction}, \textit{Methods}, \textit{Results}) are kept as level-2 (\#\#); others are demoted to level-3 (\#\#\#). Sections with fewer than 10 words are discarded unless whitelisted.

Second, we merge broken line wraps while preserving structural elements such as headers, empty lines, checklist markers, or simple equation labels. Paragraphs are separated by a single newline for downstream processing.

Records with invalid annotations, offset errors, unreadable content, or insufficient structure (e.g., missing paragraphs or section headings) were excluded from downstream processing. The reproducible public pipeline mirrors this logic (see \ref{sec:code-availability}).

\subsection{Text Chunking}

To accommodate the input length limitations of transformer-based language models, each preprocessed Markdown document was segmented into smaller, semantically coherent chunks using a recursive, token-aware chunking strategy. This choice was motivated in part by recent findings from the Chunk Twice, Embed Once study\cite{amiri2025chunk}, which found that recursive, token-aware chunking strategy\cite{langchain2022} consistently outperformed other chunking methods in chemistry-focused RAG systems. In addition, they found that retrieval-optimized models such as Intfloat E5 variants significantly outperformed domain-specific models like SciBERT\cite{beltagy2019scibert} and ChemBERTa\cite{chithrananda2020chemberta} on chemistry-focused retrieval benchmarks. Based on these findings, we used intfloat/e5-large-v2~\cite{wang2022text} for downstream embedding because of its strong performance on semantic search benchmarks.

Chunking was implemented using HuggingFace's AutoTokenizer, configured for e5-large-v2. Special tokens were excluded from token-length calculations. Our pipeline applies a hierarchical split: first on paragraph boundaries (\textbackslash n\textbackslash n), then on sentence-ending punctuation, and finally on whitespace if needed. Maximum chunk length was set to 200 tokens, with a 20-token overlap to preserve context. Chunks under 100 tokens were merged with adjacent segments.

Documents with empty content or tokenization/encoding errors were skipped during chunk generation.

\subsection{Embedding Generation}

We generated dense vector embeddings for each text chunk using the intfloat/e5-large-v2 model \cite{wang2022text} via the HuggingFace sentence-transformers library. Chunks were tokenized with the model's native tokenizer (max length: 512 tokens), prefixed with "passage:" following the model's fine-tuning schema, and truncated as needed. Each resulting embedding is a 1024-dimensional, Euclidean-norm (L2)-normalized float32 vector suitable for downstream retrieval tasks. 

\subsection{License Screening}
\label{sec:licence-screening}
To support conservative compliance review and downstream auditing, we implemented a multi-source license screening process for the chemistry corpus. For each full-text paper, the digital object identifier (DOI), when available, was extracted from the external\_ids field. That DOI was used to query the Unpaywall, OpenAlex, and Crossref APIs in parallel to obtain license metadata for each paper. Unpaywall aggregates repository and publisher licenses, OpenAlex standardizes venue-level indicators, and Crossref provides publisher-reported terms. Retrieved licenses were normalized to Software Package Data Exchange (SPDX)-compatible categories (e.g., Creative Commons Attribution (CC-BY), Creative Commons Zero (CC0)).

To improve both recall and reliability, we combined all three sources. A final license label for each paper was assigned only if at least two sources agreed on the license type, the third source was either missing or did not contradict the others, and there were no conflicting license reports. Only papers with metadata-based license signals that met our conservative study inclusion rule were retained in the internal screened corpus: cc-by, Creative Commons Attribution-ShareAlike (CC-BY-SA), cc0, public domain, government works, as well as non-commercial variants such as Creative Commons Attribution-NonCommercial (CC-BY-NC) and Creative Commons Attribution-NonCommercial-ShareAlike (CC-BY-NC-SA). Papers with restrictive, conflicting, indeterminate licensing terms, or with insufficient agreement between sources were excluded. This rule provides a conservative and auditable basis for internal corpus construction by retaining only records with consistent multi-source metadata support. It strengthens transparency and reproducibility for downstream auditing, while public redistribution decisions remain tied to the accuracy and version-specific relevance of upstream license records. Coverage and retained-corpus outcomes are reported in Section~\ref{sec:data-records}.

\subsection{Auxiliary Abstract Dataset for TL;DR and Subfield Enrichment}
\label{sec:enrichment-dataset-generation}

To generate optional derivative annotations for a subset of the screened corpus, we assembled an auxiliary corpus of 19{,}992 chemistry abstracts from S2ORC with CC-BY-compatible license metadata for model development. Each abstract was paired with two structured annotations: (i) a TL;DR-style summary (1--2 sentences) that captures the core material or method, a key chemical finding, and at least one numeric result with standardized units; and (ii) a multi-label subfield classification selected from a controlled vocabulary of 17 major chemistry subfields plus a fallback Others category (18 classes total).

These annotations were automatically generated using a system-prompted GPT-4o model designed for structured information extraction. The supervision prompt required valid JSON output only, constrained TL;DR summaries to 1--2 sentences and no more than 50 words, required at least one explicit numeric result with standard units when available, and restricted field assignments to a controlled chemistry vocabulary. For abstract-dependent enrichment, abstracts shorter than 100 characters were flagged as too short for text-quality purposes, and abstracts shorter than 1,000 characters were excluded from abstract embedding generation, subfield prediction, and TL;DR generation. Full prompt text and schema details are provided in the Supplementary Materials (Section~\ref{sec:chemextract-prompt}). TL;DR summaries enable dense semantic embeddings that improve retrieval precision and context efficiency during generation, while subfield labels support sharded indexing and prompt conditioning for domain-aware outputs. GPT-4o was used only for supervision; no proprietary APIs are required at inference time, supporting workflow-level reproducibility for the released downstream pipeline.

To assess annotation quality, a random sample of 45 abstracts was manually reviewed by domain experts Sarah Mostafapour and Jamile Jafari. Their evaluation confirmed generally reasonable alignment between the abstracts, generated summaries, and subfield labels (see Section~\ref{sec:annotation-validation}). Release details for publicly available code and derivative resources are provided in Section~\ref{sec:data-availability}.

\subsection{Abstractive Summarization Pipeline}
\label{sec:abstractive-summarization}

To transform complex and structurally varied chemistry abstracts into concise, structured TL;DR-style summaries, we fine-tuned a domain-adapted abstractive model using the sshleifer/distilbart-cnn-12-6~\cite{distilbartcnn2025} checkpoint, a distilled version of Bidirectional and Auto-Regressive Transformers (BART) optimized for efficiency. Training was conducted on a CC-BY enrichment dataset (Section~\ref{sec:enrichment-dataset-generation}), using an 80/10/10 train/validation/test split. Abstracts were tokenized to a maximum input length of 1{,}024 tokens, and output summaries were capped at 128 tokens. Fine-tuning was performed with the Hugging Face Seq2SeqTrainer on a single NVIDIA Ray Tracing Texel eXtreme (RTX)~3090 graphics processing unit (GPU) using mixed precision, AdamW, a learning rate of 2e-5 and an effective batch size of 16 (per-device batch size of 4 with gradient accumulation over 4 steps) for 5 epochs. Evaluation, checkpoint saving, and logging were performed every 1{,}000, 1{,}000, and 500 steps, respectively. A fixed random seed (42) was applied across Python, NumPy, and PyTorch, with deterministic CUDA Deep Neural Network (CuDNN) settings enabled where applicable. Recall-Oriented Understudy for Gisting Evaluation metrics (ROUGE-1, ROUGE-2, ROUGE-L, and ROUGE-Lsum) and Bidirectional Encoder Representations from Transformers Score (BERTScore) were computed using the Hugging Face evaluate framework with standard preprocessing. Full training and evaluation settings are documented in the Supplementary Materials (Section~\ref{sec:summsup}).

Benchmark results are reported in Section~\ref{sec:tldr-benchmark-results}, with complete metric definitions and evaluation details provided in Supplementary Materials Table~\ref{tab:lit2vec_bench}. Access to trained model checkpoints and inference scripts is described in Section~\ref{sec:code-availability}. Finally, using the fine-tuned model, we generated TL;DR-style summaries for all papers with valid abstracts in our corpus.

\vspace{1em}
\subsection{Topic Classification Pipeline} \label{sec:topic-classification}

To support scalable organization and domain-specific retrieval of chemical knowledge, we developed a reproducible multi-label topic classification pipeline. Each abstract is embedded with the intfloat/e5-large-v2 SentenceTransformer, yielding 1{,}024-dimensional unit-normalized vectors. These embeddings are fed to a two-layer multi-layer perceptron (MLP; 256 units per layer, rectified linear unit (ReLU), dropout = 0.3) with a sigmoid output over 18 subfields (including \textit{Others}); abstracts may be assigned to multiple subfields to reflect interdisciplinary overlap. Training used batch normalization, weighted binary cross-entropy to address class imbalance, Adam with a learning rate of 1e-3, a batch size of 32, and up to 50 epochs with early stopping and reduce-on-plateau scheduling. Model selection used 5-fold cross-validation on the pooled training and validation data. A full description of the architecture, class-imbalance weighting, and training procedure is provided in Section~\ref{sec:supp-topic-classification}.

Pointers to datasets, trained weights, and an interactive demo are provided in Sections~\ref{sec:data-availability} and \ref{sec:code-availability}. Benchmark results are reported in Section~\ref{sec:subfield-benchmark-results}. Finally, we applied the trained classifier to all papers with valid abstracts, assigning one or more subfield labels to each entry in the corpus.

\subsection{Record Structure}

Each record is stored as a JSON object with required top-level fields for schema versioning, corpus identity, bibliographic metadata, abstract text, reconstructed full text, paragraph text units, paragraph embeddings, and license metadata from Unpaywall, Crossref, and OpenAlex. Optional enrichment fields include an abstract embedding, a TL;DR summary, and predicted subfield scores. Paragraph and embedding arrays are aligned by construction: each paragraph has a corresponding 1024-dimensional \textit{float32} embedding generated with intfloat/e5-large-v2, and both arrays have identical length. Full schema definitions, nested field dictionaries, and example records are provided in the Supplementary Materials (Section~\ref{sec:schema}).

\subsection{Validation Workflow}

We performed a comprehensive validation of the screened study corpus's structure, metadata, semantics, alignment, and licensing across 582,683 records to ensure reliability for downstream research and reproducible reuse. Validation was conducted through three complementary pipelines: (i) schema and structural checks, (ii) content and metadata quality assessment, and (iii) alignment, reproducibility, and licensing verification. Core checks required the presence and correct typing of mandatory top-level fields, valid identifier and key patterns, fixed-length 1024-dimensional embedding vectors, paragraph--embedding alignment, and basic format constraints for metadata, dates, and license fields. Each pipeline generated structured, timestamped reports with pass/warn/fail outcomes and detailed diagnostics, enabling full auditability. Complete validation rules and implementation details are provided in the Supplementary Materials (Sections~\ref{sec:subfield-mapping} to~\ref{sec:Consistency-validation}).

\section{Results}

\subsection{Corpus Yield and Coverage}
\label{sec:data-records}

\begin{figure}[H]
    \centering
    \begin{subfigure}[b]{0.37\textwidth}
        \centering
        \includegraphics[width=\textwidth]{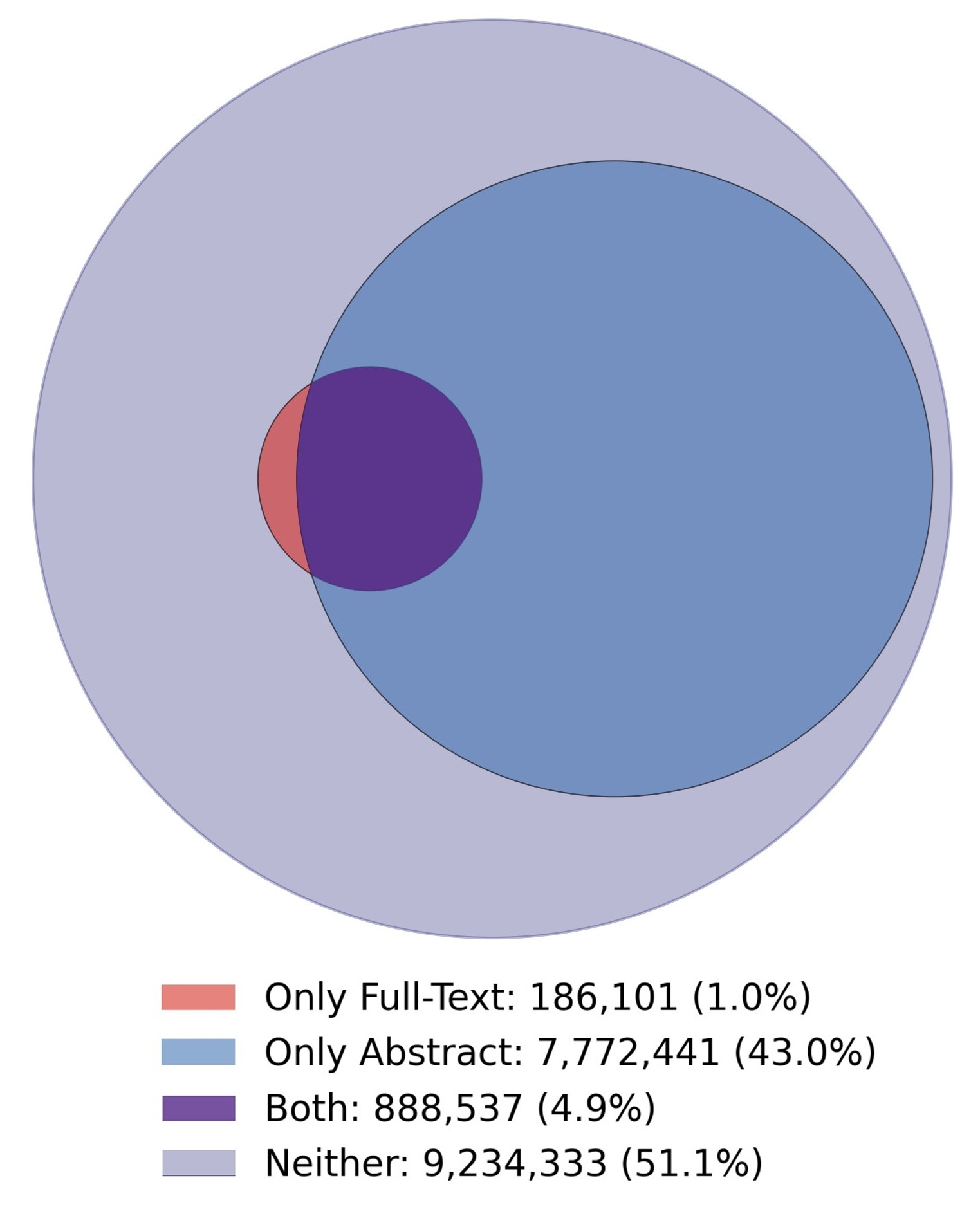}
    \end{subfigure}
    \hfill
    \begin{subfigure}[b]{0.61\textwidth}
        \centering
        \includegraphics[width=\textwidth]{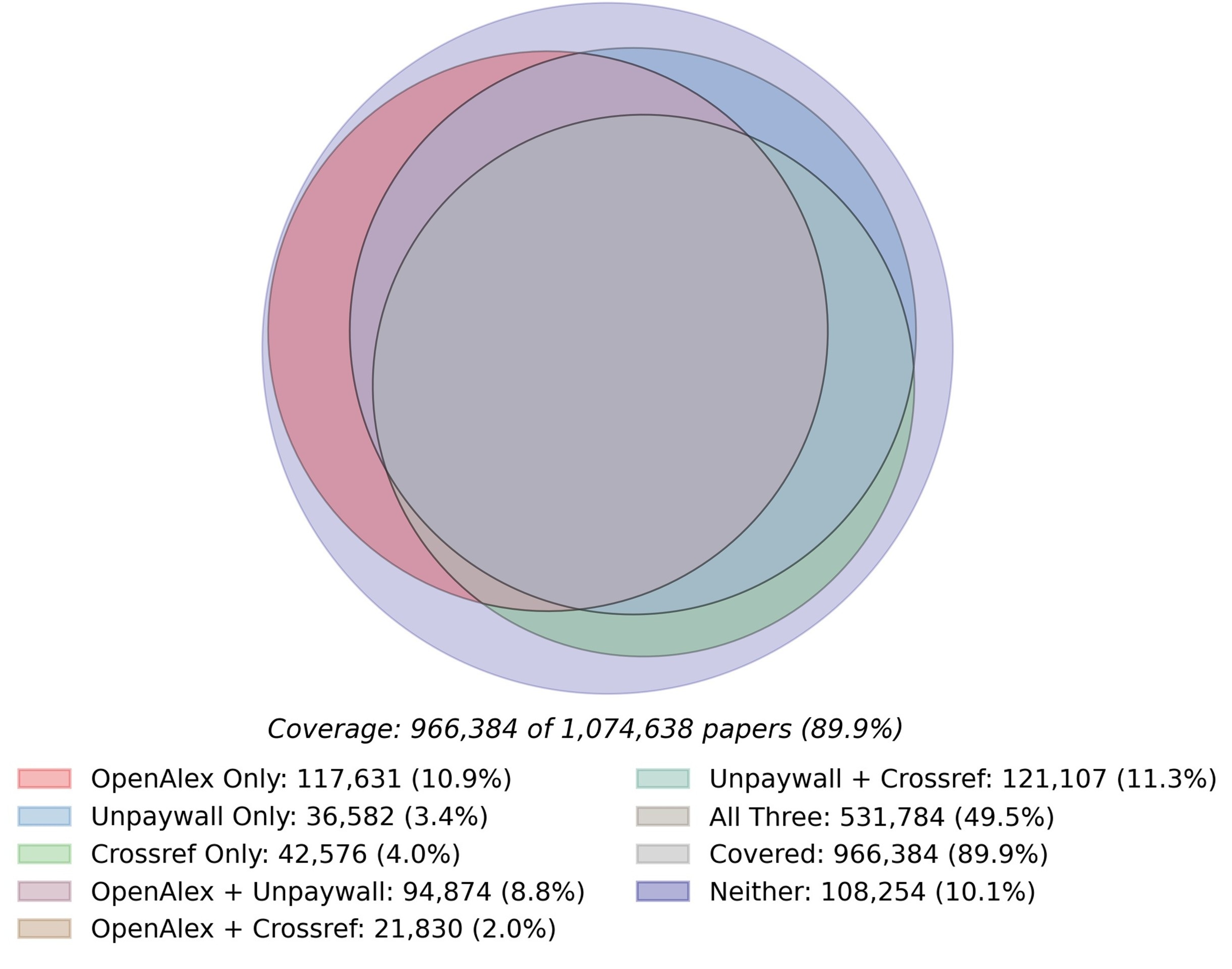}
    \end{subfigure}
    \caption{(a) Overlap between available full-text and abstract records for chemistry-labeled S2ORC papers. Most records contain only abstracts (43.0\%), while 4.9\% contain both and 1.0\% contain only full text. (b) Coverage of license metadata sources (Unpaywall, OpenAlex, Crossref) for the chemistry full-text subset (N = 1{,}074{,}638). Overall, 89.9\% of papers have license information from at least one source, with 49.5\% covered by all three.}
    \label{fig:chemistry-coverage-and-license-sources}
\end{figure}

Starting from the S2AG 2024-12-31 papers release, our chemistry filter identified 18{,}621{,}044 documents, corresponding to approximately 9.3\% of the ingested corpus. Alignment by corpus\_id yielded 8{,}660{,}979 abstracts (46.5\% of the chemistry subset) and 1{,}074{,}638 full-text records (5.8\% of the chemistry subset). As shown in Fig.~\ref{fig:chemistry-coverage-and-license-sources}(a), 48.9\% of chemistry-labeled records contained either an abstract or full text, 4.9\% contained both, and 1.0\% contained full text without an abstract.

Of the aligned full-text records, 966{,}946 were successfully converted to the normalized Markdown representation used for downstream processing, and 963{,}697 produced valid chunked outputs after token-aware segmentation. As shown in Fig.~\ref{fig:chemistry-coverage-and-license-sources}(b), 966{,}384 papers (89.9\%) had license metadata from at least one of Unpaywall, OpenAlex, or Crossref, and 531{,}784 (49.5\%) were covered by all three. Applying the multi-source agreement rule retained 582{,}683 full-text papers (60.4\% of the embedding-generated set) in the final screened corpus used throughout this study.

For a combined view of corpus coverage and license-source overlap, see Fig.~\ref{fig:chemistry-coverage-and-license-sources}.

\subsection{Technical Validation Outcomes}
\label{sec:technical-validation}

For a concise overview of the validation outcomes, see Table~\ref{tab:validation-summary}.

\begin{table}[th]
    \caption{Technical validation summary showing pass, warning, and non-complete rates (relative to the stated criteria) for each validation area.}
    \label{tab:validation-summary}
    \renewcommand{\arraystretch}{1.2} 
    \centering
    \footnotesize
    \begin{tabular*}{\textwidth}{@{\extracolsep{\fill}}p{3cm}p{5cm}p{7cm}}
    \toprule
    \textbf{Validation Area} & \textbf{Key Checks} & \textbf{Summary of Results} \\
    \midrule
    Schema \& Structure & JSON schema compliance, field typing, embedding format & 79\% pass (460{,}397); 21\% did not satisfy full enrichment/schema completeness criteria (122{,}286) — mainly due to missing abstracts or abstract-dependent fields omitted for abstracts below the embedding eligibility threshold \\
    Metadata & Completeness of title, authors, venue, year, DOI, license & 73\% pass (423{,}519); 27\% warn (159{,}164) — most warnings from missing venue, or publication date. \\
    Subfield Labels & Vocabulary match, valid confidence scores & 79\% pass (460{,}397); 21\% did not satisfy full enrichment/schema completeness criteria (122{,}286) — primarily attributable to missing or short upstream abstracts under the abstract-embedding eligibility threshold. \\
    Text Quality & Length limits, Unicode validity, abstract–full-text alignment & 68\% pass (398{,}064); 32\% warn (184{,}619) — $\sim$72\% strong alignment; most records free of quality flags. \\
    Chunking & Token size limits, Unicode integrity, paragraph–embedding mapping & 61\% pass (354{,}064); 39\% warn (228{,}619) — warnings only for naturally short final chunks. \\
    Embeddings & Presence, size, reproducibility & 100\% pass (582{,}683) — perfect cosine similarity agreement. \\
    IDs \& Integrity & ID consistency & 100\% pass (582{,}683) — no mismatches. \\
    Licensing & Cross-source open-access verification & 100\% pass (582{,}683) — 88\% CC BY; no conflicts. \\
    Summaries & ROUGE, BERTScore (evaluative) & Moderate lexical overlap; strong semantic agreement. \\
    \bottomrule
    \end{tabular*}
\end{table}
    
    \begin{figure}[H]
        \centering
        \includegraphics[width=\textwidth]{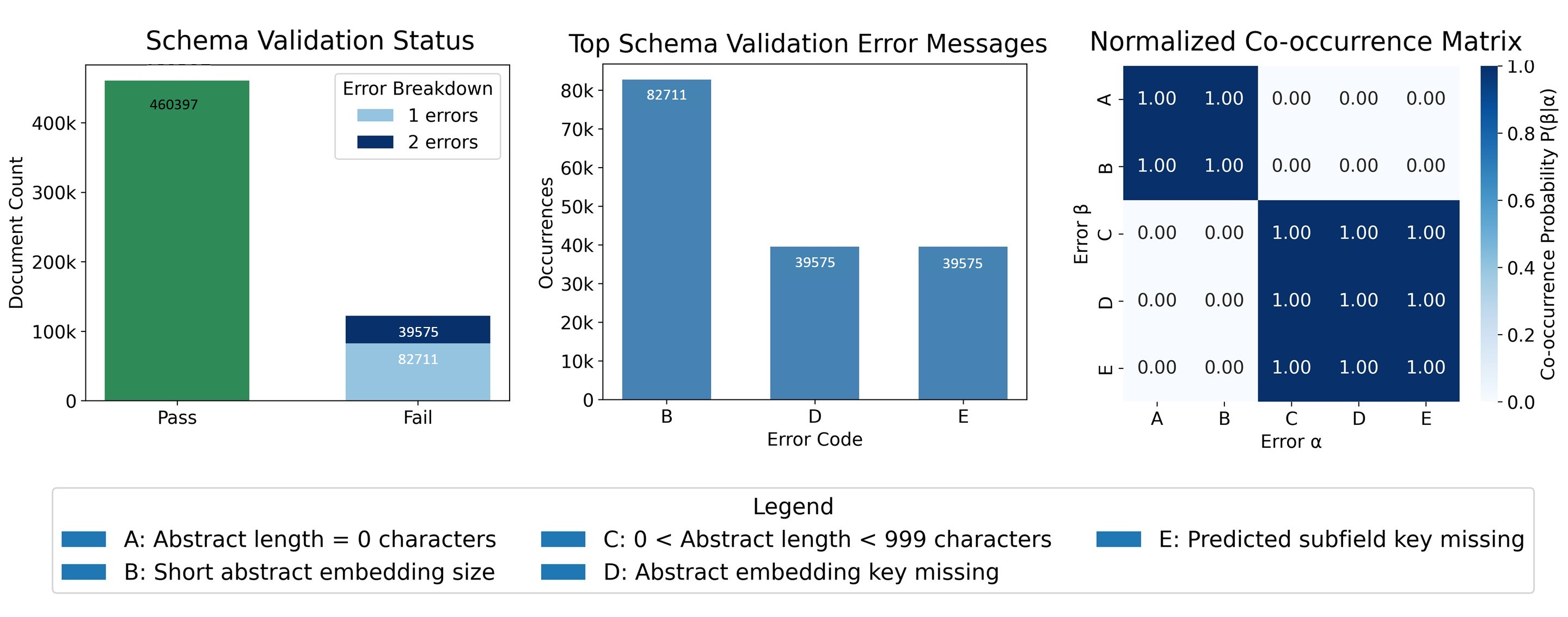}
        \caption{Schema validation results for 582{,}683 records in the chemistry full-text subset. 
        \textbf{Left:} Pass/not-fully-complete distribution under schema/enrichment criteria, with non-complete cases broken down by the number of errors per record. 
        \textbf{Center:} Most frequent non-completeness drivers are (i) short abstract embedding size (82{,}711 occurrences), (ii) missing abstract-embedding key (39{,}575), and (iii) missing predicted subfield annotation (39{,}575). 
        \textbf{Right:} Normalized co-occurrence matrix showing two tight clusters: empty abstracts always co-occur with short abstract embedding size, while short abstracts ($<\!1000$ characters) always co-occur with both missing abstract-embedding keys and missing predicted subfield annotations. These non-complete cases are primarily attributable to missing or short upstream abstracts under the abstract-length policy and are not indicative of embedding instability or licensing conflicts.}
        \label{fig:schema-validation-results}
    \end{figure}

\begin{figure}[H]
    \centering
    \includegraphics[width=0.85\textwidth]{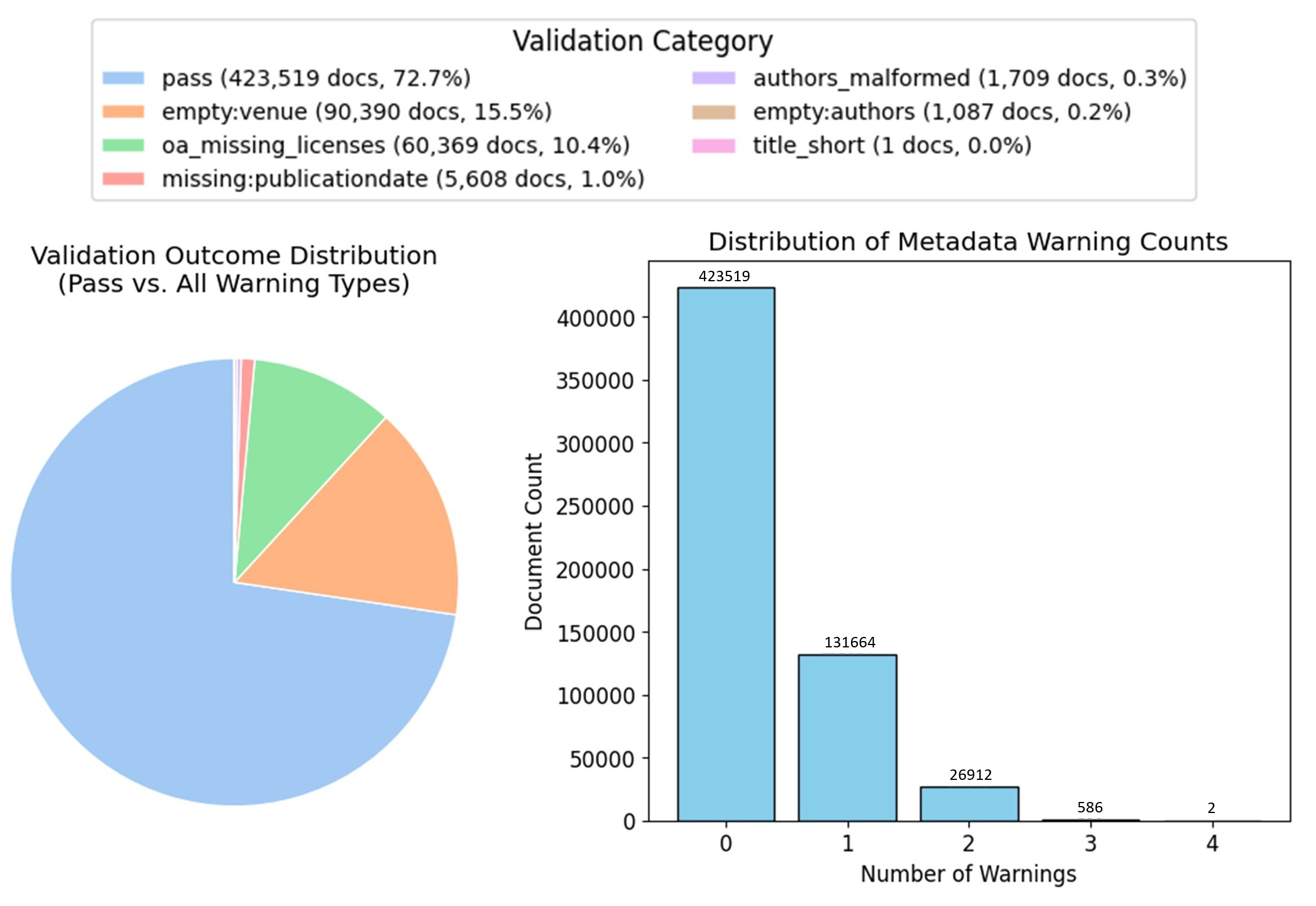}
    \caption{Metadata validation results for 582{,}683 records in the chemistry full-text subset. 
    \textbf{Left:} Distribution of validation outcomes by category, showing that 72.7\% of records pass all checks. 
    The most common warnings are missing venue information (15.5\%), missing open-access license metadata (10.4\%), and missing publication date (1.0\%). 
    Less frequent issues include malformed author fields (0.3\%), missing authors (0.2\%), and short titles ($<5$ characters, rare). 
    \textbf{Right:} Histogram of the number of warnings per record, with most affected records containing only a single warning.}
    \label{fig:metadata-validation}
\end{figure}

\subsection{Schema and Structural Validation}
 Of the $582{,}683$ records, $460{,}397$ ($79\%$) passed all checks (Fig.~\ref{fig:schema-validation-results}-left), while $122{,}286$ ($21\%$) did not satisfy full enrichment/schema completeness criteria due to either one ($67.6\%$) or two ($32.4\%$) co-occurring structural errors (Fig.~\ref{fig:schema-validation-results}-middle). These cases arise primarily from expected upstream abstract absence or abstract-length policy constraints rather than from embedding unreliability, licensing conflicts, or instability of the reconstruction workflow. 

Two main error clusters emerged (Fig.~\ref{fig:schema-validation-results}-right):
\begin{itemize}
    \item \textbf{A$\leftrightarrow$B:} Empty abstracts paired with short or invalid abstract embeddings.
    \item \textbf{C$\leftrightarrow$D$\leftrightarrow$E:} Short abstracts co-occurring with missing abstract embeddings and missing predicted subfields.
\end{itemize}

Review of source records confirmed that these issues originated upstream, where some documents lacked abstracts entirely or contained malformed text segments due to PDF-to-text conversion errors in the S2ORC (see Fig.~\ref{fig:chemistry-coverage-and-license-sources}(a)). Accordingly, records with missing abstracts or abstracts below the enrichment threshold lacked abstract embeddings, predicted subfields, and TL;DR summaries by design.

Future versions of the reconstruction workflow may address these structural gaps by sourcing valid abstracts from additional upstream repositories, thereby recovering high-quality records and reducing schema validation non-complete cases under these criteria.

\subsection{Metadata Validation}
Bibliographic completeness and consistency were assessed for titles, authors, venues, publication dates, identifiers (e.g., DOI), fields of study, and open-access status. As shown in Fig.~\ref{fig:metadata-validation}-left, $423{,}519$ records ($72.7\%$) passed without warnings. The most common warnings were missing venue information ($15.5\%$), missing license details ($10.4\%$), and missing publication dates ($1.0\%$). Warnings for malformed or missing author fields were rare ($\leq 0.3\%$).

The distribution of the number of warnings per record is shown in Fig.~\ref{fig:metadata-validation}-right. Most affected records contained only a single warning ($21.6\%$ of the corpus), while cases with three or more warnings were extremely rare ($<0.1\%$). Overall, the screened study corpus demonstrates high metadata completeness, with deficiencies largely confined to a small subset of records affected by upstream source gaps.

Notably, the license warnings primarily reflect gaps in the original S2ORC metadata rather than true absence of licensing evidence. We substantially improved record-level license provenance by enriching the corpus with metadata from OpenAlex, Crossref, and Unpaywall (see Section~\ref{sec:licence-screening}), providing a stronger and more complete basis for auditing and downstream compliance review.

\begin{figure}[H]
    \centering
    \begin{minipage}[t]{0.68\textwidth}
        \vspace{0pt}
        \centering
        \includegraphics[width=0.9\linewidth]{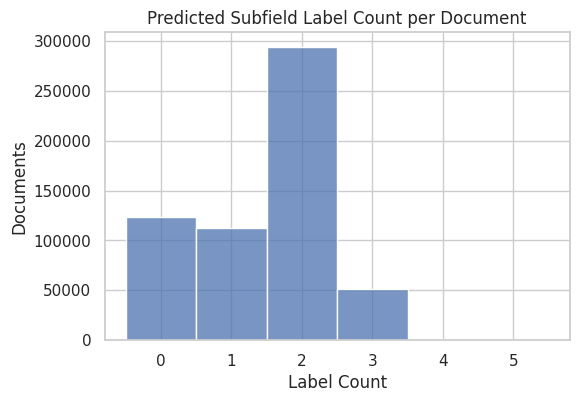}
    \end{minipage}\hfill
    \begin{minipage}[t]{0.28\textwidth}
        \vspace{0pt}
        \caption{Distribution of the number of predicted disciplinary subfield labels per document for records that passed schema validation. 
        Most documents are assigned two subfields, followed by those with zero or one label, and a smaller portion with three labels. 
        This reflects the multi-label nature of the classifier, which allows documents to be categorized into multiple overlapping chemistry subfields.}
        \label{fig:subfield-distribution}
    \end{minipage}
\end{figure}

\subsection{Subfield Validation}

Predicted disciplinary subfields were evaluated against a controlled chemistry vocabulary, with confidence scores constrained to the range $[0,\,1]$. All $122{,}286$ subfield cases that did not satisfy full enrichment/schema completeness criteria ($21\%$) resulted from missing predictions caused by the abstract-dependent embedding policy\footnote{These are design-expected omissions for short abstracts, not model failures on otherwise eligible records.}. Records passing without warnings totaled $460{,}397$ ($79\%$). As shown in Fig.~\ref{fig:subfield-distribution}, label distributions were stable across the corpus, with two subfields occurring most frequently.

\begin{figure}[H]
    \centering
    \includegraphics[width=\textwidth]{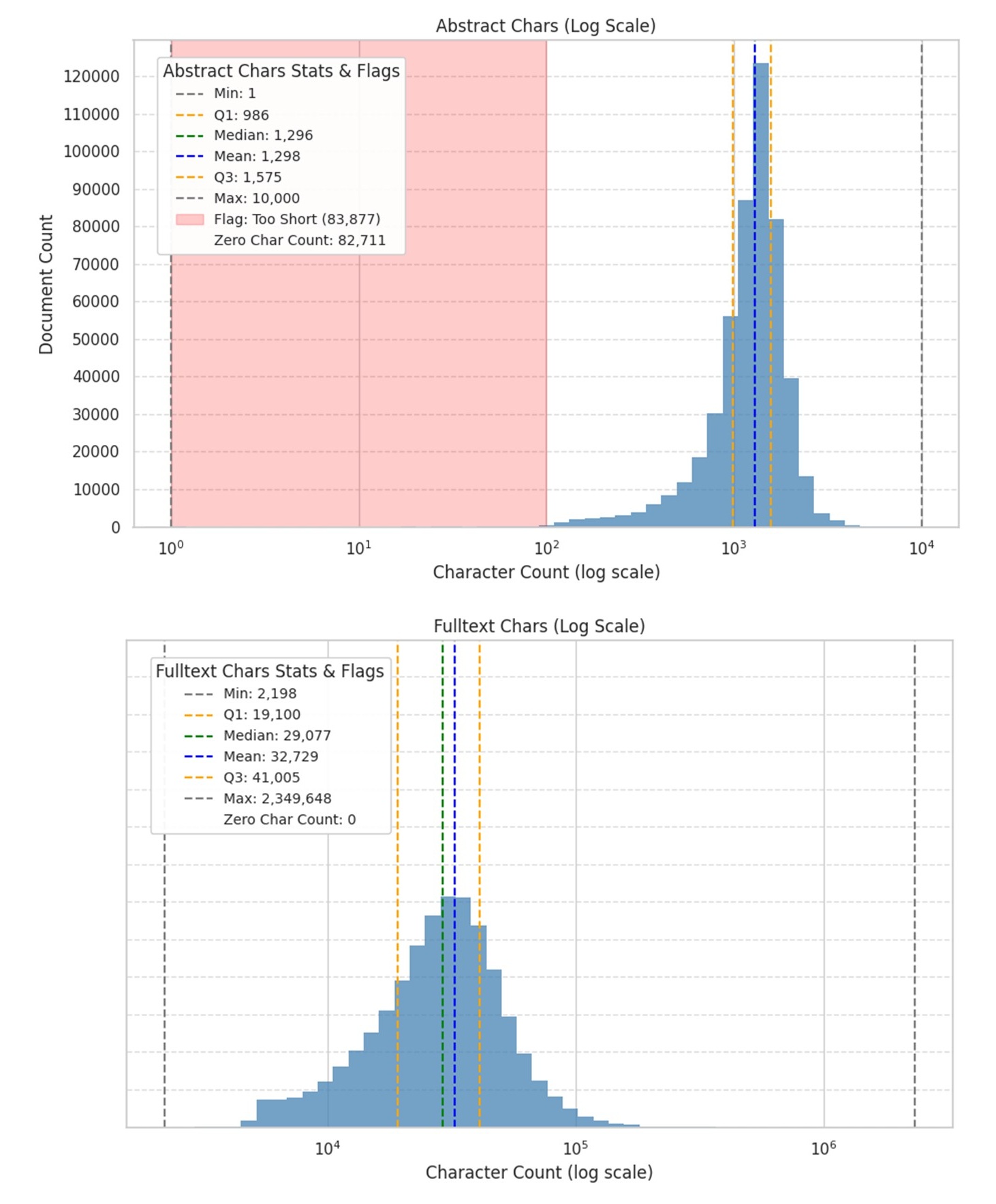}
    \caption{Character length distributions for abstracts (\textbf{top}) and full texts (\textbf{bottom}) in the chemistry full-text subset, shown on a log scale. 
    For abstracts, the median length is 1{,}296 characters, with 83{,}877 flagged as too short ($<100$ characters) and 82{,}711 having zero characters. 
    For full texts, the median length is 29{,}077 characters, with no zero-length cases. 
    Dashed vertical lines indicate quartiles and mean values.}
    \label{fig:abstract-alignment}
\end{figure}

\begin{figure}[H]
    \centering
    \includegraphics[width=\textwidth]{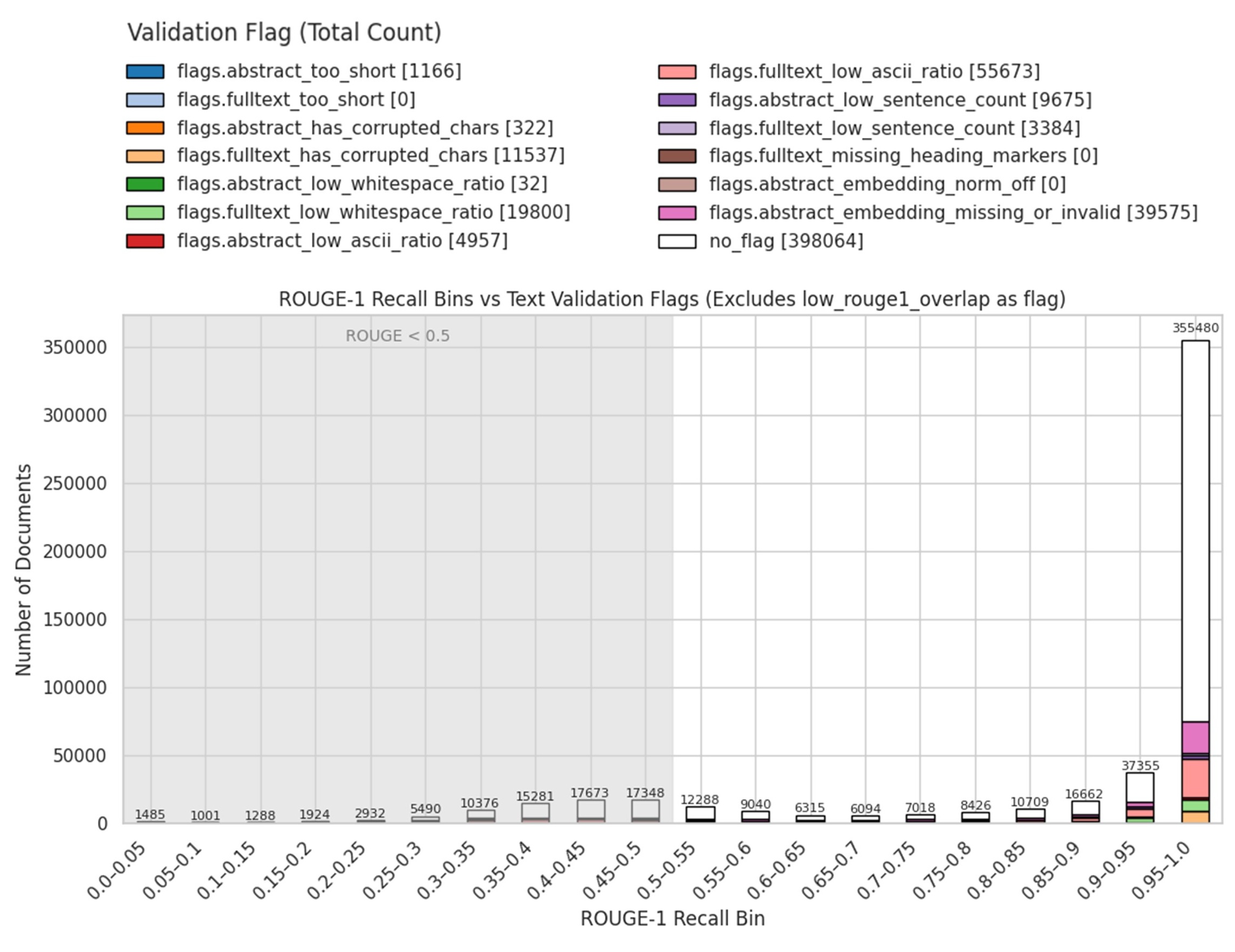}
    \caption{Distribution of ROUGE-1 recall alignment scores between abstracts and full texts in the chemistry full-text subset, grouped into bins of width 0.05. 
    Most documents have high alignment, with the largest peak in the 0.95–1.0 range. 
    Bars are color-coded by text validation flags, such as short abstracts, corrupted characters, low whitespace ratio, low American Standard Code for Information Interchange (ASCII) ratio, low sentence count, or missing/invalid abstract embeddings. 
    The shaded area on the left indicates records with ROUGE-1 recall scores below 0.5.}
    \label{fig:abstract-alignment2}
\end{figure}

\subsection{Textual Content Quality}

We assessed abstract and full-text quality using multiple criteria, including character length, sentence count, Unicode integrity, ASCII/whitespace ratios, language identification, and abstract--full-text alignment (ROUGE-1 recall). As shown in Fig.~\ref{fig:abstract-alignment}, abstracts were generally well-formed (median $= 1{,}296$ characters), although $83{,}877$ ($38\%$) were under $100$ characters, including $82{,}711$ empty abstracts---consistent with upstream source gaps. Full texts were consistently substantive (median $= 29{,}077$ characters; $\mathrm{IQR} = 19{,}100$--$41{,}005$), with outliers exceeding one million characters occurring rarely ($<0.1\%$) and typically associated with unusually long supplementary sections.

Alignment analysis (Fig.~\ref{fig:abstract-alignment2}) showed that approximately $355{,}000$ records ($72\%$) achieved strong lexical alignment between abstracts and the introductory portions of their corresponding full texts ($\mathrm{ROUGE}\text{-}1 \geq 0.95$). Fewer than $10\%$ of records had low alignment ($<0.5$), typically in combination with other quality flags. Most records ($398{,}064$) had no content-quality flags.

These results confirm that the majority of records have strong textual integrity and that flagged cases can be programmatically identified and excluded from analyses sensitive to textual quality.

\begin{figure}[H]
    \centering
    \begin{minipage}[t]{0.72\textwidth}
        \vspace{0pt}
        \centering
        \includegraphics[width=0.98\linewidth]{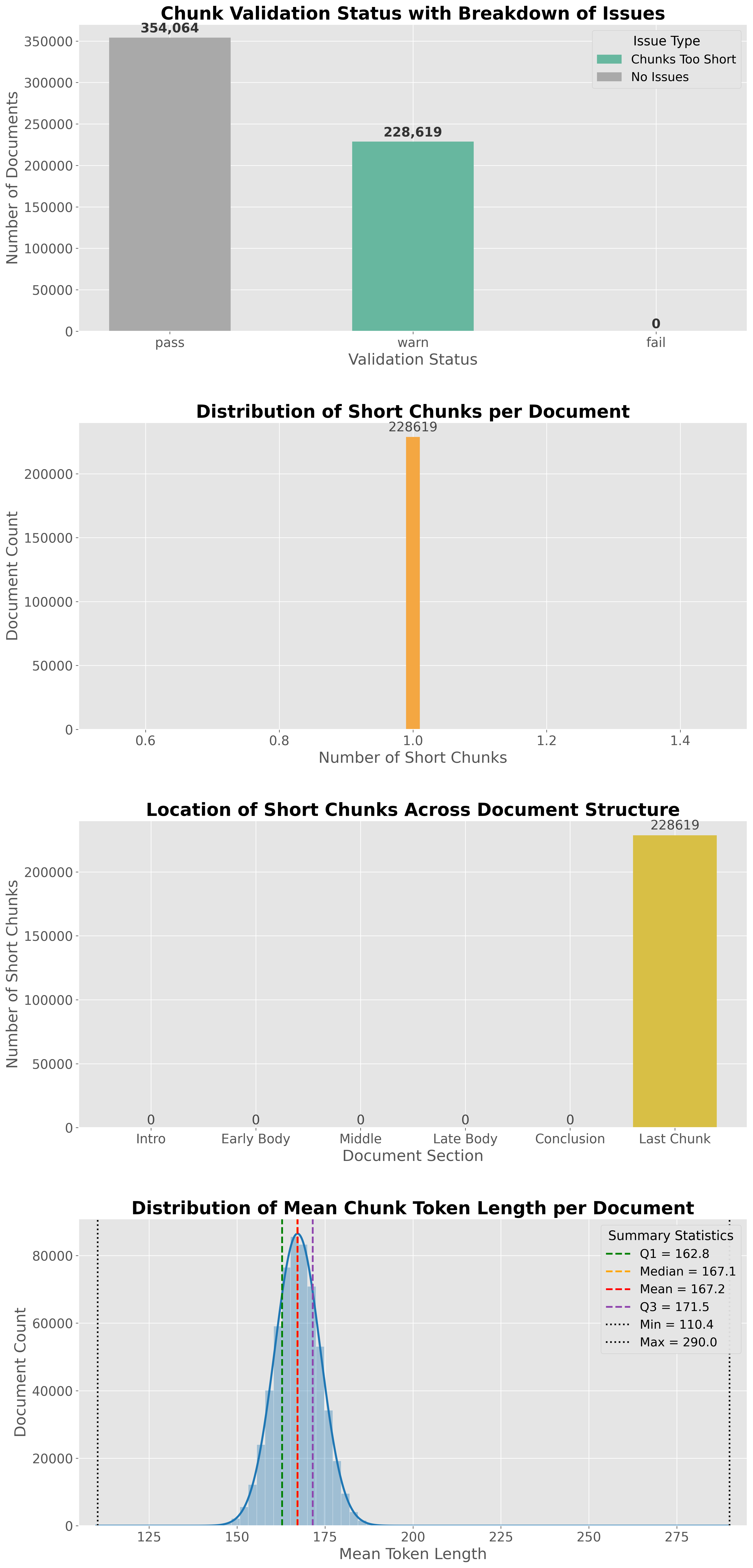}
    \end{minipage}\hfill
    \begin{minipage}[t]{0.25\textwidth}
        \vspace{0pt}
        \caption{Paragraph-level chunk validation for the chemistry full-text subset. 
        \textbf{First row:} Validation status showing that 61\% of documents pass without issues, while 39\% contain short chunks. 
        \textbf{Second row:} All affected documents contain exactly one short chunk. 
        \textbf{Third row:} All short chunks occur in the final chunk of the document. 
        \textbf{Forth row:} Distribution of mean chunk token length per document, with a median of 167 tokens and most values falling between 162 (Q1) and 171 (Q3).}
        \label{fig:chunking-validation}
    \end{minipage}
\end{figure}

\subsection{Chunk Level Validation}

For paragraph-level modeling, tokenized chunks and their embeddings were validated for token-length bounds, Unicode integrity, one-to-one paragraph--embedding mapping, dimensionality, finiteness, and near-unit vector norms. As shown in Fig.~\ref{fig:chunking-validation}-top-left, $354{,}064$ documents ($60.8\%$) passed without warnings; the remaining $228{,}619$ ($39.2\%$) received a single ``too short chunk'' warning. 

Fig.~\ref{fig:chunking-validation}-top-right shows that each affected document contained exactly one short chunk. Fig.~\ref{fig:chunking-validation}-bottom-left confirms these short chunks occur exclusively in the final paragraph position, typically containing $80$--$100$ tokens. Fig.~\ref{fig:chunking-validation}-bottom-right illustrates overall chunking consistency, with mean paragraph lengths tightly centered around $\sim 167$ tokens and a narrow interquartile range.

\begin{figure}[H]
    \centering
    \includegraphics[width=\textwidth]{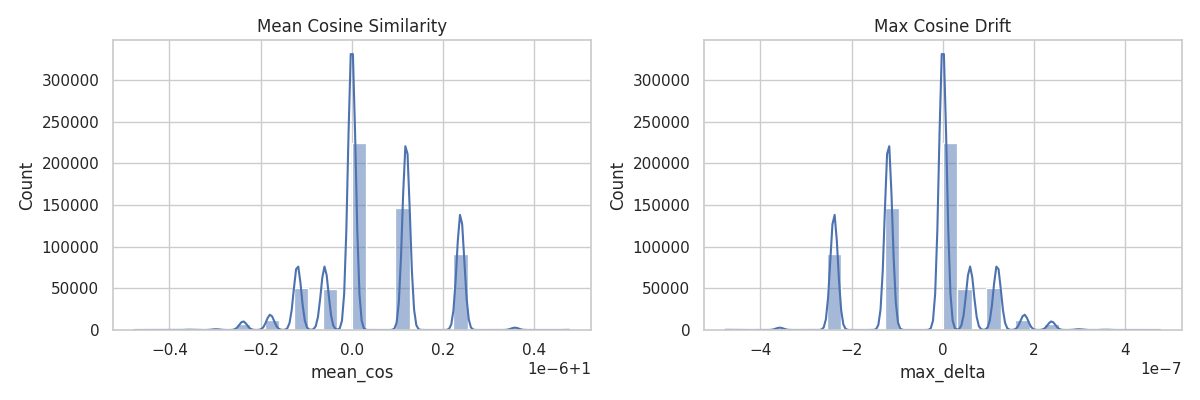}
    \caption{Embedding reproducibility analysis for the chemistry full-text subset. 
    \textbf{Left:} Distribution of mean cosine similarity between regenerated embeddings and their stored counterparts, with values tightly clustered around 1.0, indicating high reproducibility. 
    \textbf{Right:} Distribution of maximum cosine drift (largest per-token deviation), showing that even the largest observed differences are on the order of $10^{-7}$, confirming numerical stability across regeneration.}
    \label{fig:embedding-reproducibility}
\end{figure}

\subsection{Embedding Reproducibility}

Each paragraph embedding was validated for both structural integrity and semantic consistency. Checks confirmed correct dimensionality, finite values, and near-unit normalization. A strict one-to-one correspondence between paragraph IDs and embedding IDs was enforced, with any mismatches or anomalies flagged. As shown in Fig.~\ref{fig:embedding-reproducibility}, all $582{,}683$ records passed without errors (missing $=0.0\%$, extra $=0.0\%$, invalid $=0.0\%$). 

Cosine similarity between regenerated and stored embeddings was effectively perfect (mean $=1.000000$; min $=0.9999999$), with maximum numerical drift ($\pm4.77\times10^{-7}$) well within floating-point precision limits. Paragraph--embedding alignment was exact (Pearson's $r=1.0$), confirming that embeddings are fully reproducible.

\subsection{Identifier Integrity}
Internal identifiers (e.g., filename\_id, corpus\_id) and external identifiers (Crossref/Unpaywall/OpenAlex DOIs) were cross-validated and all records passed the test. 

\begin{figure}[H]
    \centering
    \includegraphics[width=\textwidth]{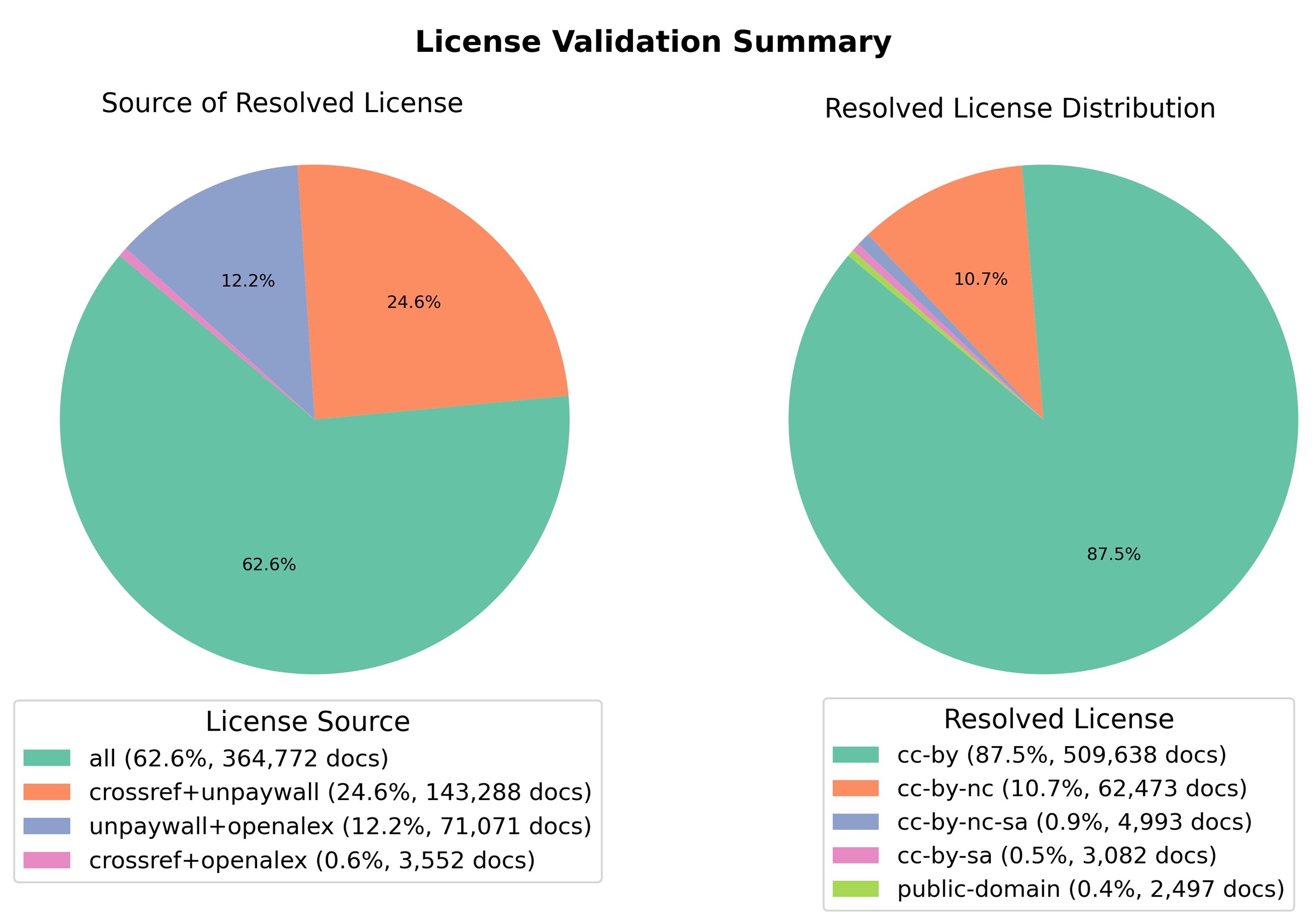}
    \caption{License validation results for 582{,}683 chemistry full-text records. 
    \textbf{Left:} Sources from which license information was resolved, showing that most records (62.6\%) were resolved by agreement across all three metadata sources (Crossref, Unpaywall, OpenAlex), followed by Crossref+Unpaywall (24.6\%), Unpaywall+OpenAlex (12.2\%), and Crossref+OpenAlex (0.6\%). 
    \textbf{Right:} Distribution of resolved license types, with CC-BY dominating (87.5\%), followed by CC-BY-NC (10.7\%), CC-BY-NC-SA (0.9\%), CC-BY-SA (0.5\%), and public domain (0.4\%).}
    \label{fig:licensing-validation}
\end{figure}

\subsection{Licensing Verification}

Licensing metadata from Crossref, Unpaywall, and OpenAlex were harmonized to a controlled vocabulary (e.g., CC-BY, CC-BY-NC). A record was considered a pass under the licensing validation step only when it satisfied metadata agreement under the study rule, meaning that at least two sources agreed on the open-access license without conflict. As shown in Fig.~\ref{fig:licensing-validation}-left, license resolution under this metadata-agreement rule was supported by agreement from all three sources for $62.6\%$ of records, from Crossref and Unpaywall for $24.6\%$, from Unpaywall and OpenAlex for $12.2\%$, and from Crossref and OpenAlex for $0.6\%$.

The resolved license distribution (Fig.~\ref{fig:licensing-validation}-right) is dominated by CC-BY ($87.5\%$; $509{,}638$ records), followed by CC-BY-NC ($10.7\%$; $62{,}473$ records). Other licenses---including CC-BY-NC-SA, CC-BY-SA, and Public~Domain---account for less than $2\%$ combined. No license conflicts were identified within the metadata sources under the study rule.

\begin{figure}[H]
    \centering
    \includegraphics[width=\textwidth]{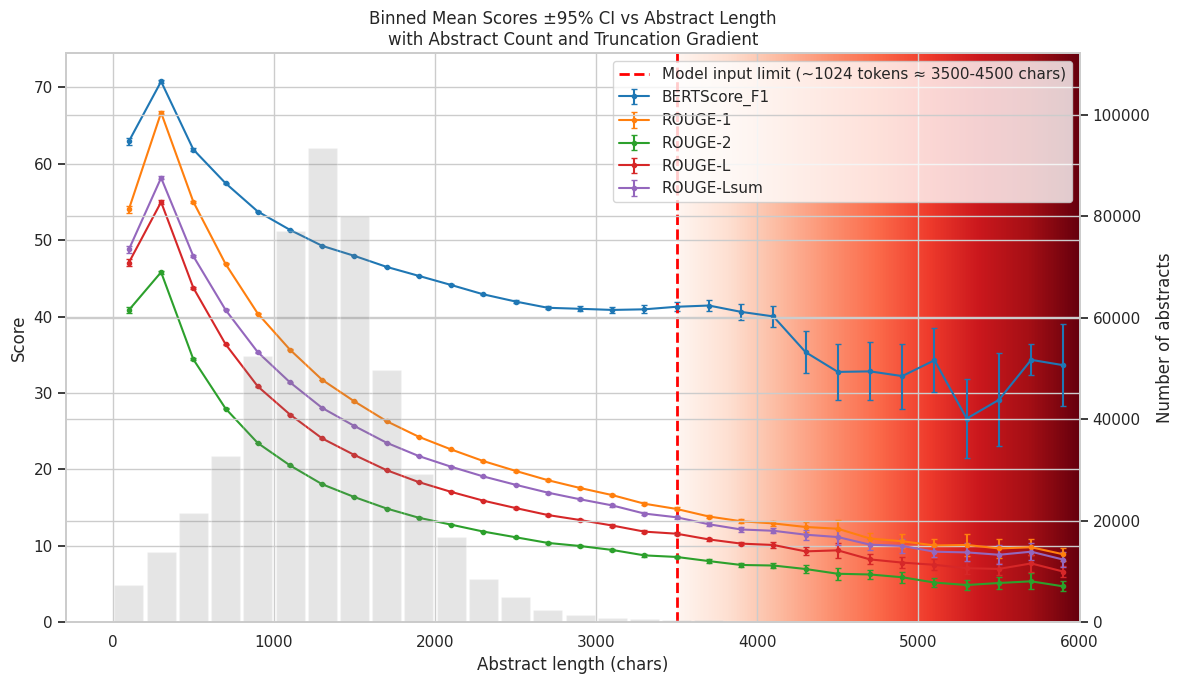}
    \caption{Relationship between abstract length and machine-generated summary evaluation scores for records with available summaries. 
    Lines show binned mean scores with 95\% confidence intervals for BERTScore\_F1, ROUGE-1, ROUGE-2, ROUGE-L, and ROUGE-Lsum. 
    The histogram (grey bars) indicates the number of abstracts per length bin, while the red gradient shows the proportion of abstracts exceeding the model's input limit ($\sim$1024 tokens $\approx$ 3500--4500 characters, dashed vertical line). 
    Scores generally peak at shorter lengths and decline as abstracts become longer, particularly beyond the model input limit.}
    \label{fig:summary-evaluation}
\end{figure}

\subsection{Automated Summary Evaluation}

For records with available machine-generated summaries, we compared summaries to their corresponding abstracts using ROUGE-1/2/Lsum and BERTScore metrics (Fig.~\ref{fig:summary-evaluation}). Median ROUGE values (\textsc{ROUGE-1}:~$32.0$; \textsc{ROUGE-2}:~$18.1$; \textsc{ROUGE-Lsum}:~$28.2$) indicate moderate lexical overlap, while BERTScore results demonstrate higher semantic capture (median recall:~$67.8$; precision:~$36.3$; $F_{1}$:~$49.8$). These results suggest that, although the summaries are not highly similar at the surface lexical level, they retain substantial semantic content from the abstracts.

\subsection{TL;DR Summarization Benchmark Results}
\label{sec:tldr-benchmark-results}

Table~\ref{tab:lit2vec_bench} shows that lit2vec\_tldr\_finetuned outperforms both extractive and zero-shot abstractive baselines across ROUGE and BERTScore metrics. Compared with bart\_large\_cnn, the fine-tuned model improves ROUGE-1 by +19.38, ROUGE-2 by +15.98, ROUGE-L by +18.38, and BERTScore F1 by +4.02, while decoding 36.6\% faster than that larger baseline. Although extractive methods such as lead3 and lexrank remain faster in absolute runtime, they consistently underperform on summary quality.

Qualitative inspection showed that the fine-tuned model typically preserves key experimental methods, midpoint numeric results, and brief application statements from the source abstracts. In high-performing examples, outputs closely match the GPT-4o-generated reference summaries in structure and specificity, whereas moderate examples retain the core findings with minor redundancy or phrasing issues (Section~\ref{sec:supplementary-summaries}). These results support the use of the fine-tuned DistilBART model as the main TL;DR generator in the Lit2Vec enrichment workflow.

\subsection{Subfield Classification Benchmark Results}
\label{sec:subfield-benchmark-results}

On the held-out test set at decision threshold tau = 0.5, the classifier achieved micro-F1 = 0.81, weighted-F1 = 0.80, and macro-F1 = 0.75. Performance was strongest for frequent subfields such as \textit{Biochemistry} (F1 = 0.92), \textit{Medicinal Chemistry} (F1 = 0.82), and \textit{Materials Science} (F1 = 0.80), whereas lower-resource domains such as \textit{Chemical Engineering} (F1 = 0.53) remained more difficult. The full per-label benchmark is reported in Supplementary Table~\ref{tab:s-classification-report}.

These results indicate that the embedding-based MLP captures major chemistry subfields reliably when adequate labeled support is available, while performance degrades for rarer classes. In practice, this makes the classifier suitable for broad corpus organization and retrieval support, with the main limitations concentrated in underrepresented labels.

\subsection{Technical Validation Summary}
Across validation areas, the screened study corpus showed strong structural integrity, metadata completeness, textual quality, embedding reproducibility, and consistent metadata agreement under the study rule across 582,683 records. Across all checks, the corpus demonstrated high reliability: $79\%$ of records passed strict JSON Schema validation; metadata completeness exceeded $72\%$, with gaps largely limited to venue, license, and publication date fields; over $68\%$ of records showed strong abstract--full-text alignment; chunking and embedding checks achieved near-perfect reproducibility; and licensing showed metadata agreement under the study rule, with $100\%$ of retained records passing the licensing-validation rule and $88\%$ of records screened as CC-BY. 

Edge cases---primarily short or missing abstracts and naturally short final chunks---are well characterized and can be programmatically filtered, supporting a robust and reproducible reconstruction workflow. The core validated contribution of this work is therefore the reconstruction workflow, together with the screened study corpus it produces under the stated access and licensing constraints. Enrichment layers should be understood as optional and inherently subset-dependent rather than universal outputs of the pipeline. In particular, abstract-related omissions are best interpreted as expected consequences of the validation and enrichment policy, especially where source metadata are sparse or incomplete, rather than as arbitrary processing failures. Remaining issues are thus tied mainly to upstream source quality, not to irreproducible downstream steps.

\subsection{Example Applications}

We demonstrate Lit2Vec on three reference tasks: \emph{Trend Analysis}, \emph{Semantic Paper Recommendation}, and \emph{Retrieval-Augmented Generation}. These examples illustrate representative downstream uses of the corpus and associated derivative resources.

\subsubsection{RAG Pipeline}
\label{sec:RAG-Pipeline}

To support scientific question answering grounded in published literature, we implemented a simple RAG system that integrates dense retrieval over paragraph-level embeddings with answer synthesis via an LLM. This system enables the generation of responses that are explicitly supported by source documents, with paragraph-level citations included inline.

Valid embeddings are converted into NumPy arrays and indexed using FAISS with an IndexFlatL2 structure. This enables efficient approximate nearest-neighbor search over paragraph-level vectors. Paragraphs and their corresponding metadata are stored in memory for lookup and retrieval.

To perform retrieval, the system encodes the user's query into an embedding using the intfloat/e5-large-v2 model from the Sentence-Transformers library~\cite{wang2022text}. This query vector is normalized and searched against the FAISS index to retrieve the top-$k$ most semantically relevant paragraphs. Each retrieved result includes the paragraph text, a paragraph-level identifier, and associated metadata (title, authors, year, DOI).

The retrieved paragraphs are assembled into a structured context block, where each paragraph is labeled with its unique identifier (e.g., 123456P2). This block, along with the user's question, is provided as a prompt to an OpenAI GPT-4 model (gpt-4o) using the chat completion API.

The system instructs the model to answer the question in a coherent paragraph format while citing supporting statements using paragraph IDs in parentheses. This approach ensures that each part of the generated answer can be traced back to its evidence source.

In addition to the generated answer, the system outputs a reference list citing the original documents that contributed supporting paragraphs. Each reference includes the full title, authors, publication year, and DOI (if available), associated with the paragraph-level identifiers used in the inline citations.

The full implementation is provided in a Jupyter notebook at \href{https://github.com/Bocklitz-Lab/Lit2Vec-code/tree/main/example_tasks/RAG}{example\_tasks/RAG} on GitHub.

\subsubsection{Semantic Paper Recommendation Pipeline}
\label{sec:Recommendation-Pipeline}

We developed a lightweight semantic recommendation system designed to support literature discovery by identifying papers that are semantically related to a reference document or a free-text scientific query. The system is implemented in Python and leverages vector representations of scientific abstracts to perform similarity-based retrieval and ranking.

During data loading, the system filters out any records with missing or malformed embeddings. Specifically, it retains only those entries where the abstract\_embedding field contains a non-empty, one-dimensional array of numeric values. This ensures that only valid documents are included in the subsequent indexing and retrieval steps.

For semantic retrieval, we use the FAISS library~\cite{johnson2019billion} to build an efficient nearest-neighbor index. The abstract embeddings are L2-normalized and indexed using an IndexFlatIP structure, which enables fast cosine similarity search via inner products.

The system supports two primary retrieval modes. In the first mode, a known paper (specified by its corpus\_id) is used as a query, and the most similar papers are returned based on abstract-level semantic similarity. In the second mode, a free-text query is encoded into an embedding using the intfloat/e5-large-v2 model from the Sentence-Transformers library~\cite{wang2022text}. This query embedding is then used to retrieve relevant documents from the FAISS index.

After retrieving a set of candidate papers, the system optionally applies a set of filters. These include filtering by publication year, enforcing a minimum confidence threshold for a predicted subfield label (e.g., ensuring that the score for ``Biochemistry'' exceeds 0.6), and selecting only open-access papers.

To improve diversity among the top-ranked results, we implement maximal marginal relevance (MMR)\footnote{Maximal Marginal Relevance}~\cite{carbonell1998use} reranking. MMR balances relevance to the query and dissimilarity among the results, thereby promoting coverage of multiple subtopics. This is particularly useful when the user query is broad or when diversity in retrieved papers is desired.

For each retrieved document, the system returns its title, publication year, authors, venue, DOI (if available), and cosine similarity score relative to the query. If present, a Too Long; Didn't Read (TL;DR) summary is also included. 

The full implementation is provided in a Jupyter notebook at \href{https://github.com/Bocklitz-Lab/Lit2Vec-code/tree/main/example_tasks/recomendation_system}{recomendation\_system} on GitHub.

\subsubsection{Trend Analysis}
\label{sec:trend-analysis}

One of the most immediate applications enabled by Lit2Vec is fine-grained temporal trend analysis across chemical subfields. The dataset's structured full text, standardized Markdown formatting, and paragraph-level segmentation allow targeted extraction of experimental and methodological mentions over time. In addition, its paragraph-level embeddings support both lexical and semantic filtering with minimal infrastructure.

To demonstrate this, we extract and analyze trends in the use of surface-enhanced Raman spectroscopy (SERS) substrates and excitation wavelengths. Using regular expressions applied to paragraphs from relevant sections (e.g., Methods, Instrumentation, Results), we identify mentions of gold and silver nanoparticles co-occurring with SERS. We also track mentions of Raman excitation at 785~nm and 532~nm, restricting attention to sentences containing Raman-relevant context and instrumentation cues (e.g., ``laser'', ``excitation'').

Documents are grouped by publication year using metadata fields (year, published, publicationDate), and binary flags are computed to indicate the presence of specific technologies per paper. Temporal smoothing is applied using a 3-year centered moving average.

\begin{figure}[h]
    \centering
    \begin{minipage}[t]{0.72\textwidth}
        \vspace{0pt}
        \centering
        \includegraphics[width=\linewidth]{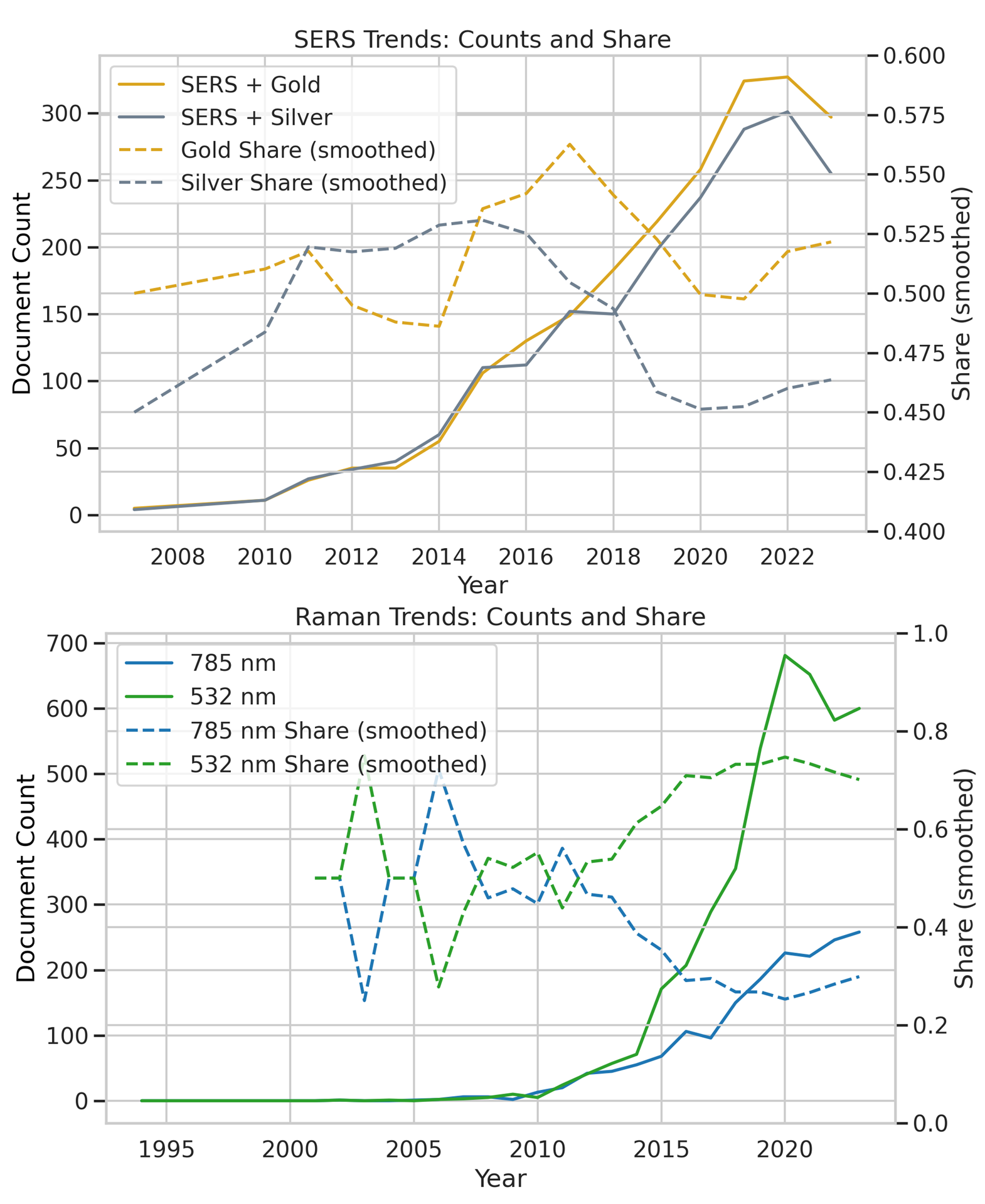}
    \end{minipage}\hfill
    \begin{minipage}[t]{0.25\textwidth}
        \vspace{0pt}
        \caption{Temporal trends in SERS and Raman spectroscopy usage across 582,683 full-text chemistry papers in Lit2Vec. (Top) Document counts and smoothed shares for gold and silver nanoparticle use in SERS. (Bottom) Document counts and normalized shares for 785~nm and 532~nm excitation in Raman experiments. Trend analysis uses paragraph-level filtering and metadata-driven grouping, without fine-tuned models.}
        \label{fig:trend-analysis}
    \end{minipage}
\end{figure}

As shown in Figure~\ref{fig:trend-analysis}, these methods capture meaningful domain signals. For instance, 532~nm excitation has become dominant in Raman experiments over the last decade, although 785~nm is resurging slightly in recent years. In SERS applications, gold and silver substrates are nearly balanced in recent years, with a slight shift toward gold after 2020.

Because Lit2Vec includes paragraph-level embeddings, the same trend analysis workflow can be extended to embedding-based retrieval using tools like FAISS, or combined with domain-specific classifiers or keyword expansion models. These capabilities make Lit2Vec suitable for bibliometric studies, citation-aware recommendation systems, and retrospective analyses of research trends across multiple chemistry subdomains.

A complete implementation of this example is available at \href{https://github.com/Bocklitz-Lab/Lit2Vec-code/tree/main/example_tasks/trend_analysis}{example\_tasks/trend\_analysis} on GitHub.

\section{Discussion}

Lit2Vec should be interpreted not simply as a static corpus release, but as a reproducible workflow for constructing a retrieval-ready chemistry text resource from public upstream sources using conservative, metadata-based license screening. This distinction matters because many existing chemistry text resources are limited to abstracts, insufficiently documented, or not prepared for semantic retrieval at the paragraph level. The present results show that, under the stated access and licensing constraints, the workflow can reconstruct a large screened corpus with stable structure, explicit provenance, and reproducible internal representations for downstream analysis. In that sense, the central validated contribution is the combination of the reconstruction pipeline and the screened study corpus it produces, rather than any single downstream artifact in isolation.

The technical validation results support the reliability of this workflow for downstream use. Across 582{,}683 records, the corpus showed strong schema integrity, substantial metadata completeness, generally high textual quality, exact identifier consistency, reproducible embeddings, and consistent metadata agreement under the study rule for license resolution. Taken together, these checks suggest that the dataset is reliable enough for retrieval, recommendation, trend analysis, and related literature-mining tasks, because the main operational requirements for such uses are addressed jointly rather than one at a time. Equally important, the edge cases are well characterized: short or missing abstracts, naturally short final chunks, and sparse metadata appear in identifiable patterns and can be programmatically filtered when a downstream task requires stricter inclusion criteria. These cases therefore indicate bounded limitations of the available source material and enrichment policy, not instability of the reconstruction workflow itself.

The enrichment results add a second layer of interpretation. Automated summary evaluation shows only moderate lexical overlap between generated TL;DRs and source abstracts, but substantially stronger semantic agreement, indicating that the summaries are more useful as compact semantic surrogates than as extractive replicas. This is consistent with the intended role of the summaries in retrieval and browsing workflows, where concise semantic compression is often more valuable than token-level similarity. The benchmark results further show that the fine-tuned lit2vec\_tldr\_finetuned model is the most practical main summarization component in the workflow, outperforming the tested baselines across ROUGE and BERTScore while remaining operationally efficient enough for large-scale enrichment.

The same interpretation applies to the subfield classifier. Its held-out performance indicates that the embedding-based MLP is suitable for broad organization of the corpus and for retrieval support across major chemistry domains, especially where labeled support is adequate. At the same time, weaker performance on rarer classes means that these labels should be treated as useful navigational metadata rather than as uniformly definitive annotations for all subfields. More broadly, both the TL;DR summaries and predicted subfields should be understood as optional enrichment layers that improve usability for eligible subsets of records, not as universal outputs that define the validity of the corpus itself.

The example applications help clarify why these validation and enrichment results matter in practice. The retrieval-augmented generation, semantic recommendation, and trend-analysis demonstrations show that paragraph-level chunking, dense embeddings, and license-screened metadata are sufficient to support operational downstream workflows rather than only offline validation. In particular, the resource can support evidence-grounded retrieval, similarity-based literature discovery, and longitudinal analysis of technical trends with relatively modest additional infrastructure. This practical usability is an important outcome of the study because it shows that the corpus is not only structurally well-formed, but also aligned with concrete chemistry AI and bibliometric use cases.

Several limitations remain and should shape how Lit2Vec is reused. First, a substantial subset of records has missing abstracts or abstracts below the abstract-enrichment threshold, which limits abstract-level embeddings, summaries, and subfield annotations. However, many of these omissions are expected consequences of upstream source gaps and the enrichment policy, especially where S2ORC records lack abstracts entirely or provide text too short for the abstract-dependent enrichment steps, and should not be interpreted as arbitrary downstream processing failures. Second, exact regeneration remains upstream-dependent: the study corpus is not fully redistributed, and faithful reconstruction depends on continued access to the pinned source release, archived metadata snapshots, and compatible external services. Relatedly, redistribution constraints remain intrinsic to the problem setting, so the contribution is a transparent and reproducible workflow under legal screening rather than an unrestricted public mirror of all source-derived content. Additional uncertainty comes from noisy or imperfect source labels for interdisciplinary documents, possible bias in GPT-4o-supervised pseudo-labels and TL;DR targets, temporal drift in live metadata after collection, and the text-only scope of the resource, which omits figures, chemical structures, and tabular data that are often scientifically important.

Future work should therefore focus on extending coverage and tightening reproducibility without overstating what can be redistributed. The most immediate directions are to recover missing abstracts from additional upstream repositories where legally possible, improve classification performance for rare labels, and expand the framework beyond text-only content to modalities such as figures, tables, or structured chemical representations. Reproducibility could also be strengthened further through more extensively pinned upstream snapshots, archived metadata resolutions, and continued release of metadata, provenance, validation, and other low-risk reproducibility artifacts, alongside any separately vetted permissive subsets that clearly support public redistribution. These steps would not change the core contribution of Lit2Vec, but would increase the completeness and practical reach of the workflow for chemistry-focused AI research.

\section{Conclusions}

Lit2Vec provides a reproducible workflow for constructing and validating a chemistry corpus from S2ORC using conservative, metadata-based license screening, together with schema documentation, reconstruction code, metadata/provenance artifacts, and validation resources that support workflow-level reproducibility. The resulting study corpus combines structured full text, token-aware paragraph chunks, dense embeddings, and optional enrichment layers in a form that is suitable for retrieval-oriented chemistry AI workflows under explicit access and licensing constraints.

The release is intentionally scoped: the full reconstructed corpus is not publicly redistributed, and some enrichment outputs remain subset-dependent because they rely on abstract availability and upstream metadata quality. Within those limits, Lit2Vec establishes a transparent foundation for workflow-level reproducibility and for building downstream systems grounded in license-screened scientific text.

\subsection{Applications and limitations}

The full reconstructed corpus used in this study (582{,}683 records; 277~GB compressed, $\sim$825~GB uncompressed) is not publicly redistributed at present because it is derived from upstream sources with heterogeneous licensing and redistribution conditions. The released materials support workflow-level reproducibility rather than unrestricted corpus redistribution. Researchers seeking to reuse the pipeline should use the released reconstruction workflow to rebuild the corpus from publicly available upstream datasets and metadata services, following the providers' published terms of use, licensing conditions, and normal download policies. Optional API credentials may improve download speed, but are not required.

\textbf{Data Structure and Reconstruction}: The internal corpus representation uses standalone JSON records with text, embeddings, and metadata. The released schema documents this structure so that users can reconstruct compatible records from upstream data and use compatible local records or separately vetted permissive subsets with the same field conventions. Example loading code for reconstructed JSON files:
\begin{verbatim}
    from datasets import load_dataset

    dataset = load_dataset(
         "json",
         data_files={"train": "reconstructed_records/*.jsonl"},
         trust_remote_code=True,)
    
    # Peek at one example
    sample = dataset["train"][0]
    print(sample["metadata"])  # Raw JSON string with title, authors, etc.
    print(sample["predicted_subfield"])  # List of {label, score}
    
    # Extract title from metadata
    import json
    metadata = json.loads(sample["metadata"])
    print("Title:", metadata.get("title"))
    
    # Display subfield predictions
    print("Predicted subfields:")
    for sub in sample["predicted_subfield"]:
        print(f"  {sub['label']}: {sub['score']:.3f}")
    
\end{verbatim}
Loading the full reconstructed corpus in memory requires approximately 900~GB of RAM. Most users should use streaming or sharded loading after reconstructing it locally.
All paragraph vectors are 1024-dimensional, float32, and L2-normalised. A FAISS-based retrieval baseline is provided in section \ref{sec:RAG-Pipeline}. The average paragraph length (167 tokens) allows multiple hits to fit within an 8k-token LLM context window.

The corpus supports tasks such as semantic retrieval, domain-specific LLM fine-tuning, chemistry knowledge graph construction, and subfield prediction. For modeling, predicted\_subfield should be treated as multi-label.

\textbf{Quality Flags and Filtering}: Each record includes validation flags such as short\_abstract, missing\_embeddings, and too\_short\_chunk. For high-recall applications like retrieval, flagged records may be retained. For LLM training or evaluation, users may exclude records with hard fail statuses (see Section~\ref{sec:technical-validation}).

\textbf{Compute Considerations}: A flat inner-product index over the full set of 29M paragraph vectors requires $\sim$120~GB RAM. Optimized variants (e.g., IVF-PQ or HNSW) can reduce this to $\leq$40~GB with minimal recall loss. Full-corpus embedding or fine-tuning benefits from $\geq$40~GB GPUs, though chunk-wise streaming is feasible on consumer-grade hardware.

\textbf{Licensing and Attribution}: Licensing metadata is harmonised from OpenAlex, Unpaywall, and Crossref. Approximately 87.5\% of records in the screened study corpus are CC-BY and 10.7\% are CC-BY-NC. Each JSON includes explicit license fields (unpaywall\_license, crossref\_license, openalex\_license). Users must verify source-specific terms before reconstructing, sharing, or redistributing any derivative outputs.

\noindent
\textbf{Citation}: Users should cite this manuscript and the corresponding code or derivative-resource release when using Lit2Vec materials.

\section{Declarations}

\subsection{Availability of data and materials}
\label{sec:data-availability}

Third parties can deterministically reproduce the screened study corpus and the reported analyses by using the pinned S2AG 2024-12-31 release together with the archived reconstruction manifests, frozen license-resolution metadata, released code, and analysis scripts provided with this work. Minor differences should arise only when later upstream releases are substituted or live metadata services are queried instead of the archived snapshot.

The full reconstructed Lit2Vec corpus analyzed in this study is not publicly redistributed because it is derived from upstream sources with heterogeneous licensing and redistribution conditions. Reproducibility is therefore supported through public release of the reconstruction and validation code, workflow documentation, machine-readable schema and example records, record-level reconstruction manifests, license-screening and license-resolution outputs, technical validation reports, figure- and table-level source data, and legally redistributable derivative resources generated in this work.

All released materials are available at \href{https://huggingface.co/datasets/Bocklitz-Lab/Lit2Vec-dataset}{https://huggingface.co/datasets/Bocklitz-Lab/Lit2Vec-dataset}. The released code repository is available at \href{https://github.com/Bocklitz-Lab/Lit2Vec-code}{Bocklitz-Lab/Lit2Vec-code}. Together, these resources provide the acquisition, preprocessing, chunking, embedding, filtering, and license-screening workflow required to reconstruct Lit2Vec-style records from the same pinned upstream release and associated metadata snapshot. Optional API credentials may improve throughput for some steps but are not required for reproduction.

We do not redistribute the full reconstructed full-text corpus, restricted abstracts, paragraph-level corpus exports derived from non-redistributable source content, embedding vectors derived from non-redistributable corpus content, or other source-derived outputs whose redistribution is not permitted.

The released schema documents the internal record structure used in this study, including both publicly releasable metadata/provenance fields and internal-or-local reconstruction fields such as text, chunk, embedding, and summary fields. Researchers can use this schema to generate compatible local records after reconstructing the corpus from the same public upstream sources. The schema documents these structures for compatibility and reproducibility, but does not by itself imply general public redistribution of all field contents.

Each reconstructed record retains machine-readable license metadata in the fields unpaywall\_license, crossref\_license, and openalex\_license. These fields are provided to support auditing and downstream compliance checks, and users should verify record-level reuse conditions before redistribution or derivative use.

A limited set of auxiliary resources used for model training and supervision is publicly available only where redistribution was separately scoped as appropriate for release:

\begin{itemize}
    \item \textbf{TL;DR Supervision Dataset} (19,992 CC-BY abstracts with GPT-4o-generated summaries)\\
    \href{https://huggingface.co/datasets/Bocklitz-Lab/lit2vec-tldr-bart}{Bocklitz-Lab/lit2vec-tldr-bart}

    \item \textbf{Subfield Classification Dataset} (multi-label annotations for 18 chemistry subfields)\\
    \href{https://huggingface.co/datasets/Bocklitz-Lab/lit2vec-subfield-classifier}{Bocklitz-Lab/lit2vec-subfield-classifier}
\end{itemize}

These released metadata, schema, and validation artifacts support inspection, auditing, and verification of the workflow outputs. Full regeneration depends on access to the same pinned upstream release and archived metadata snapshot distributed with this work.

\subsection{Code availability}
\label{sec:code-availability}
All code used to generate the screened study corpus and reproduce the analyses is open-source and publicly available. A reproducible version of the workflow is released as a simplified, single-machine implementation. While the original pipeline was designed for high-throughput processing (e.g., parallel MongoDB filtering, distributed embedding inference), the released version faithfully replicates the core acquisition, preprocessing, chunking, embedding, and license-screening steps for local reconstruction from pinned upstream resources. Input/output formats, chunking logic, and embedding normalization remain identical. This code release should not be interpreted as a public redistribution of the general corpus or of broad text-derived stores.

\vspace{0.5em}

\noindent\textbf{Main pipeline repository:}  
\begin{itemize}
    \item \href{https://github.com/Bocklitz-Lab/Lit2Vec-code}{Bocklitz-Lab/Lit2Vec-code} --- contains the complete processing pipeline, including:
    \begin{itemize}
        \item \href{https://github.com/Bocklitz-Lab/Lit2Vec-code/blob/main/src/PaperProcessor.py}{PaperProcessor.py}: Parses full-text JSON into structured Markdown with normalized headers and section-aware grouping.
        \item \href{https://github.com/Bocklitz-Lab/Lit2Vec-code/blob/main/src/ChunkProcessor.py}{ChunkProcessor.py}: Performs recursive, token-aware chunking of text for transformer compatibility.
        \item \href{https://github.com/Bocklitz-Lab/Lit2Vec-code/blob/main/src/EmbeddingProcessor.py}{EmbeddingProcessor.py}: Generates embeddings using intfloat/e5-large-v2.
        \item \href{https://github.com/Bocklitz-Lab/Lit2Vec-code/blob/main/src/LicenseValidator.py}{LicenseValidator.py}: Filters reusable content based on license metadata from Unpaywall, OpenAlex, and Crossref.
    \end{itemize}
\end{itemize}

\vspace{0.5em}

\noindent\textbf{Optional enrichment repositories:}
\begin{itemize}
    \item \href{https://github.com/Bocklitz-Lab/lit2vec-tldr-bart}{lit2vec-tldr-bart} --- scripts for TL;DR summarization, including model configs, training metrics, and evaluation logic.
    \item \href{https://github.com/Bocklitz-Lab/lit2vec-subfield-classifier}{lit2vec-subfield-classifier} --- multi-label classification pipeline for scientific subfields, including training, class balancing, and inference.
\end{itemize}

\vspace{0.5em}

\noindent\textbf{Example downstream applications:}
\begin{itemize}
    \item \textit{RAG Pipeline}:  
    \href{https://github.com/Bocklitz-Lab/Lit2Vec-code/tree/main/example_tasks/RAG}{example\_tasks/RAG} --- dense paragraph-level retrieval using FAISS (IndexFlatL2) and answer synthesis with gpt-4o, including inline paragraph ID citations and auto-generated references.
    
    \item \textit{Semantic Paper Recommendation}:  
    \href{https://github.com/Bocklitz-Lab/Lit2Vec-code/tree/main/example_tasks/recomendation_system}{example\_tasks/recommendation\_system} --- cosine similarity search over abstract embeddings using FAISS (IndexFlatIP); supports both paper-to-paper and free-text queries (encoded with intfloat/e5-large-v2), with optional filters and MMR reranking.
    
    \item \textit{Trend Analysis}:  
    \href{https://github.com/Bocklitz-Lab/Lit2Vec-code/tree/main/example_tasks/trend_analysis}{example\_tasks/trend\_analysis} --- regex- and embedding-friendly pipelines for analyzing temporal trends in SERS substrates and Raman excitation wavelengths, using centered 3-year smoothing.
\end{itemize}

\vspace{0.5em}

\noindent\textbf{Trained model checkpoints and interactive demos:}
\begin{itemize}
    \item TL;DR Summarization:
    \begin{itemize}
        \item \href{https://huggingface.co/Bocklitz-Lab/lit2vec-tldr-bart}{lit2vec-tldr-bart} (model)
        \item \href{https://huggingface.co/spaces/Bocklitz-Lab/lit2vec-tldr-bart-space}{lit2vec-tldr-bart-space} (demo)
    \end{itemize}
    \item Subfield Classifier:
    \begin{itemize}
        \item \href{https://huggingface.co/Bocklitz-Lab/lit2vec-subfield-classifier}{lit2vec-subfield-classifier} (model)
        \item \href{https://huggingface.co/spaces/Bocklitz-Lab/lit2vec-subfield-classifier-space}{lit2vec-subfield-classifier-space} (demo)
    \end{itemize}
\end{itemize}

\vspace{0.5em}

\subsection{Competing interests}
The authors declare no competing interests.

\subsection{Funding}
This work is supported by the BMFTR, funding program Photonics Research Germany (13N15466 (LPI-BT1-FSU), 13N15710 (LPI-BT3-FSU), 13N15715 (LPI-BT4-FSU)) and is integrated into the Leibniz Center for Photonics in Infection Research (LPI). The LPI initiated by Leibniz IPHT, Leibniz-HKI, Friedrich Schiller University Jena and Jena University Hospital is part of the BMFTR national roadmap for research infrastructures.

\subsection{Authors' contributions}
Mahmoud Amiri was responsible for the conceptualization and design of the pipeline, software implementation, data curation, validation, analysis, visualization, and preparation of the original manuscript draft. Prof. Dr. Thomas Bocklitz provided supervision, guidance on methodological design, and critical revision of the manuscript. Sara Mostafapourghasrodashti and Jamile Mohammad Jafari contributed to the validation of the study. Author roles follow the CRediT taxonomy; for more detail, see Table~\ref{tab:credit}.

\renewcommand{\arraystretch}{1}
\begin{table}[h!]
    \centering
    \footnotesize
    \caption{Author contributions according to the CRediT taxonomy.}
    \label{tab:credit}
    \begin{tabular}{p{4cm}p{2.2cm}p{2.5cm}p{1.8cm}p{1.8cm}}
    \hline
    \textbf{Role} & \textbf{Mahmoud A.} & \textbf{Thomas B.} & \textbf{Sara M.} & \textbf{Jamile M.J.} \\
    \hline
    Conceptualization & X &  &  &  \\
    Methodology & X & X &  &  \\
    Software & X &  &  &  \\
    Validation & X &  & X & X \\
    Formal Analysis & X &  &  &  \\
    Investigation & X &  &  &  \\
    Data Curation & X &  &  &  \\
    Visualization & X &  &  &  \\
    Writing -- Original Draft & X &  &  &  \\
    Writing -- Review \& Editing & X & X & X & X \\
    Supervision &  & X &  &  \\
    Project Administration &  & X &  &  \\
    \hline
    \end{tabular}
\end{table}

All authors approved the final version of the manuscript.

\subsection{Acknowledgements}
We thank the Allen Institute for AI and the Semantic Scholar team for providing the well-documented S2ORC dataset. We also acknowledge the open-source contributions of Hugging Face, SentenceTransformers, and MongoDB, which made the development of a scalable pipeline possible.


\nocite{*}
\bibliographystyle{unsrtnat}
\bibliography{references_inline}

\section{Supplementary Materials}

\subsection{Related Work}
\label{sec:related-work}
Early efforts in biomedical text mining focused on manually annotated, task-specific datasets. The CHEMDNER corpus~\cite{krallinger2015chemdner} established a benchmark for chemical named entity recognition (NER), featuring approximately 84{,}000 chemical mentions across 10{,}000 PubMed~\cite{pubmed2025} abstracts. The BioCreative V Chemical--Disease Relation (CDR) corpus~\cite{li2016cdr} extended this direction to relation extraction, annotating 1{,}500 abstracts with chemical--disease pairs. NLM-Chem~\cite{islamaj2021nlmchem} further advanced the field by providing high-quality chemical entity annotations in full-text articles. In parallel, domain-specific corpora emerged from patents and other non-journal sources. Notably, the ChEMU 2020 corpus~\cite{he2020chemu} comprises approximately 1{,}500 annotated text segments drawn from around 170 chemical patent documents, with fine-grained entity annotations---such as the roles of chemicals in reactions---and detailed event extraction over procedural steps.

More recently, initiatives have begun to unify multiple NLP tasks within chemistry. ChemRxivQuest~\cite{feng2023chemrxivquest} automatically generates approximately 970 validated question--answer pairs from 155 ChemRxiv preprints across 17 subfields, enabling the development of chemistry-specific question answering (QA) systems. ChemNLP~\cite{choudhary2023chemnlp} provides an open-source library that aggregates and processes large-scale chemical literature, including around 1 million abstracts from arXiv and 100{,}000 compound records from PubChem, with pipelines for tasks such as named entity recognition, classification, clustering, summarization, and text generation.

Recent efforts have also focused on creating large-scale corpora for training general-purpose chemical foundation models. ChemPile~\cite{mirza2025chempile} is a 250~GB, multimodal dataset comprising over 75~billion tokens across educational materials, scientific articles, structured chemical representations (e.g., Simplified Molecular Input Line Entry System (SMILES), Self-Referencing Embedded Strings (SELFIES), International Chemical Identifier (InChI), International Union of Pure and Applied Chemistry (IUPAC)), executable code, reasoning traces, and molecular images.

Broad scientific text collections, such as the Semantic Scholar Open Research Corpus (S2ORC)~\cite{lo2019s2orc, kinney2023semantic}, the PubMed Central Open Access subset~\cite{pmcoa2022}, and CORE~\cite{knoth2023core}, contain millions of articles across disciplines, including chemistry. These corpora have played a pivotal role in enabling large-scale text mining, citation analysis, document classification, and pretraining of general-purpose language models. 

Despite their individual contributions, existing corpora for chemical NLP face common limitations that hinder their integration into modern RAG and semantic search workflows. These include: (i) incomplete or inconsistent full-text coverage; (ii) lack of paragraph-level, token-aware semantic segmentation suitable for transformer-based models; (iii) absence of precomputed dense embeddings for immediate use in retrieval systems; (iv) restrictive or heterogeneous licensing conditions that complicate downstream reuse and distribution; and (v) insufficient metadata normalization across chemistry subdomains. Furthermore, general-purpose scientific datasets lack chemical-specific structuring, while even large-scale efforts like ChemPile are optimized for model pretraining rather than real-time retrieval or question answering. These gaps collectively limit reproducibility, scalability, and legal certainty for building open, domain-aware AI systems in chemistry.

\begin{table*}[t]
    \centering
    \renewcommand{\arraystretch}{1}
    \footnotesize
    \begin{tabular}{p{3.5cm}p{6cm}p{4cm}}
    \hline
    \textbf{Corpus} & \textbf{Description} & \textbf{Limitations} \\
    \hline

    \textbf{CHEMDNER}~\cite{krallinger2015chemdner} &
    84K mentions in 10K PubMed abstracts; strong NER benchmark. &
    Abstracts only; limited license; no embeddings. \\ \hline

    \textbf{CDR}~\cite{li2016cdr} &
    1.5K abstracts with chemical–disease relations; expert-annotated. &
    Small; abstracts only; no embeddings. \\ \hline

    \textbf{NLM-Chem}~\cite{islamaj2021nlmchem} &
    Full-text NER in PMC; good chemical coverage. &
    NER-only; no embeddings or structure. \\ \hline

    \textbf{ChEMU 2020}~\cite{he2020chemu} &
    1.5K patent segments with NER/reaction events. &
    Small; patent-focused; no embeddings. \\ \hline

    \textbf{ChemRxivQuest}~\cite{feng2023chemrxivquest} &
    970 QA pairs from 155 preprints in 17 subfields. &
    Small; QA-only; no full-text. \\ \hline

    \textbf{ChemNLP}~\cite{choudhary2023chemnlp} &
    1M abstracts + 100K PubChem records for NLP. &
    No full-text; mixed licensing; no embeddings. \\ \hline

    \textbf{ChemPile}~\cite{mirza2025chempile} &
    250~GB multimodal data: text, SMILES, code, images. &
    No retrieval structure; no embeddings. \\ \hline

    \textbf{S2ORC}\cite{lo2019s2orc, kinney2023semantic}, \textbf{PMC-OA}\cite{pmcoa2022}, \textbf{CORE}\cite{knoth2023core} &
    Large general corpora with partial full-text and metadata. &
    Not chemistry-specific; license variability. \\ \hline

    \textbf{Lit2Vec} &
    Lit2Vec workflow / screened study corpus constructed from S2ORC; code, schema, and legally safe derivatives released. &
    Full reconstructed corpus not publicly redistributed; text-only. \\ \hline

    \end{tabular}
    \caption{Comparison of chemistry and general scientific corpora by scope and features.}
    \label{tab:chem_corpora_comparison}
\end{table*}

\subsection{System Prompt for Summary Generation and Field Classification}
\label{sec:chemextract-prompt}
The pipeline assembled a high-quality corpus of approximately 19{,}992 chemistry abstracts, each licensed under CC-BY terms. Next, we used GPT-4o with a dedicated system prompt to generate high-quality summaries and field classifications. The system prompt instructed the model as follows:

{\footnotesize\sffamily
\setlength{\parindent}{0pt}
\vspace{2em}
{You are an information-extraction model for chemistry research.} \\
{TASK:} Read chemistry abstracts, output valid JSON only (no explanations, Markdown, or extra text), precisely matching the schema below.

\vspace{1em}

\textbf{General Rules:}
\begin{itemize}[leftmargin=1.5em]
    \item Always include numeric value units, using standard units only.
    \item Don't invent or guess values or details not explicitly stated.
    \item Use only standard units: \%,
    $^\circ$C, K, atm, bar, Pa, kPa, MPa, Torr, psi; h, min, s, ms, µs, ns; L, mL, µL; mol, mmol, µmol, mol L$^{-1}$ (M); g, mg, µg, kg; ppm, ppb, wt \%, vol \%, mol \%; kJ mol$^{-1}$, J mol$^{-1}$, eV; W m$^{-2}$, mW cm$^{-2}$; A cm$^{-2}$, mA cm$^{-2}$, µA cm$^{-2}$; V, mV; S cm$^{-1}$, µS cm$^{-1}$; F g$^{-1}$, mAh g$^{-1}$, mAh cm$^{-2}$; nm, µm, \AA; cm$^{-1}$; quantum yield (\%), etc.
\end{itemize}

\vspace{1em}
\textbf{Summary (1--2 sentences, $\leq$ 50 words):}
\begin{itemize}[leftmargin=1.5em]
    \item Clearly state material/method and main chemical problem.
    \item Clearly state key numeric results with midpoint values and brief significance or application.
    \item Use plain English, active voice, capital first letter, period at the end.
    \item Do not include citations, funding.
    \item The summary must always include at least one explicit numeric midpoint with its unit, unless no numeric data is available. Do not generalize or omit the number if it's in the abstract.
\end{itemize}

\vspace{1em}
\textbf{Field Classification:}
\begin{itemize}[leftmargin=1.5em]
    \item "field\_classification" must be an array of one or more strings.
    \item Each string must exactly match one of the approved main fields below.
    \item If the abstract clearly covers multiple domains, include multiple strings.
    \item {Approved Main Fields (must match exactly):} \\
    \{Catalysis, Organic Chemistry, Polymer Chemistry, Inorganic Chemistry, Materials Science, Analytical Chemistry, Physical Chemistry, Biochemistry, Environmental Chemistry, Energy Chemistry, Medicinal Chemistry, Chemical Engineering, Supramolecular Chemistry, Radiochemistry \& Nuclear Chemistry, Forensic \& Legal Chemistry, Food Chemistry, Chemical Education\}
\end{itemize}

\vspace{1em}
Output Schema:
\begin{verbatim}
{
  "field_classification": [],
  "summary": ""
}
\end{verbatim}

\vspace{1em}
OUTPUT \\
Return only the pure JSON, matching this style. \\
No explanations. No extra text. No formatting.

\vspace{1em}
\textbf{Example}

Input: \\
Abstract: We synthesized a TiO$_2$/g-C$_3$N$_4$ heterojunction that efficiently converts CO$_2$ to CO under visible light ($\lambda > 420$ nm, 100 mW cm$^{-2}$). At 25 $^\circ$C and 1 atm, the catalyst delivered a CO production rate of 115–125 µmol g$^{-1}$ h$^{-1}$ with 85–90 \% CO selectivity over 5 h in the presence of 0.10 M triethanolamine (TEOA) sacrificial agent (pH 7). The apparent quantum yield was 15 $\pm$ 1 \%, and 93 \% of the initial activity was retained after five cycles, indicating good stability. Photoluminescence and time-resolved spectroscopy confirmed suppressed charge recombination. This work highlights a robust, metal-free approach for solar-driven CO production from CO$_2$.

Output:
\begin{verbatim}
{
  "field_classification": ["Catalysis", "Materials Science"],
  "summary": "TiO₂/g-C₃N₄ heterojunction photocatalyst reduces CO₂ to CO under visible light, achieving 87.5 % 
       CO selectivity and 120 µmol g⁻¹ h⁻¹ production rate, with 15 % quantum yield and 93 % activity retention 
       after five cycles."
}
\end{verbatim}
}

\subsection{Subfield Classifier Resources and Label Mapping}
\label{sec:subfield-mapping}

The subfield classifier was trained using chemistry-related categories (this set includes 17 chemistry subfields and a fallback Others class, for a total of 18 labels), each mapped to a unique index as shown in the table below. These indices correspond to the model's output classes during training and evaluation.

\renewcommand{\arraystretch}{1}
\begin{center}
\footnotesize
\begin{tabular}{ll ll ll}
\toprule
\textbf{Subfield} & \textbf{Index} & \textbf{Subfield} & \textbf{Index} & \textbf{Subfield} & \textbf{Index} \\
\midrule
Catalysis & 0 & Physical Chemistry & 6 & Supramolecular Chemistry & 12 \\
Organic Chemistry & 1 & Biochemistry & 7 & Radiochemistry \& Nuclear Chemistry & 13 \\
Polymer Chemistry & 2 & Environmental Chemistry & 8 & Forensic \& Legal Chemistry & 14 \\
Inorganic Chemistry & 3 & Energy Chemistry & 9 & Food Chemistry & 15 \\
Materials Science & 4 & Medicinal Chemistry & 10 & Chemical Education & 16 \\
Analytical Chemistry & 5 & Chemical Engineering & 11 & Others & 17 \\
\bottomrule
\end{tabular}
\end{center}

\subsection{Annotation Validation}
\label{sec:annotation-validation} 

Two domain experts (\emph{Jamile} and \emph{Sara}) independently validated a stratified subset of 25 abstracts each, assessing (i) multi-label field classification and (ii) TL;DR summary quality. Overall ratings were strongly positive. For field classification, average scores were $\mu_{\text{Jamile}} = 4.63 \pm 0.65$ (median $=5$) and $\mu_{\text{Sara}} = 4.88 \pm 0.45$ (median $=5$). For summaries, averages were $\mu_{\text{Jamile}} = 4.38 \pm 0.92$ and $\mu_{\text{Sara}} = 5.00 \pm 0.00$ (all scores on a 1--5 scale).

Experts also provided targeted corrective feedback, revising 28--48\% of items (Jamile) and 8--12\% (Sara), suggesting localized improvements rather than systematic issues. Cross-expert overlap comprised 5 items for field ratings and 4 for summary ratings (aligned by normalized titles). Because overlap was limited and one rater produced a degenerate distribution (all 5's for summaries), inter-rater reliability measures such as Cohen's $\kappa$ and correlation coefficients were undefined.

As an alternative, we report per-expert distributions alongside agreement rates. For field classification, exact agreement was 60\% and agreement within one point was 80\%. For summaries, exact agreement was 75\% and agreement within one point was 100\%. These results indicate broad consistency across experts despite small overlap. 

The full validation results and analysis scripts are provided in our GitHub repository: \href{https://github.com/Bocklitz-Lab/Lit2Vec-code/tree/main/annotation_validation}{annotation\_validation}.

\subsection{Abstractive Summarization: Training, Evaluation, and Benchmarks}
\label{sec:summsup}

This section provides full implementation and evaluation details for the abstractive summarization pipeline described in Section~\ref{sec:abstractive-summarization} of the main text. Benchmark results are reported in Table~\ref{tab:lit2vec_bench}.

We used the curated dataset described in Section~\ref{sec:enrichment-dataset-generation}, comprising 19{,}992 chemistry research abstracts licensed under CC-BY. We adopt the predefined train/validation/test splits provided with the dataset (80/10/10). Inputs are truncated to a maximum of 1{,}024 tokens and targets to 128 tokens. Tokenization uses the Hugging Face tokenizer (modern text\_target API).

We fine-tuned sshleifer/distilbart-cnn-12-6 using the Hugging Face Seq2SeqTrainer on a single NVIDIA GeForce Ray Tracing Texel eXtreme (RTX)~3090 graphics processing unit (GPU) with mixed precision (16-bit floating point (FP16) enabled when Compute Unified Device Architecture (CUDA) is available). The optimizer used was AdamW with a learning rate of $2\times10^{-5}$. Training was performed for 5~epochs with a per-device batch size of~4 and gradient accumulation over~4 steps, resulting in an effective batch size of~16. Evaluation, saving, and logging occurred at fixed intervals, with eval\_steps~$=1000$, save\_steps~$=1000$, and logging\_steps~$=500$. A fixed random seed (42) was applied across Python, NumPy, and PyTorch to ensure reproducibility, with deterministic CuDNN settings enabled where applicable.

During evaluation, the per-device batch size was set to~4. The ``Sec.'' column in Table~\ref{tab:lit2vec_bench} reports the total wall-clock time required to summarize the test set using these settings on a single RTX~3090 GPU.

ROUGE-1, ROUGE-2, ROUGE-L, and ROUGE-Lsum scores were computed using the Hugging Face evaluate package with stemming enabled. Prior to scoring, predictions and references were stripped of whitespace and segmented into sentences using NLTK to match standard ROUGE evaluation practices. All ROUGE scores are reported as percentages, where higher values indicate better overlap with the reference summaries. BERTScore (precision, recall, and F1) was computed in a separate evaluation pass using the evaluate implementation with the RoBERTa-large model and IDF weighting. These scores are also reported in Table~\ref{tab:lit2vec_bench} alongside ROUGE metrics.

\begin{table*}[h]
\centering
\footnotesize
\begin{tabularx}{\textwidth}{
  L
  S[table-format=2.2]
  S[table-format=2.2]
  S[table-format=2.2]
  S[table-format=2.2]
  S[table-format=2.2]
  S[table-format=2.2]
  S[table-format=2.2]
  S[table-format=3.2]
}
\toprule
\textbf{Model} & \textbf{R-1} & \textbf{R-2} & \textbf{R-L} & \textbf{ROUGE-Lsum} & \textbf{BERT P} & \textbf{BERT R} & \textbf{BERT F1} & \textbf{Sec.} \\
\midrule
lit2vec\_tldr\_finetuned        & \bfseries 56.38 & \bfseries 31.00 & \bfseries 44.28 & \bfseries 45.64 & \bfseries 90.81 & \bfseries 92.14 & \bfseries 91.46 & 255.16 \\
distilbart\_cnn\_12\_6\_zeroshot & 37.67 & 15.57 & 26.68 & 30.54 & 87.14 & 88.02 & 87.57 & 318.94 \\
bart\_large\_cnn                & 37.00 & 15.02 & 25.90 & 30.01 & 87.12 & 87.80 & 87.44 & 402.62 \\
lexrank                         & 35.65 & 14.78 & 24.73 & 28.57 & 85.74 & 88.79 & 87.22 & 15.05 \\
lead3                           & 33.74 & 13.79 & 23.53 & 27.44 & 85.45 & 88.06 & 86.71 & \bfseries 13.01 \\
scibert\_extractive             & 33.74 & 13.79 & 23.53 & 27.44 & 85.45 & 88.06 & 86.71 & 24.07 \\
\bottomrule
\end{tabularx}
\caption{Test-set performance across abstractive and extractive summarization models. ROUGE and BERTScore metrics are reported as percentages (higher is better). Sec. indicates total evaluation runtime on a single NVIDIA GeForce RTX~3090 under the default generation settings used by the training script.}
\label{tab:lit2vec_bench}
\end{table*}

The results in Table~\ref{tab:lit2vec_bench} show that our fine-tuned model (lit2vec\_tldr\_finetuned) consistently outperforms all baseline methods across both ROUGE and BERTScore metrics. Compared to the strongest zero-shot abstractive baseline (distilbart\_cnn\_12\_6\_zeroshot), our model improves ROUGE-1 by +18.71 and BERT F1 by +3.89, demonstrating the effectiveness of domain-specific fine-tuning. While extractive methods such as lead3 and lexrank offer significantly faster inference times (13--15 seconds), they lag behind in all quality metrics, highlighting the limitations of purely extractive summarization for dense scientific text. Our model achieves a strong balance of accuracy and efficiency, decoding 36.6\% faster than the larger bart\_large\_cnn baseline while yielding substantially better output quality. Overall, the trade-off between quality and speed favors the fine-tuned DistilBART model for integration in production chemistry pipelines.

\subsection{Qualitative Examples of Abstractive Summarization Performance}
\label{sec:supplementary-summaries}

To qualitatively assess the performance of our fine-tuned summarization model, we present three representative examples from the test set. Each example includes the original abstract, the reference summary generated using GPT-4o (used as the supervision target during fine-tuning), and the predicted summary produced by our fine-tuned DistilBART model. The examples illustrate cases of strong, moderate, and weak performance based on how well the predicted summary aligns with the key information, structure, and factual content of the reference.

\vspace{1em}
\textbf{Example 1: High-Quality Prediction}

\textit{Abstract:} \\
Ultraviolet B (UVB; 290--320~nm) irradiation-induced lipid peroxidation induces inflammatory responses that lead to skin wrinkle formation and epidermal thickening. Peroxisome proliferator-activated receptor (PPAR) $\alpha/\gamma$ dual agonists have the potential to be used as anti-wrinkle agents because they inhibit inflammatory response and lipid peroxidation. In this study, we evaluated the function of 2-bromo-4-(5-chloro-benzo[d]thiazol-2-yl) phenol (MHY 966), a novel synthetic PPAR $\alpha/\gamma$ dual agonist, and investigated its anti-inflammatory and anti-lipid peroxidation effects. The action of MHY 966 as a PPAR $\alpha/\gamma$ dual agonist was also determined in vitro by reporter gene assay. Additionally, 8-week-old melanin-possessing hairless mice 2 (HRM2) were exposed to 150~mJ/cm\textsuperscript{2} UVB every other day for 17 days and MHY 966 was simultaneously pre-treated every day for 17 days to investigate the molecular mechanisms involved. MHY 966 was found to stimulate the transcriptional activities of both PPAR $\alpha$ and $\gamma$. In HRM2 mice, we found that the skins of mice exposed to UVB showed significantly increased pro-inflammatory mediator levels (NF-κB, iNOS, and COX-2) and increased lipid peroxidation, whereas MHY 966 co-treatment down-regulated these effects of UVB by activating PPAR $\alpha$ and $\gamma$. Thus, the present study shows that MHY 966 exhibits beneficial effects on inflammatory responses and lipid peroxidation by simultaneously activating PPAR $\alpha$ and $\gamma$. The major finding of this study is that MHY 966 demonstrates potential as an agent against wrinkle formation associated with chronic UVB exposure.

\textit{Reference Summary (GPT-4o):} \\
MHY 966, a novel synthetic PPAR $\alpha/\gamma$ dual agonist, reduces skin inflammation and lipid peroxidation induced by UVB exposure, showing potential as an anti-wrinkle agent.

\textit{Predicted Summary (Fine-tuned model):} \\
MHY 966, a novel synthetic PPAR $\alpha/\gamma$ dual agonist, reduces UVB-induced lipid peroxidation and inflammatory responses in HRM2 mice exposed to 150~mJ/cm\textsuperscript{2} UVB over 17 days, showing potential as an anti-wrinkle agent.

\vspace{1em}
\textbf{Example 2: Moderate-Quality Prediction}

\textit{Abstract:} \\
Recent clinical trials of small interfering RNAs (siRNAs) highlight the need for robust delivery technologies that will facilitate the successful application of these therapeutics to humans. Arguably, cell targeting by conjugation to cell-specific ligands provides a viable solution to this problem. Synthetic RNA ligands (aptamers) represent an emerging class of pharmaceuticals with great potential for targeted therapeutic applications. For targeted delivery of siRNAs with aptamers, the aptamer-siRNA conjugate must be taken up by cells and reach the cytoplasm. To this end, we have developed cell-based selection approaches to isolate aptamers that internalize upon binding to their cognate receptor on the cell surface. Here we describe methods to monitor for cellular uptake of aptamers. These include: (1) antibody amplification microscopy, (2) microplate-based fluorescence assay, (3) a quantitative and ultrasensitive internalization method ("QUSIM") and (4) a way to monitor for cytoplasmic delivery using the ribosome inactivating protein-based (RNA-RIP) assay. Collectively, these methods provide a toolset that can expedite the development of aptamer ligands to target and deliver therapeutic siRNAs in vivo.

\textit{Reference Summary (GPT-4o):} \\
The study addresses the targeted delivery of siRNAs using aptamers, with methods to monitor cellular uptake and cytoplasmic delivery of aptamer-siRNA conjugates, potentially enhancing therapeutic applications.

\textit{Predicted Summary (Fine-tuned model):} \\
Cell-based selection approaches for aptamer ligands for targeted delivery of small interfering RNAs are developed using cell-based methods like antibody amplification microscopy and microplate-based fluorescence assays to monitor cellular uptake and cytoplasmic delivery, aiding in the development of aptamer-siRNA conjugates for therapeutic applications.

\vspace{1em}
\textbf{Example 3: Lower-Quality Prediction}

\textit{Abstract:} \\
This study evaluated the chemical composition, antioxidant, anti-inflammatory and anticancer activities of a \textit{Euphorbia hirta} L. extract. The antioxidant activities of whole \textit{E. hirta} ethanol extract were determined by electron spin resonance spectrophotometric analysis of 1,1-diphenyl-2-picryl-hydrazyl (DPPH), hydroxyl, and alkyl radical levels and by using an online high-performance liquid chromatography (HPLC)-2,2'-azino-bis(3-ethylbenzothiazoline-6-sulfonic acid) assay. The \textit{E. hirta} ethanol extract (0.5~mg/mL) exhibited DPPH-scavenging activity of 61.19\% $\pm$ 0.22\%, while the positive control (0.5~mg/mL ascorbic acid) had 100\% $\pm$ 0.22\% activity. The concentration of the extract required to trap 50\% of DPPH (IC50) was 0.205~mg/mL. Online HPLC analysis of the extract also showed strong antioxidant activity. The anti-inflammatory activity of the \textit{E. hirta} extract was assessed in lipopolysaccharide-induced RAW 264.7 macrophages. The anti-inflammatory activity was highest in the presence of 200~µg/mL \textit{E. hirta} extract, and nitric oxide production was decreased significantly ($p < 0.05$). The extract also showed selective anticancer activity at a concentration of 100~µg/mL ($p < 0.05$). These results indicated that \textit{E. hirta} may warrant further investigation for the development of antioxidant, anti-inflammatory, and anticancer herbal medications.

\textit{Reference Summary (GPT-4o):} \\
The study evaluated \textit{Euphorbia hirta} extract for antioxidant activity, with a DPPH-scavenging rate of 61.19\% at 0.5~mg/mL and an IC50 of 0.205~mg/mL, and observed significant anti-inflammatory and anticancer activities in cell models.

\textit{Predicted Summary (Fine-tuned model):} \\
\textit{Euphorbia hirta} L. extract exhibits strong antioxidant and anti-inflammatory activities, with 61.19\% DPPH scavenging activity at 0.205~mg/mL and selective anticancer activity at 100~µg/mL, indicating potential for herbal medicine development.

\subsection{Topic Classification: Training, Evaluation, and Benchmarks}
\label{sec:supp-topic-classification}

The classifier was trained on the enrichment dataset described in Section~\ref{sec:enrichment-dataset-generation}, comprising 39{,}900 chemistry texts (abstracts and corresponding TL;DR summaries) and multi-label subfield annotations. The taxonomy covers 18 labels, as described in Section~\ref{sec:subfield-mapping}.

We encode each text using the intfloat/e5-large-v2 SentenceTransformer~\cite{wang2022text}, optimized for semantic similarity. Each document is mapped to a 1{,}024-dimensional, unit-normalized vector $\mathbf{x}\in\mathbb{R}^{1024}$ (i.e., $\lVert \mathbf{x}\rVert_2 = 1$). No additional text normalization beyond the embedding model's default preprocessing is applied.

Embeddings serve as input to a feedforward multi-layer perceptron (MLP) implemented in TensorFlow~\cite{abadi2016tensorflow}. The network has two hidden layers of 256 units each with ReLU activations, batch normalization for training stability, and dropout (rate $=0.3$) after each hidden layer. The output layer uses elementwise sigmoid activations to produce per-label probabilities over $C=18$ subfields:
\[
\hat{\mathbf{y}} = \sigma\!\left(W_3\,\phi\!\big(\mathrm{BN}_2(\cdot)\big) + \mathbf{b}_3\right), \quad
\phi(\mathbf{z})=\max(0,\mathbf{z}), \quad \hat{\mathbf{y}}\in[0,1]^C.
\]
Multi-label assignments are obtained by thresholding $\hat{\mathbf{y}}$ at a global decision threshold $\tau=0.5$.

Training uses weighted binary cross-entropy to mitigate label imbalance. Let $y_{\ell}\in\{0,1\}$ and $\hat{y}_{\ell}\in[0,1]$ denote the ground-truth indicator and predicted probability for label $\ell\in\{1,\dots,C\}$. With $N$ training samples and $P_{\ell}$ positives for label $\ell$, we set a per-label weight
\[
w_{\ell} = \frac{N - P_{\ell}}{P_{\ell}} \quad \text{(ratio of negatives to positives).}
\]
The loss for one instance is
\[
\mathcal{L} = \frac{1}{C}\sum_{\ell=1}^{C} w_{\ell}\,\Big[-\,y_{\ell}\log(\hat{y}_{\ell}) - (1-y_{\ell})\log(1-\hat{y}_{\ell})\Big].
\]

Optimization uses Adam (learning rate $=10^{-3}$), mini-batches of size 32, for up to 50 epochs. We employ early stopping (patience $=5$) on validation loss and reduce the learning rate on plateau (factor $=0.5$, patience $=3$). Model parameters yielding the best validation loss are retained.

We use 5-fold cross-validation on the pooled training and validation data (non-stratified due to the multilabel setting). The final test results are reported at a fixed decision threshold of $\tau=0.5$. Following standard practice, we report micro-averaged F1 (aggregating decisions across labels), macro-averaged F1 (unweighted mean across labels), weighted-average F1 (weighted by label support), and samples-average F1 (averaged over instances).

\begin{figure}[H]
    \centering
    \includegraphics[width=\textwidth]{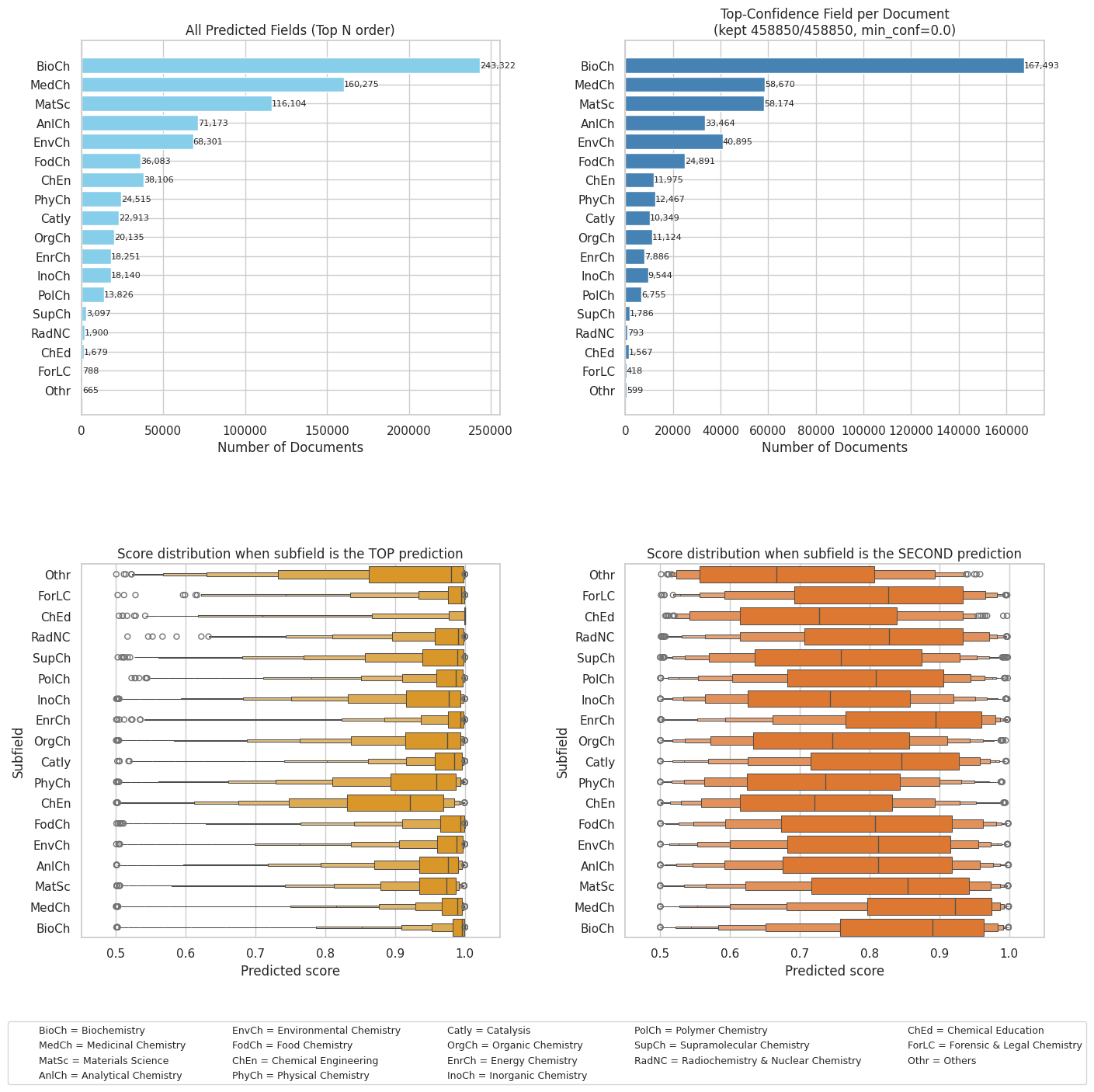}
    \caption{
        Analysis of predicted research subfields from document-level classification.
        \textbf{Top row:} (left) counts of all predicted subfields across documents and 
        (right) counts of only the top-confidence subfield per document.  
        \textbf{Bottom row:} distribution of prediction scores for each subfield when it appears as 
        (left) the top-predicted category and (right) the second-predicted category (if applicable).
        Subfield names are abbreviated (legend) for clarity.
    }
    \label{fig:s-subfield-analysis}
\end{figure}

Table~\ref{tab:s-classification-report} provides the per-label precision, recall, F1-score, and support on the held-out test set (threshold $\tau=0.5$). Consistent with the main text, frequent subfields (e.g., \textit{Biochemistry}, \textit{Medicinal Chemistry}, \textit{Materials Science}) achieve higher F1-scores, while underrepresented domains (e.g., \textit{Chemical Engineering}, \textit{Supramolecular Chemistry}) remain challenging.

\renewcommand{\arraystretch}{1}
\begin{table}[ht]
\footnotesize
\centering
\caption{Per-label classification report on the test set (threshold $\tau=0.5$). Macro average is the unweighted mean across labels; micro average aggregates decisions across labels; weighted average weights by label support; samples average is the average across instances.}
\label{tab:s-classification-report}
\begin{tabular}{lrrrr}
\toprule
Label & Precision & Recall & F1-score & Support \\
\midrule
Catalysis                          & 0.83 & 0.78 & 0.80 & 197 \\
Organic Chemistry                  & 0.85 & 0.60 & 0.70 & 245 \\
Polymer Chemistry                  & 0.80 & 0.65 & 0.72 & 120 \\
Inorganic Chemistry                & 0.84 & 0.62 & 0.71 & 203 \\
Materials Science                  & 0.82 & 0.78 & 0.80 & 917 \\
Analytical Chemistry               & 0.87 & 0.60 & 0.71 & 633 \\
Physical Chemistry                 & 0.78 & 0.53 & 0.63 & 240 \\
Biochemistry                       & 0.89 & 0.94 & 0.92 & 2106 \\
Environmental Chemistry            & 0.84 & 0.75 & 0.79 & 508 \\
Energy Chemistry                   & 0.75 & 0.83 & 0.79 & 166 \\
Medicinal Chemistry                & 0.88 & 0.76 & 0.82 & 1343 \\
Chemical Engineering               & 0.75 & 0.41 & 0.53 & 413 \\
Supramolecular Chemistry           & 0.68 & 0.68 & 0.68 & 34 \\
Radiochemistry \& Nuclear Chemistry & 0.71 & 0.60 & 0.65 & 20 \\
Forensic \& Legal Chemistry        & 0.62 & 0.81 & 0.70 & 16 \\
Food Chemistry                     & 0.83 & 0.83 & 0.83 & 282 \\
Chemical Education                 & 0.85 & 0.85 & 0.85 & 20 \\
Others                             & 0.88 & 0.79 & 0.83 & 19 \\
\midrule
Micro avg                          & 0.85 & 0.77 & 0.81 & 7482 \\
Macro avg                          & 0.80 & 0.71 & 0.75 & 7482 \\
Weighted avg                       & 0.85 & 0.77 & 0.80 & 7482 \\
Samples avg                        & 0.87 & 0.79 & 0.80 & 7482 \\
\bottomrule
\end{tabular}
\end{table}

\begin{sidewaysfigure}
    \centering
    \includegraphics[width=\textheight]{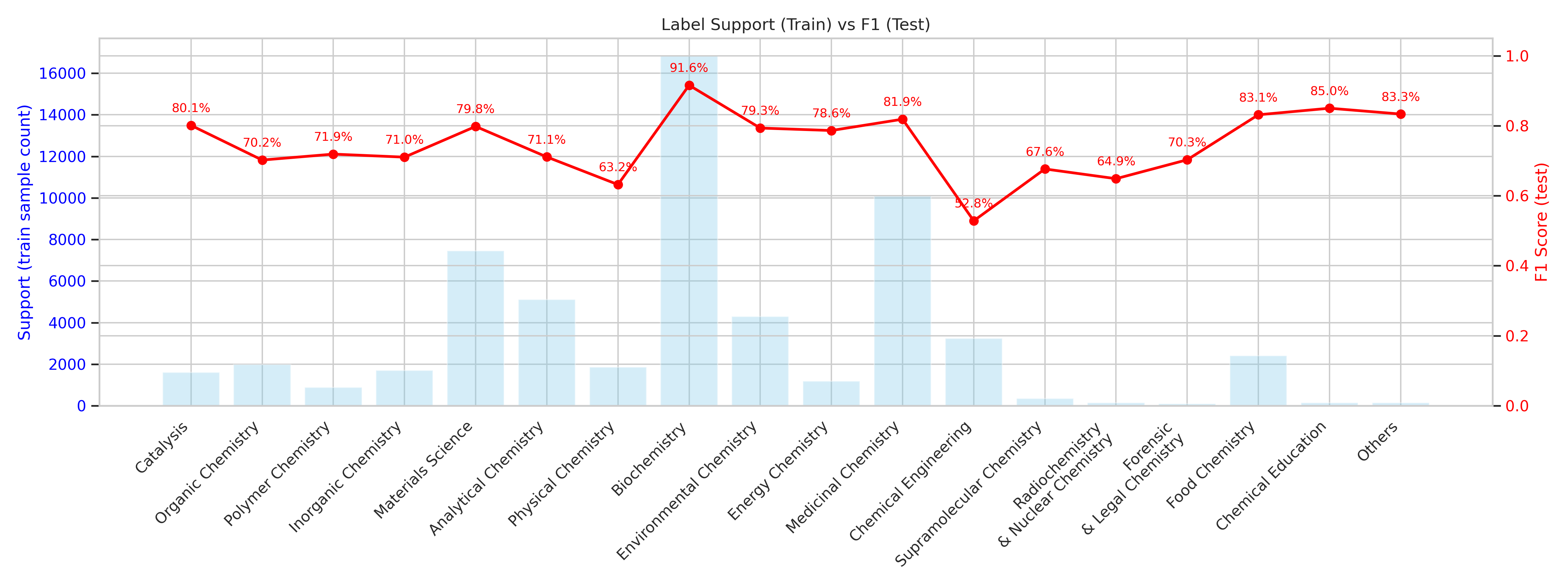}
    \caption{Label support in the training set (bars, left axis) versus F1-score on the test set (red line, right axis) for each subfield. Frequent subfields such as \textit{Biochemistry} and \textit{Medicinal Chemistry} achieve high F1-scores, whereas underrepresented subfields including \textit{Chemical Engineering} and \textit{Supramolecular Chemistry} remain challenging.}
    \label{fig:s-support-vs-f1}
\end{sidewaysfigure}

Figure~S\ref{fig:s-subfield-analysis} summarizes the distribution of predicted subfields and confidence scores. Figure~S\ref{fig:s-support-vs-f1} relates label support in the training set to test-set F1-scores, highlighting the impact of representation on performance.

At inference, each abstract is embedded as above and passed through the MLP to obtain per-label probabilities. Unless otherwise specified, labels with $\hat{y}_{\ell}\geq\tau$ (with $\tau=0.5$) are assigned. For applications that prefer higher precision (e.g., curated knowledge graphs), users may raise $\tau$ or apply label-specific thresholds calibrated on validation data.

\subsection{Data Records: Schema and Field Dictionaries}
\label{sec:schema}

\renewcommand{\arraystretch}{1}
\begin{tabular}{p{4.2cm}p{10.8cm}}
\toprule
\footnotesize
\textbf{Key} & \textbf{Description} \\
\midrule
\_id (string) & Internal unique identifier (optional). \\
corpusid (integer) & Mirror of corpus\_id for traceability (optional). \\
externalids (object) & External IDs (any subset): DOI, PubMed, PubMedCentral, ArXiv, MAG, DBLP, ACL, CorpusId. \\
url (string) & Canonical landing page (e.g., Semantic Scholar). \\
title (string) & Full paper title. \\
authors (array) & List of \{authorId (string|int), name (string)\}. \\
venue (string|null) & Journal or conference name. \\
publicationvenueid (string|null) & Venue identifier if available. \\
year (integer) & Publication year. \\
referencecount, citationcount, influentialcitationcount (integer) & Citation metrics. \\
isopenaccess (boolean) & Open access flag. \\
s2fieldsofstudy (array) & \{category (string), source (string)\}. \\
publicationtypes (string|array|null) & Publication type(s). \\
publicationdate (string|null) & ISO 8601 date. \\
journal (object|null) & \{name (string), volume (string|null), pages (string|null)\}. \\
\bottomrule
\label{tab:metadata}
\end{tabular}

The internal corpus used in this study was represented as individual JSON records named \{corpus\_id\}.json. We release the machine-readable JSON Schema and example record structure so that researchers can reconstruct compatible local records without requiring redistribution of the full corpus.

For an overview of the top-level record fields, see Table~\ref{tab:data-records}.

\renewcommand{\arraystretch}{1}
\begin{table}[H]
\centering
\footnotesize
\begin{tabular}{p{3.3cm}p{2.5cm}p{9.2cm}}
\toprule
\textbf{Field} & \textbf{Type} & \textbf{Description} \\
\midrule
schema\_version & string & Schema version (e.g., "1.0"). \\
corpus\_id & integer & Primary identifier (S2ORC CorpusId) for traceability and file naming. \\
metadata & object & Bibliographic metadata from S2ORC including title, authors, venue, year, URLs, and external IDs; see Table \ref{tab:metadata}. \\
abstract & string & Original abstract text from S2ORC. \\
fulltext & string & Reconstructed full text (plain text). \\
paragraphs & array[string] & Paragraph-level text units used for retrieval (aligned with embeddings by index). \\
embeddings & array[array[float]] & Paragraph embeddings (model: intfloat/e5-large-v2; 1024-D, float32). \\
abstract\_embedding & array[float] & Embedding for the abstract (same model/dtype; length 1024). \\
predicted\_subfield & object & Map from subfield label to confidence score in $[0,1]$ (17 chemistry subfields + Others; label mapping in section \ref{sec:subfield-mapping}). \\
tldr & string & Two-sentence abstractive summary (DistilBART-CNN adapted to chemistry). \\
unpaywall\_license & object|null & License metadata from Unpaywall. \\
crossref\_license & object|null & License metadata from Crossref. \\
openalex\_license & object|null & License metadata from OpenAlex. \\
\bottomrule
\end{tabular}
\caption{Top-level fields for each JSON record. Complete nested field dictionaries and a machine-readable schema are provided in the Supplementary Information.}
\label{tab:data-records}
\end{table}

At the top level, a record is an object with the following required keys:
schema\_version, corpus\_id, metadata, abstract, fulltext, paragraphs, embeddings, unpaywall\_license, crossref\_license, openalex\_license.
Optional keys: , abstract\_embedding, tldr, predicted\_subfield.

The field paragraphs is an array of strings of length $N \geq 1$, and embeddings is a parallel array of length $N$, where each element is an array of 1024 \textit{float32} values generated with the intfloat/e5-large-v2 model; by construction, index $i$ in embeddings corresponds to index $i$ in paragraphs. The field abstract\_embedding is a single array of 1024 \textit{float32} values (optional), while predicted\_subfield is an object mapping strings to confidence scores in the range $[0,1]$. The license metadata fields unpaywall\_license, crossref\_license, and openalex\_license are represented as objects when available, or as null otherwise.

Each license source (Unpaywall, Crossref, OpenAlex) is represented as an object (or null if unavailable). We preserve the upstream structure for provenance. Typical keys include: license or best\_oa\_location.license (string), url (string), and where provided, embargo/host/type fields. Users should consult the upstream license URL to verify reuse terms.

Paragraph text is produced by a token-aware recursive splitter with overlap to preserve coherence in downstream retrieval. For each paragraph $p_i$, we store an embedding $e_i \in \mathbb{R}^{1024}$ computed with intfloat/e5-large-v2. Embeddings are serialized as arrays of \textit{float32}. We guarantee:
\begin{enumerate}
  \item $\lvert paragraphs \rvert = \lvert embeddings \rvert = N$.
  \item $e_i$ encodes paragraphs[i] (same index).
  \item Model/dtype are fixed across all records (1024-D, float32).
\end{enumerate}

\noindent\textbf{Truncated example JSON record:}
\begin{lstlisting}[basicstyle=\ttfamily\color{black}\small]
{
  "schema_version": "1.0",
  "corpus_id": 37254803,
  "metadata": { "title": "...", "year": 2016,
    "externalids": { "DOI": "10.xxxx/xxxxx" }, "url": "https://..." },
  "abstract": "Epigallocatechin gallate ...",
  "fulltext": "# Protective effect ...",
  "paragraphs": [37254803P0:"passage: ...", "..."],
  "embeddings": [37254803P0:[0.0123, -0.0456, ...], "..."],
  "abstract_embedding": [0.0345, -0.0789, 0.0123, "..."],
  "predicted_subfield": {"Biochemistry": 0.997, "Medicinal Chemistry": 0.759},
  "tldr": "EGCG reduces lipid peroxidation ...",
  "unpaywall_license": {"best_oa_location": {"license": "cc-by"}, "..."},
  "crossref_license": {"license": "http://creativecommons.org/licenses/by/4.0/", "..."},
  "openalex_license": null
}
\end{lstlisting}

A complete, untruncated example JSON record is provided in the code repository at examples/record\_full.json. It illustrates the full nested structures for metadata and license objects, the alignment between paragraphs and embeddings, and the numerical precision of stored vectors, without implying that the full corpus is redistributed.

\subsection{Schema \& Structural Validation}
\label{sec:Schema-validation}
We developed a schema-level validation framework to ensure all dataset records conform to a consistent, machine-readable standard. This is critical for maintaining data integrity, supporting reliable downstream processing, and ensuring compatibility with external tools. The validator enforces the Draft 2020-12 JSON Schema specification for Lit2Vec-style records, with fully deterministic and reproducible validation outcomes.
In large datasets, small inconsistencies—such as missing fields, incorrect types, or mis-sized arrays—can cause downstream errors, misinterpretations, or ingestion failures. Schema validation addresses this by enforcing structural consistency, semantic correctness (e.g., plausible publication years), and interoperability with standard workflows and repositories. It also ensures auditability through reproducible, timestamped error reporting. Tables~\ref{tab:schema-validation-flags} and \ref{tab:schema-validation-examples} summarize the validation rules and provide output examples, respectively.

\renewcommand{\arraystretch}{1}
\footnotesize
\begin{longtable}{p{3cm} p{5cm} p{6cm}}
\caption{Schema Validation Rules and Diagnostic Flags for Lit2Vec-Style Records} \label{tab:schema-validation-flags} \\
\hline
\textbf{Validation Rule} & \textbf{Description} & \textbf{Flag (on failure)} \\
\hline
\endfirsthead

\hline
\textbf{Validation Rule} & \textbf{Description} & \textbf{Flag (on failure)} \\
\hline
\endhead

\hline \multicolumn{3}{r}{Continued on next page} \\
\endfoot

\hline
\endlastfoot
Top-level Required Fields & 
Must include: schema\_version, corpus\_id, metadata, abstract, paragraphs, embeddings, tldr, abstract\_embedding, predicted\_subfield. & 
missing\_<field> — Raised if any required field is missing (e.g., missing\_abstract). \\

Additional Properties & 
Extra fields not defined in the schema are disallowed at all levels (top-level and nested). & 
additional\_property\_<field> — Raised when unexpected field is present (e.g., additional\_property\_validation\_info). \\

Type Enforcement & 
Fields must conform to expected types: strings, integers, booleans, arrays, or objects. & 
type\_mismatch\_<field> — Type does not match schema definition (e.g., string provided where object expected). \\

Fixed-Length Arrays & 
embeddings and abstract\_embedding must be numeric arrays of exactly length 1024. & 
too\_short\_<field> or too\_long\_<field> — Array length is incorrect (e.g., too\_short\_abstract\_embedding). \\

Paragraph Key Pattern & 
Keys in paragraphs and embeddings must match regex \verb|^\d+P\d+$|. & 
pattern\_violation\_<field> — Raised when key format is invalid (e.g., \verb|1Para2|). \\

Field Format Constraints & 
Fields like metadata.url must be valid URIs; publicationdate must be ISO 8601 date. & 
invalid\_format\_<field> — Raised for malformed dates or URLs (e.g., invalid\_format\_metadata\_url). \\

Nullability Rules & 
Some fields may be null (e.g., publicationvenueid), others must not be. & 
type\_mismatch\_<field> — Raised if null is used in a non-nullable field. \\

Required Metadata Fields & 
metadata must include: title, authors (with name), venue, year. & 
missing\_<field> — Raised for any missing required metadata subfield (e.g., missing\_metadata\_title). \\

Numeric Ranges in Metadata & 
year must be between 1800–2100; citation/reference counts must be $\geq 0$. & 
value\_below\_minimum\_<field> or value\_above\_maximum\_<field> — Value out of bounds (e.g., value\_below\_minimum\_metadata\_year). \\

Authors Field Structure & 
Each author must include name; no extra fields allowed. & 
missing\_name — Name missing in author object; additional\_property\_<field> — Extra field present. \\

Abstract and TLDR & 
abstract and tldr must be strings. & 
type\_mismatch\_<field> — Raised if value is not a string (e.g., array given instead). \\

Paragraph Content & 
Each key in paragraphs must map to a string value. & 
type\_mismatch\_paragraphs — Raised when a paragraph value is not a string. \\

Controlled Vocabulary & 
predicted\_subfield must contain only known keys, with float values in the range [0.0, 1.0]. & 
type\_mismatch\_predicted\_subfield, value\_below\_minimum\_predicted\_subfield, or value\_above\_maximum\_predicted\_subfield. \\

Validation Outcome & 
A record is marked "pass" only if it satisfies all schema constraints without triggering any flags. Otherwise, it is marked "fail". & 
Not a flag — Outcome is tracked via status: "pass" or "fail"; includes summary and error count. \\
\hline
\end{longtable}

\renewcommand{\arraystretch}{0.8}
\begin{table}[H]
    \centering
    \footnotesize    
    \caption{Example schema validation outputs: failing (left) and passing (right).}
    \label{tab:schema-validation-examples}
    \begin{tabular}{p{8cm} p{6cm}}
    \toprule
    \textbf{Example Failing Record} & \textbf{Example Passing Record} \\
    \midrule
    \begin{minipage}[t]{\linewidth}
    \raggedright\ttfamily
    "schema\_validation": \{\\
    \quad "status": "fail",\\
    \quad "summary": "1 schema validation error(s) found.",\\
    \quad "metrics": \{\\
    \quad\quad "validation\_errors": [\\
    \quad\quad\quad \{\\
    \quad\quad\quad\quad "message": "[] is too short",\\
    \quad\quad\quad\quad "field": "abstract\_embedding",\\
    \quad\quad\quad\quad "schema\_path": "properties/abstract\_embedding/minItems",\\
    \quad\quad\quad\quad "validator": "minItems"\\
    \quad\quad\quad \}\\
    \quad\quad ],\\
    \quad\quad "flags": \{\\
    \quad\quad\quad "too\_short\_abstract\_embedding": true\\
    \quad\quad \}\\
    \quad \},\\
    \quad "checked\_at": "2025-07-30T23:41:30.136585"\\
    \}
    \end{minipage}
    &
    \begin{minipage}[t]{\linewidth}
    \raggedright\ttfamily
    "schema\_validation": \{\\
    \quad "status": "pass",\\
    \quad "summary": "Document conforms to the JSON schema.",\\
    \quad "metrics": \{\\
    \quad\quad "validation\_errors": [],\\
    \quad\quad "flags": \{\}\\
    \quad \},\\
    \quad "checked\_at": "2025-07-30T12:08:22.875185"\\
    \}
    \end{minipage}
    \\ 
    \bottomrule
    \end{tabular}
    \end{table}

\subsection{Metadata Validation}
\label{sec:Metadata-validation}

To ensure structural integrity, completeness, and semantic correctness of the dataset's bibliographic records, we implemented a strict metadata validation pipeline. Each record is checked against required field definitions, expected types, semantic rules (e.g., date consistency, author structure), and dataset-specific constraints. The pipeline distinguishes between hard errors (e.g., type mismatches) and soft warnings (e.g., empty strings or missing licenses), enabling both robust schema enforcement and fine-grained data curation.

Validation covers presence and typing of critical fields such as title, authors, year, venue, publicationdate, and s2fieldsofstudy, among others. Fields like authors are recursively validated to ensure proper structure (e.g., non-empty names and valid author IDs), while open-access metadata is cross-checked against expected license fields. Publication dates are parsed in a timezone-aware fashion and compared with the declared year. Discrepancies and future-dated entries are flagged. Tables~\ref{tab:Metadata-validation-flags} and \ref{tab:Metadata-validation-examples} summarize the validation rules and provide output examples, respectively.

\renewcommand{\arraystretch}{1}
\footnotesize
\begin{longtable}{p{4.5cm} p{5.5cm} p{4.5cm}}
\caption{Metadata Validation Rules and Diagnostic Flags\label{tab:Metadata-validation-flags}}\\
\hline
\textbf{Validation Rule} & \textbf{Description} & \textbf{Flag} \\
\hline
\endfirsthead

\hline
\textbf{Validation Rule} & \textbf{Description} & \textbf{Flag} \\
\hline
\endhead

\hline \multicolumn{3}{r}{Continued on next page} \\
\endfoot

\hline
\endlastfoot

Missing Required Fields & Required metadata fields are missing. & missing:\textless field\textgreater \\

Type Mismatch & Field type does not match expected type (e.g., string vs int). & type:\textless field\textgreater:\textless found\_type\textgreater\_to\_\textless expected\_type\textgreater \\

Empty Fields & Field is present but empty (e.g., empty string, list, or dict). & empty:\textless field\textgreater \\

Title Too Short & Title has fewer than 5 characters (possible placeholder). & title\_short \\

Malformed Author Entries & Author missing name or ID, or not a valid dict. & authors\_malformed \\

Publication Year Mismatch & Parsed publicationdate year disagrees with year field. & year\_vs\_date \\

Future Publication Date & Publication date is after current Coordinated Universal Time (UTC) date. & date\_in\_future \\

Invalid Date Format & publicationdate is unparseable. & date\_bad\_format \\

Year Out of Range & year is outside 1800–(current\_year + 1). & year\_out\_of\_range \\

Missing Chemistry FOS & No s2fieldsofstudy entry with category "Chemistry". & fos\_no\_chemistry \\

Empty External IDs & All external ID fields (e.g., DOI) are empty or missing. & externalids\_empty \\

Invalid Publication Types & One or more publicationtypes items are not valid non-empty strings. & pubtypes\_bad\_item \\

Missing OA Licenses & isopenaccess = true but no license fields are present. & oa\_missing\_licenses \\
\hline

\end{longtable}

\renewcommand{\arraystretch}{0.8}
\begin{table}[H]
    \centering
    \footnotesize  
    \caption{Example metadata validation outputs: warning (left) and passing (right).}
    \label{tab:Metadata-validation-examples}
    \begin{tabular}{p{7cm} p{7cm}}
    \toprule
    \textbf{Warning Record} & \textbf{Passing Record} \\
    \midrule
    \begin{minipage}[t]{\linewidth}
    \raggedright\ttfamily
    "metadata\_validation": \{\\
    \quad "status": "warn",\\
    \quad "summary": "Metadata field \& content validation",\\
    \quad "metrics": \{\\
    \quad\quad "error\_count": 0,\\
    \quad\quad "warning\_count": 1,\\
    \quad\quad "errors": [],\\
    \quad\quad "warnings": [\\
    \quad\quad\quad "empty:venue"\\
    \quad\quad ],\\
    \quad\quad "diagnostics": \{\\
    \quad\quad\quad "missing\_fields": [],\\
    \quad\quad\quad "wrong\_type\_fields": [],\\
    \quad\quad\quad "empty\_fields": [\\
    \quad\quad\quad\quad "venue"\\
    \quad\quad\quad ],\\
    \quad\quad\quad "malformed\_authors": [],\\
    \quad\quad\quad "title\_short": false,\\
    \quad\quad\quad "year\_mismatch": false,\\
    \quad\quad\quad "future\_date": false,\\
    \quad\quad\quad "date\_issue": null,\\
    \quad\quad\quad "no\_chemistry\_field": false,\\
    \quad\quad\quad "oa\_missing\_licenses": []\\
    \quad\quad \}\\
    \quad \},\\
    \quad "checked\_at": "2025-07-30T12:10:51.411572"\\
    \}
    \end{minipage}
    &
    \begin{minipage}[t]{\linewidth}
    \raggedright\ttfamily
    "metadata\_validation": \{\\
    \quad "status": "pass",\\
    \quad "summary": "Metadata field \& content validation",\\
    \quad "metrics": \{\\
    \quad\quad "error\_count": 0,\\
    \quad\quad "warning\_count": 0,\\
    \quad\quad "errors": [],\\
    \quad\quad "warnings": [],\\
    \quad\quad "diagnostics": \{\\
    \quad\quad\quad "missing\_fields": [],\\
    \quad\quad\quad "wrong\_type\_fields": [],\\
    \quad\quad\quad "empty\_fields": [],\\
    \quad\quad\quad "malformed\_authors": [],\\
    \quad\quad\quad "title\_short": false,\\
    \quad\quad\quad "year\_mismatch": false,\\
    \quad\quad\quad "future\_date": false,\\
    \quad\quad\quad "date\_issue": null,\\
    \quad\quad\quad "no\_chemistry\_field": false,\\
    \quad\quad\quad "oa\_missing\_licenses": []\\
    \quad\quad \}\\
    \quad \}\\
    \}
    \end{minipage}
    \\
    \bottomrule
    \end{tabular}
    \end{table}
    
\subsection{License Validation}
\label{sec:License-validation}

To support record-level license provenance review and machine-readable compliance screening, we implemented an automated license validation pipeline that normalizes, compares, and resolves licensing information from three metadata sources: Crossref, Unpaywall, and OpenAlex. Licensing inconsistencies, vague descriptors, or non-standard license formats are common in large bibliographic datasets, and can impact downstream reuse, redistribution, and compliance. Our validator applies a flexible parsing strategy that extracts the most explicit license statement from each source, normalizes it to a canonical identifier (e.g., cc-by, cc0, public-domain), and determines a resolved license by checking for agreement across sources.

Non-informative values (e.g., unknown, other-oa) are excluded from agreement checks. When licenses from multiple sources agree, the record is treated as having stronger metadata support under the study's screening rule. Conflicts between informative licenses are explicitly captured using conflict expressions (e.g., conflict:cc-by\_vs\_closed). A record passes validation only when the resolved license is considered open-access, confirmed by at least two independent sources, and free from conflicts under the metadata-agreement rule. All other outcomes—including restrictive licenses (e.g., cc-by-nd) or source disagreement—result in a "fail" status. This validation step provides compliance evidence for the study rule; it does not by itself establish a legal guarantee for public redistribution of the exact reconstructed text.

Each validation result includes the resolved license, contributing sources, raw normalized inputs, a conflict indicator, and a UTC timestamp. The validator supports parallel processing for large-scale evaluation and produces reproducible, auditable outputs. Tables~\ref{tab:license-validation-flags} and \ref{tab:license-validation-examples} summarize the validation rules and provide output examples, respectively.

\renewcommand{\arraystretch}{1}
\footnotesize
\begin{longtable}{p{3.5cm} p{5cm} p{5.5cm}}
\caption{License Validation Rules and Conflict Flags} \label{tab:license-validation-flags} \\
\hline
\textbf{Validation Rule} & \textbf{Description} & \textbf{Flag (on failure)} \\
\hline
\endfirsthead

\hline
\textbf{Validation Rule} & \textbf{Description} & \textbf{Flag (on failure)} \\
\hline
\endhead

\hline \multicolumn{3}{r}{Continued on next page} \\
\endfoot

\hline
\endlastfoot

Raw Field Extraction & License fields are extracted from raw or nested metadata under Crossref, Unpaywall, and OpenAlex. & missing\_license\_<source> — Raised when no usable license is found in source. \\

Normalization to Canonical ID & Raw values are mapped to canonical forms using direct mappings or regex (e.g., cc-by, cc0, public-domain). & unknown\_license — Raised if license is unrecognized or ambiguous. \\

Source Agreement Logic & If two or more sources provide the same informative license, that license is accepted as resolved. & license\_conflict — Raised when two or more informative sources disagree. \\

Three-Way Conflict Detection & If all three sources provide different informative licenses, a multi-way conflict is recorded. & license\_conflict — Set to true and conflict expression recorded (e.g., conflict:cc-by\_vs\_closed\_vs\_cc0). \\

Resolved License Recording & The final resolved license is stored explicitly in resolved\_license. & Not a flag — Used for downstream compliance filters. \\

Open License Whitelist & Licenses must be from an accepted set (e.g., cc0, cc-by, cc-by-sa). & fail — If resolved license is missing, restrictive, or not from allowed list. \\

Conflict-Free Validation & Final license must be agreed upon by at least two sources and be conflict-free. & fail — If only one informative source or a conflict is detected. \\
\hline

\end{longtable}

\renewcommand{\arraystretch}{0.8}
\begin{table}[H]
    \centering
    \footnotesize  
    \caption{Example license validation outputs: failing (left) and passing (right).}
    \label{tab:license-validation-examples} 
    \begin{tabular}{p{7cm} p{7cm}}
    \toprule
    \textbf{Failing Record} & \textbf{Passing Record} \\
    \midrule
    \begin{minipage}[t]{\linewidth}
    \raggedright\ttfamily
    "license\_validation": \{\\
    \quad "status": "fail",\\
    \quad "summary": "Checks licence consistency and open-access compliance.",\\
    \quad "metrics": \{\\
    \quad\quad "resolved\_license": "conflict:closed\_vs\_cc-by-nc",\\
    \quad\quad "license\_source": "unpaywall+openalex",\\
    \quad\quad "license\_conflict": true,\\
    \quad\quad "input\_licenses": \{\\
    \quad\quad\quad "crossref": "missing",\\
    \quad\quad\quad "unpaywall": "closed",\\
    \quad\quad\quad "openalex": "cc-by-nc"\\
    \quad\quad \}\\
    \quad \},\\
    \quad "checked\_at": "2025-07-30T23:41:30Z"\\
    \}
    \end{minipage}
    &
    \begin{minipage}[t]{\linewidth}
    \raggedright\ttfamily
    "license\_validation": \{\\
    \quad "status": "pass",\\
    \quad "summary": "Checks licence consistency and open-access compliance.",\\
    \quad "metrics": \{\\
    \quad\quad "resolved\_license": "cc-by",\\
    \quad\quad "license\_source": "crossref+unpaywall",\\
    \quad\quad "license\_conflict": false,\\
    \quad\quad "input\_licenses": \{\\
    \quad\quad\quad "crossref": "cc-by",\\
    \quad\quad\quad "unpaywall": "cc-by",\\
    \quad\quad\quad "openalex": "missing"\\
    \quad\quad \}\\
    \quad \},\\
    \quad "checked\_at": "2025-07-30T12:08:22Z"\\
    \}
    \end{minipage}
    \\
    \bottomrule
    \end{tabular}
    \end{table}
    

\subsection{Text Validation}
\label{sec:Text-validation}

To verify the structural integrity, linguistic quality, and semantic consistency of the textual components in the screened study corpus, we implemented an automated text validation pipeline. This validator operates on both the abstract and the fulltext fields of each record. It performs a comprehensive suite of checks spanning length requirements, Unicode integrity, formatting quality, language detection, semantic overlap, and embedding consistency.

The pipeline first applies character and sentence-length thresholds to identify truncated or underspecified content. It then scans the text for corrupted characters (e.g., Unicode replacement characters), invisible formatting symbols, and control characters, reporting examples when found. Additional formatting quality checks include whitespace density and ASCII letter share, both of which help identify structural anomalies such as excessive markup or malformed text.

Language detection is applied to the beginning of the full text to verify alignment with the expected language (typically English), and semantic alignment between the abstract and body is measured using ROUGE-1 recall between the abstract and the first 2,000 characters of the full text. If an abstract\_embedding vector is present, its L2 norm is compared against an expected range to ensure numerical consistency.

Each record is assigned a validation status of "pass", "warn", or "fail" depending on which checks are triggered. Failures are reserved for critical issues (e.g., missing embeddings or extremely short full text), while warnings denote potentially problematic but non-blocking anomalies. Diagnostic flags and metrics are included in the output for auditability and triage. Tables~\ref{tab:text-validation-flags} and \ref{tab:text-validation-examples} summarize the validation rules and provide output examples, respectively.

\renewcommand{\arraystretch}{1}
\footnotesize
\begin{longtable}{p{4.2cm} p{5.3cm} p{5cm}}
\caption{Text Validation Rules and Diagnostic Flags} \label{tab:text-validation-flags} \\
\hline
\textbf{Validation Rule} & \textbf{Description} & \textbf{Flag (on failure)} \\
\hline
\endfirsthead

\hline
\textbf{Validation Rule} & \textbf{Description} & \textbf{Flag (on failure)} \\
\hline
\endhead

\hline \multicolumn{3}{r}{Continued on next page} \\
\endfoot

\hline
\endlastfoot

Abstract Character Length & Abstract must be at least 100 characters. & abstract\_too\_short \\

Full Text Character Length & Full text must be at least 1000 characters. & fulltext\_too\_short \\

Abstract Sentence Count & Must contain at least 2 sentence-ending punctuation marks. & abstract\_low\_sentence\_count \\

Full Text Sentence Count & Must contain at least 50 sentence-ending punctuation marks. & fulltext\_low\_sentence\_count \\

Heading Markers in Full Text & Detects Markdown or numbered headings (e.g., \# Introduction). & fulltext\_missing\_heading\_markers \\

Corrupted Characters & Unicode replacement char (U+FFFD) present in text. & abstract\_has\_corrupted\_chars, fulltext\_has\_corrupted\_chars \\

Whitespace Density & Non-whitespace density must exceed threshold (abstract: 0.75, full text: 0.83). & abstract\_low\_whitespace\_ratio, fulltext\_low\_whitespace\_ratio \\

ASCII Letter Share & ASCII letter proportion must exceed threshold (abstract: 0.7, full text: 0.75). & abstract\_low\_ascii\_ratio, fulltext\_low\_ascii\_ratio \\

Language Detection & Must match expected language (e.g., English) with $\geq$ 0.9 confidence. & language\_mismatch\_or\_low\_confidence \\

Semantic Overlap (ROUGE-1) & ROUGE-1 recall between abstract and full text (first 2000 chars) must be $\geq$ 0.5. & low\_rouge1\_overlap \\

Abstract Embedding Presence & Must be present and valid numeric vector. & abstract\_embedding\_missing\_or\_invalid \\

Abstract Embedding Norm & Must be within tolerance of expected L2 norm (default: 1.0 ± 1e-3). & abstract\_embedding\_norm\_off \\

Validation Status & Status is "fail" if any critical flags are raised; otherwise "warn" or "pass". & Not a flag — See status field \\
\hline
\end{longtable}

\renewcommand{\arraystretch}{0.8}
\begin{table}[H]
    \centering
    \footnotesize 
    \caption{Example text validation output: warning (left).} 
    \label{tab:text-validation-examples}
    \begin{tabular}{p{7cm} p{7cm}}
    \toprule
    \textbf{Warning Record} & \textbf{(Placeholder for Passing Record)} \\
    \midrule
    \begin{minipage}[t]{\linewidth}
    \raggedright\ttfamily
    "text\_validation": \{\\
    \quad "status": "warn",\\
    \quad "flags": \{\\
    \quad\quad "abstract\_too\_short": false,\\
    \quad\quad "fulltext\_too\_short": false,\\
    \quad\quad "abstract\_has\_corrupted\_chars": false,\\
    \quad\quad "fulltext\_has\_corrupted\_chars": false,\\
    \quad\quad "abstract\_low\_whitespace\_ratio": false,\\
    \quad\quad "fulltext\_low\_whitespace\_ratio": false,\\
    \quad\quad "abstract\_low\_ascii\_ratio": false,\\
    \quad\quad "fulltext\_low\_ascii\_ratio": false,\\
    \quad\quad "abstract\_low\_sentence\_count": false,\\
    \quad\quad "fulltext\_low\_sentence\_count": false,\\
    \quad\quad "fulltext\_missing\_heading\_markers": false,\\
    \quad\quad "language\_mismatch\_or\_low\_confidence": false,\\
    \quad\quad "low\_rouge1\_overlap": true,\\
    \quad\quad "abstract\_embedding\_norm\_off": false,\\
    \quad\quad "abstract\_embedding\_missing\_or\_invalid": false\\
    \quad \},\\
    \quad "metrics": \{\\
    \quad\quad "abstract\_chars": 1344,\\
    \quad\quad "fulltext\_chars": 15561,\\
    \quad\quad "abstract\_bad\_chars": \{\\
    \quad\quad\quad "valid": 1145,\\
    \quad\quad\quad "corrupted": 0,\\
    \quad\quad\quad "control": 0,\\
    \quad\quad\quad "formatting": 0,\\
    \quad\quad\quad "unassigned": 0,\\
    \quad\quad\quad "whitespace": 199\\
    \quad\quad \},\\
    \quad\quad "fulltext\_bad\_chars": \{\\
    \quad\quad\quad "valid": 13300,\\
    \quad\quad\quad "corrupted": 0,\\
    \quad\quad\quad "control": 0,\\
    \quad\quad\quad "formatting": 0,\\
    \quad\quad\quad "unassigned": 0,\\
    \quad\quad\quad "whitespace": 2261\\
    \quad\quad \},\\
    \quad\quad "abstract\_ws\_ratio": 0.8519345238095238,\\
    \quad\quad "fulltext\_ws\_ratio": 0.8547008547008547,\\
    \quad\quad "abstract\_ascii\_ratio": 0.8050595238095238,\\
    \quad\quad "fulltext\_ascii\_ratio": 0.80573227941649,\\
    \quad\quad "abstract\_sentence\_count": 5,\\
    \quad\quad "fulltext\_sentence\_count": 104,\\
    \quad\quad "fulltext\_heading\_markers": 9,\\
    \quad\quad "language": "en",\\
    \quad\quad "language\_score": 0.9999956877319424,\\
    \quad\quad "rouge1\_recall": 0.45145631067961167,\\
    \quad\quad "abstract\_embedding\_norm": 1.0\\
    \quad \},\\
    \quad "examples": \{\\
    \quad\quad "abstract\_problem\_chars": [],\\
    \quad\quad "fulltext\_problem\_chars": []\\
    \quad \},\\
    \quad "checked\_at": "2025-07-22T06:38:13.266722+00:00"\\
    \}
    \end{minipage}
    &
    \begin{minipage}[t]{\linewidth}
    \raggedright\ttfamily
    \end{minipage}
    \\
    \bottomrule
    \end{tabular}
    \end{table}


\subsection{Chunk-Level Validation}
\label{sec:Chunk-validation}

To ensure the internal consistency and semantic usability of paragraph-level text and embedding data, we implemented an automated chunk-level validation workflow. This process evaluates both the linguistic and vector components of each paragraph in a record to verify structure, dimensionality, and encoding correctness. These checks are critical in large-scale embedding datasets, where even minor irregularities can corrupt downstream model behavior.

The validation pipeline analyzes each document's paragraphs and their associated embedding vectors. It confirms the presence of both, measures token lengths using a consistent tokenizer (intfloat/e5-large-v2), and enforces dataset-wide length thresholds. Token lengths falling outside predefined minimum or maximum limits are flagged. Paragraph text is also inspected for Unicode compliance by counting valid, corrupted (�), control, formatting, and unassigned characters, revealing encoding issues that may not surface through standard inspection.

Embeddings are validated in parallel. Each paragraph must be paired with a numeric array of fixed length (1024). Missing embeddings or incorrect-length vectors are flagged as critical errors. The overall chunk validation outcome is determined by the presence of such errors: records with any embedding failures are marked "fail"; records with token length violations but valid embeddings are marked "warn"; records with no issues receive a "pass".

The validator scales across large corpora using multithreading or multiprocessing and provides detailed diagnostics per paragraph, aggregate metrics, and a UTC timestamp to support auditability and reproducibility. Tables~\ref{tab:chunk-validation-flags} and \ref{tab:chunk-validation-examples} summarize the validation rules and provide output examples, respectively.

\renewcommand{\arraystretch}{1}
\footnotesize
\begin{longtable}{p{4cm} p{5.5cm} p{4.5cm}}
\caption{Chunk-Level Validation Rules and Diagnostic Flags} \label{tab:chunk-validation-flags} \\
\hline
\textbf{Validation Rule} & \textbf{Description} & \textbf{Flag (on failure)} \\
\hline
\endfirsthead

\hline
\textbf{Validation Rule} & \textbf{Description} & \textbf{Flag (on failure)} \\
\hline
\endhead

\hline \multicolumn{3}{r}{Continued on next page} \\
\endfoot

\hline
\endlastfoot

Empty Paragraphs & Paragraph value is missing, empty, or not a string & empty\_chunks — Count of non-usable paragraph entries \\

Token Length Too Short & Paragraph has fewer than 100 tokens & chunks\_too\_short — Count of under-length paragraphs \\

Token Length Too Long & Paragraph has more than 300 tokens & chunks\_too\_long — Count of over-length paragraphs \\

Corrupted Characters & Paragraph contains replacement characters (�) & char\_counts.corrupted — Aggregated count per paragraph \\

Control or Formatting Characters & Paragraph includes invisible control (Cc), formatting (Cf), or unassigned (Cn) code points & char\_counts.control, formatting, unassigned — Diagnostic only \\

Missing Embedding & No vector found for paragraph key & missing\_embeddings — Count of missing embeddings \\

Invalid Embedding Shape & Vector is not a list of length 1024 & invalid\_embedding\_vectors — Count of malformed vectors \\

Validation Outcome & Embedding errors → "fail"; Token length errors only → "warn"; No issues → "pass" & Not a flag — outcome recorded in status, with human-readable summary \\
\hline
\end{longtable}

\renewcommand{\arraystretch}{0.8}
\begin{table}[H]
    \centering
    \footnotesize  
    \caption{Example chunk validation output: warning status with one chunk under the minimum token limit.}
    \label{tab:chunk-validation-examples}
    \begin{tabular}{p{14.5cm}}
    \toprule
    \textbf{Warning Record} \\
    \midrule
    \begin{minipage}[t]{\linewidth}
    \raggedright\ttfamily
    "chunk\_validation": \{\\
    \quad "status": "warn",\\
    \quad "summary": "0 chunk(s) over max tokens; 1 under min tokens.",\\
    \quad "metrics": \{\\
    \quad\quad "paragraph\_count": 36,\\
    \quad\quad "token\_length\_distribution": \{\\
    \quad\quad\quad "Q1": 147.75,\\
    \quad\quad\quad "Q2": 165.5,\\
    \quad\quad\quad "Q3": 183.25,\\
    \quad\quad\quad "min": 56,\\
    \quad\quad\quad "max": 239,\\
    \quad\quad\quad "mean": 166.38888888888889\\
    \quad\quad \},\\
    \quad\quad "chunks\_too\_long": 0,\\
    \quad\quad "chunks\_too\_short": 1,\\
    \quad\quad "empty\_chunks": 0,\\
    \quad\quad "missing\_embeddings": 0,\\
    \quad\quad "invalid\_embedding\_vectors": 0,\\
    \quad\quad "max\_token\_limit": 300,\\
    \quad\quad "min\_token\_limit": 100\\
    \quad \},\\
    \quad "paragraphs": \{\\
    \quad\quad "9991603P0": \{\\
    \quad\quad\quad "token\_length": 201,\\
    \quad\quad\quad "char\_counts": \{\\
    \quad\quad\quad\quad "valid": 864,\\
    \quad\quad\quad\quad "corrupted": 0,\\
    \quad\quad\quad\quad "control": 0,\\
    \quad\quad\quad\quad "formatting": 0,\\
    \quad\quad\quad\quad "unassigned": 0\\
    \quad\quad\quad \},\\
    \quad\quad\quad "problem\_characters": []\\
    \quad\quad \},\\
    \quad\quad "9991603P1": \{\\
    \quad\quad\quad "token\_length": 232,\\
    \quad\quad\quad "char\_counts": \{\\
    \quad\quad\quad\quad "valid": 1068,\\
    \quad\quad\quad\quad "corrupted": 0,\\
    \quad\quad\quad\quad "control": 0,\\
    \quad\quad\quad\quad "formatting": 0,\\
    \quad\quad\quad\quad "unassigned": 0\\
    \quad\quad\quad \},\\
    \quad\quad\quad "problem\_characters": []\\
    \quad\quad \},\\
    \quad\quad \vdots\\
    \quad\quad "9991603P35": \{\\
    \quad\quad\quad "token\_length": 56,\\
    \quad\quad\quad "char\_counts": \{\\
    \quad\quad\quad\quad "valid": 330,\\
    \quad\quad\quad\quad "corrupted": 0,\\
    \quad\quad\quad\quad "control": 0,\\
    \quad\quad\quad\quad "formatting": 0,\\
    \quad\quad\quad\quad "unassigned": 0\\
    \quad\quad\quad \},\\
    \quad\quad\quad "problem\_characters": []\\
    \quad\quad \}\\
    \quad \},\\
    \quad "checked\_at": "2025-07-22T06:37:27.593943+00:00"\\
    \}
    \end{minipage}
    \\
    \bottomrule
    \end{tabular}
    \end{table}

\subsection{Embedding Validation}
\label{sec:Embedding-validation}

To ensure the structural integrity and semantic fidelity of paragraph-level vector representations in Lit2Vec-style records, we implemented a dedicated embedding validation framework. This framework checks alignment between paragraph texts and their corresponding embeddings, enforces vector shape and type constraints, and confirms semantic consistency through spot-checked re-encoding. Each embedding must be a finite, normalized 1024-dimensional float vector that accurately represents the source paragraph, as verified by cosine similarity to a reference model.

The validation procedure begins by verifying that both paragraphs and embeddings fields are structured as dictionaries with matching keys. Paragraph identifiers must exist in both fields, with no missing or extra embeddings. Each vector must conform to strict structural constraints, including exact dimensionality, finiteness, and approximate unit norm. To test reproducibility, a random subset of paragraph–embedding pairs is re-encoded using a reference Sentence-Transformer model (e.g., intfloat/e5-large-v2), and cosine similarity is computed. Records pass this test only if all sampled pairs meet or exceed a defined similarity threshold.

Validation is deterministic, reproducible, and supports both sequential and parallel processing. The final output includes status, summary, detailed metrics (including alignment and similarity scores), and a UTC timestamp for auditing. Tables~\ref{tab:embedding-validation-flags} and \ref{tab:embedding-validation-examples} summarize the validation rules and provide output examples, respectively.

\renewcommand{\arraystretch}{1}
\footnotesize
\begin{longtable}{p{3.5cm} p{5cm} p{5.5cm}}
\caption{Embedding Validation Rules and Diagnostic Flags for Lit2Vec-Style Records} \label{tab:embedding-validation-flags} \\
\hline
\textbf{Validation Rule} & \textbf{Description} & \textbf{Flag (on failure)} \\
\hline
\endfirsthead

\hline
\textbf{Validation Rule} & \textbf{Description} & \textbf{Flag (on failure)} \\
\hline
\endhead

\hline \multicolumn{3}{r}{Continued on next page} \\
\endfoot

\hline
\endlastfoot

Field Type Check &
paragraphs and embeddings must be dictionaries. &
invalid\_structure — Raised when fields are not of type dict. \\

Paragraph–Embedding Alignment &
Each paragraph must have a corresponding embedding and vice versa. &
missing\_embedding, extra\_embedding — Raised for mismatched IDs. \\

Vector Length &
Each embedding must have exactly 1024 float elements. &
invalid\_shape\_embedding — Raised for incorrect dimensionality. \\

Finite Float Values &
All elements in embedding must be finite 32-bit floats. &
nonfinite\_values\_embedding — Raised if NaNs or Infs are present. \\

Unit Norm (Approximate) &
Embedding must be approximately unit-normalized within a 0.05 tolerance. &
unnormalized\_embedding — Raised if norm deviates from 1.0 beyond tolerance. \\

Cosine Similarity Threshold &
Randomly sampled paragraph–embedding pairs must match re-encoded vectors (cosine similarity ≥ threshold). &
cosine\_mismatch — Raised if any sampled vector falls below similarity threshold. \\

Pass Criteria &
All structural and reproducibility checks must pass. &
No specific flag — Outcome recorded as status: "pass" or "fail". \\
\hline
\end{longtable}

\renewcommand{\arraystretch}{0.8}
\begin{table}[H]
    \centering
    \footnotesize  
    \caption{Example embedding validation outputs: failing (left) and passing (right).}
    \label{tab:embedding-validation-examples}
    \begin{tabular}{p{7cm} p{7cm}}
    \toprule
    \textbf{Failing Record} & \textbf{Passing Record} \\
    \midrule
    \begin{minipage}[t]{\linewidth}
    \raggedright\ttfamily
    "embedding\_validation": \{\\
    \quad "status": "fail",\\
    \quad "summary": "Embedding validation complete",\\
    \quad "metrics": \{\\
    \quad\quad "paragraphs": 5,\\
    \quad\quad "embeddings": 5,\\
    \quad\quad "alignment": \{\\
    \quad\quad\quad "missing": [],\\
    \quad\quad\quad "extra": [],\\
    \quad\quad\quad "bad": ["p3"]\\
    \quad\quad \},\\
    \quad\quad "determinism": \{\\
    \quad\quad\quad "mean\_cos": 0.821,\\
    \quad\quad\quad "min\_cos": 0.765,\\
    \quad\quad\quad "max\_delta": 0.235,\\
    \quad\quad\quad "threshold": 0.01,\\
    \quad\quad\quad "passed": false\\
    \quad\quad \}\\
    \quad \},\\
    \quad "checked\_at": "2025-07-30T22:05:49.882404"\\
    \}
    \end{minipage}
    &
    \begin{minipage}[t]{\linewidth}
    \raggedright\ttfamily
    "embedding\_validation": \{\\
    \quad "status": "pass",\\
    \quad "summary": "Embedding validation complete",\\
    \quad "metrics": \{\\
    \quad\quad "paragraphs": 6,\\
    \quad\quad "embeddings": 6,\\
    \quad\quad "alignment": \{\\
    \quad\quad\quad "missing": [],\\
    \quad\quad\quad "extra": [],\\
    \quad\quad\quad "bad": []\\
    \quad\quad \},\\
    \quad\quad "determinism": \{\\
    \quad\quad\quad "mean\_cos": 0.998,\\
    \quad\quad\quad "min\_cos": 0.996,\\
    \quad\quad\quad "max\_delta": 0.004,\\
    \quad\quad\quad "threshold": 0.01,\\
    \quad\quad\quad "passed": true\\
    \quad\quad \}\\
    \quad \},\\
    \quad "checked\_at": "2025-07-30T22:08:41.208336"\\
    \}
    \end{minipage}
    \\
    \bottomrule
    \end{tabular}
    \end{table}
    

\subsection{Predicted Subfield Validation}
\label{sec:Subfield-validation}
To ensure that predicted disciplinary subfields are accurate, standardized, and suitable for downstream analysis, we implemented an automated subfield validation pipeline. This validation addresses common issues stemming from machine learning–based classification outputs or heterogeneous annotations, such as inconsistent label formats, incomplete predictions, and invalid probability scores.

Each record's predicted\_subfield field is expected to be a dictionary mapping subfield names (drawn from a fixed controlled vocabulary) to floating-point confidence scores in the range [0.0, 1.0]. The controlled vocabulary includes core and specialized areas of chemistry, such as Catalysis, Supramolecular Chemistry, Forensic Chemistry, and others.

The validator performs multiple checks:
\begin{itemize}
    \item Presence and correct data type of the predicted\_subfield field.
    \item Non-emptiness of the prediction map (optional but recommended).
    \item All labels must match entries in the controlled vocabulary.
    \item All scores must be floats within [0.0, 1.0].
\end{itemize}

Tables~\ref{tab:subfield-validation-flags} and \ref{tab:subfield-validation-examples} summarize the validation rules and provide output examples, respectively.

\renewcommand{\arraystretch}{1.5}
\footnotesize
\begin{longtable}{p{4cm} p{6cm} p{4.5cm}}
\caption{Validation Rules and Diagnostic Flags for Predicted Subfield Assignments} \label{tab:subfield-validation-flags} \\
\hline
\textbf{Validation Rule} & \textbf{Description} & \textbf{Flag (on failure)} \\
\hline
\endfirsthead

\hline
\textbf{Validation Rule} & \textbf{Description} & \textbf{Flag (on failure)} \\
\hline
\endhead

\hline \multicolumn{3}{r}{Continued on next page} \\
\endfoot

\hline
\endlastfoot

Field Presence & 
The predicted\_subfield field must exist in the record. & 
missing\_predicted\_subfield — Field is entirely absent. \\

Type Constraint & 
The field must be a dictionary. & 
not\_a\_dict — Field is not a dict (e.g., list or string given). \\

Empty Dictionary Check & 
Empty predictions are optionally disallowed. & 
empty\_predictions — No predicted subfields found. \\

Vocabulary Match & 
Each subfield label must match a known entry in the controlled vocabulary. & 
contains\_invalid\_labels — Unknown label(s) detected. \\

Score Validity & 
Each score must be a float in the range [0.0, 1.0]. & 
contains\_invalid\_scores — Value is non-float or out-of-bounds. \\

Validation Outcome & 
Determines status: "pass", "warn", or "fail" based on aggregated rule checks. & 
Not a flag — Derived from combination of failure conditions. \\
\hline
\end{longtable}

\renewcommand{\arraystretch}{0.8}
\begin{table}[H]
    \centering
    \footnotesize  
    \caption{Example predicted subfield validation outputs: failing (left), warning (center), and passing (right).}
    \label{tab:subfield-validation-examples}
    \begin{tabular}{p{5cm} p{5cm} p{5cm}}
    \toprule
    \textbf{Failing Record} & \textbf{Warning Record} & \textbf{Passing Record} \\
    \midrule
    \begin{minipage}[t]{\linewidth}
    \raggedright\ttfamily
    "predicted\_subfield\_validation": \{\\
    \quad "status": "fail",\\
    \quad "flags": \{\\
    \quad\quad "missing\_predicted\_subfield": true,\\
    \quad\quad "not\_a\_dict": false,\\
    \quad\quad "contains\_invalid\_labels": false,\\
    \quad\quad "contains\_invalid\_scores": false,\\
    \quad\quad "empty\_predictions": false\\
    \quad \},\\
    \quad "metrics": \{\\
    \quad\quad "label\_count": 0,\\
    \quad\quad "invalid\_labels": [],\\
    \quad\quad "invalid\_scores": []\\
    \quad \},\\
    \quad "checked\_at": "2025-07-30T23:41:30.136585Z"\\
    \}
    \end{minipage}
    &
    \begin{minipage}[t]{\linewidth}
    \raggedright\ttfamily
    "predicted\_subfield\_validation": \{\\
    \quad "status": "warn",\\
    \quad "flags": \{\\
    \quad\quad "missing\_predicted\_subfield": false,\\
    \quad\quad "not\_a\_dict": false,\\
    \quad\quad "contains\_invalid\_labels": true,\\
    \quad\quad "contains\_invalid\_scores": true,\\
    \quad\quad "empty\_predictions": false\\
    \quad \},\\
    \quad "metrics": \{\\
    \quad\quad "label\_count": 2,\\
    \quad\quad "invalid\_labels": [\\
    \quad\quad\quad "Quantum Wizardry"\\
    \quad\quad ],\\
    \quad\quad "invalid\_scores": [\\
    \quad\quad\quad \{"Catalysis": 1.2\}\\
    \quad\quad ]\\
    \quad \},\\
    \quad "checked\_at": "2025-07-30T23:41:31.004728Z"\\
    \}
    \end{minipage}
    &
    \begin{minipage}[t]{\linewidth}
    \raggedright\ttfamily
    "predicted\_subfield\_validation": \{\\
    \quad "status": "pass",\\
    \quad "flags": \{\\
    \quad\quad "missing\_predicted\_subfield": false,\\
    \quad\quad "not\_a\_dict": false,\\
    \quad\quad "contains\_invalid\_labels": false,\\
    \quad\quad "contains\_invalid\_scores": false,\\
    \quad\quad "empty\_predictions": false\\
    \quad \},\\
    \quad "metrics": \{\\
    \quad\quad "label\_count": 3,\\
    \quad\quad "invalid\_labels": [],\\
    \quad\quad "invalid\_scores": []\\
    \quad \},\\
    \quad "checked\_at": "2025-07-30T23:41:32.288271Z"\\
    \}
    \end{minipage}
    \\
    \bottomrule
    \end{tabular}
    \end{table}

\subsection{Summary Quality Validation}
\label{sec:Summary-validation}
To assess the quality of automatically generated summaries in the internal record set used in this study, we implemented a reproducible, multi-metric evaluation framework that measures both lexical and semantic similarity between each record's abstract and its associated tldr. This framework ensures that summaries are faithful, relevant, and information-preserving with respect to the source text.

The validation pipeline applies sentence-level tokenization to both the abstract and summary before computing two families of metrics:

\begin{itemize}
    \item \textbf{ROUGE (ROUGE-1, ROUGE-2, ROUGE-L, ROUGE-Lsum)}: Measures lexical overlap based on unigrams, bigrams, and longest common subsequences.
    \item \textbf{BERTScore (Precision, Recall, F1)}: Measures semantic similarity using contextualized embeddings from pre-trained transformer models.
\end{itemize}

These metrics are calculated using the Hugging Face evaluate library when available, with a fallback to native implementations for offline or high-throughput use. All scores are scaled to the range [0.0, 100.0]. Records with empty or invalid summaries or abstracts are assigned zero scores across all metrics. Tables~\ref{tab:summary-validation-flags} and \ref{tab:summary-validation-examples} summarize the validation rules and provide output examples, respectively.

\renewcommand{\arraystretch}{1}
\footnotesize
\begin{longtable}{p{4cm} p{6cm} p{4cm}}
\caption{Summary Evaluation Rules and Diagnostic Flags} \label{tab:summary-validation-flags} \\
\hline
\textbf{Validation Rule} & \textbf{Description} & \textbf{Flag (on failure)} \\
\hline
\endfirsthead

\hline
\textbf{Validation Rule} & \textbf{Description} & \textbf{Flag (on failure)} \\
\hline
\endhead

\hline \multicolumn{3}{r}{Continued on next page} \\
\endfoot

\hline
\endlastfoot

Non-empty Abstract and TLDR &
Both abstract and tldr must be non-empty strings. &
empty\_abstract, empty\_tldr \\

Minimum ROUGE-Lsum Threshold &
ROUGE-Lsum must exceed a configurable minimum threshold (default: 10.0). &
low\_rouge\_lsum \\

Minimum BERTScore-F1 Threshold &
BERTScore\_F1 must exceed a configurable minimum threshold (default: 30.0). &
low\_bertscore\_f1 \\

Optional Composite Score &
Composite score may be defined as the average of ROUGE-Lsum and BERTScore-F1. Records with low composite score may be flagged. &
low\_summary\_quality \\

Evaluation Fallback &
If online evaluation tools are unavailable, fall back to native libraries without compromising score fidelity. &
Not a flag — automatically handled by backend logic. \\

Evaluation Output Fields &
Evaluation must produce all 7 scores: ROUGE-1, ROUGE-2, ROUGE-L, ROUGE-Lsum, BERTScore Precision, Recall, and F1. &
missing\_score\_<metric> \\
\hline

\end{longtable}

\renewcommand{\arraystretch}{0.8}
\begin{table}[H]
    \centering
    \footnotesize  
    \caption{Example summary validation outputs: failing (left) and passing (right).}
    \label{tab:summary-validation-examples}
    \begin{tabular}{p{7cm} p{7cm}}
    \toprule
    \textbf{Failing Record} & \textbf{Passing Record} \\
    \midrule
    \begin{minipage}[t]{\linewidth}
    \raggedright\ttfamily
    "summary\_validation": \{\\
    \quad "scores": \{\\
    \quad\quad "ROUGE-1": 4.12,\\
    \quad\quad "ROUGE-2": 1.34,\\
    \quad\quad "ROUGE-L": 3.56,\\
    \quad\quad "ROUGE-Lsum": 5.89,\\
    \quad\quad "BERTScore\_Precision": 22.74,\\
    \quad\quad "BERTScore\_Recall": 24.65,\\
    \quad\quad "BERTScore\_F1": 23.44\\
    \quad \},\\
    \quad "flags": \{\\
    \quad\quad "low\_rouge\_lsum": true,\\
    \quad\quad "low\_bertscore\_f1": true\\
    \quad \},\\
    \quad "summary\_score": 14.66,\\
    \quad "evaluated\_at": "2025-07-30T13:22:18.543672"\\
    \}
    \end{minipage}
    &
    \begin{minipage}[t]{\linewidth}
    \raggedright\ttfamily
    "summary\_validation": \{\\
    \quad "scores": \{\\
    \quad\quad "ROUGE-1": 48.21,\\
    \quad\quad "ROUGE-2": 31.04,\\
    \quad\quad "ROUGE-L": 43.89,\\
    \quad\quad "ROUGE-Lsum": 46.75,\\
    \quad\quad "BERTScore\_Precision": 88.93,\\
    \quad\quad "BERTScore\_Recall": 90.14,\\
    \quad\quad "BERTScore\_F1": 89.52\\
    \quad \},\\
    \quad "flags": \{\},\\
    \quad "summary\_score": 68.13,\\
    \quad "evaluated\_at": "2025-07-30T13:25:14.781129"\\
    \}
    \end{minipage}
    \\
    \bottomrule
    \end{tabular}
    \end{table}
    

\subsection{Identifier and Consistency Validation}
\label{sec:Consistency-validation}

To ensure that each record is internally consistent and correctly linked to external bibliographic resources, we implemented an automated identifier validation pipeline. This validator performs three core consistency checks: (1) alignment of internal corpus identifiers; (2) normalization and comparison of Digital Object Identifiers (DOIs) across multiple metadata sources; and (3) alignment between text content and vector embeddings. 

Corpus identifiers are verified by comparing the filename\_id (from ingest), the top-level corpus\_id, and the nested metadata.corpusid. All three must agree for the record to pass this check. DOI consistency is assessed by collecting DOI strings from three sources—metadata.externalids.DOI, unpaywall.doi, and openalex.doi—normalizing them to canonical form, and checking for agreement. Finally, paragraph–embedding alignment ensures that each paragraph has a corresponding embedding, and vice versa, with up to five unmatched IDs reported per type. Tables~\ref{tab:consistency-validation-flags} and \ref{tab:consistency-validation-examples} summarize the validation rules and provide output examples, respectively.

\renewcommand{\arraystretch}{1}
\footnotesize
\begin{longtable}{p{3.5cm} p{5cm} p{5.5cm}}
\caption{Consistency Validation Rules and Diagnostic Flags} \label{tab:consistency-validation-flags} \\
\hline
\textbf{Validation Rule} & \textbf{Description} & \textbf{Flag (on failure)} \\
\hline
\endfirsthead

\hline
\textbf{Validation Rule} & \textbf{Description} & \textbf{Flag (on failure)} \\
\hline
\endhead

\hline \multicolumn{3}{r}{Continued on next page} \\
\endfoot

\hline
\endlastfoot

ID Agreement Check & 
The filename\_id, corpus\_id, and metadata.corpusid must all be present and equal. &
corpusid\_mismatch — Raised if values disagree. \\

Missing Metadata ID &
metadata.corpusid must be defined if other IDs are present. &
missing\_metadata\_id — Raised if missing. \\

DOI Normalization &
DOIs from metadata, unpaywall, and openalex are normalized to a canonical form. &
Not a flag — Used internally for comparison. \\

DOI Agreement Check &
DOIs from all available sources must match after normalization. &
doi\_mismatch — Raised if two or more DOIs differ. \\

DOI Coverage Check &
At least one DOI must be available across all sources. &
doi\_missing\_sources — Raised if all DOI sources are missing. \\

Missing Embeddings &
All paragraph IDs must have corresponding embedding IDs. &
missing\_embeddings — Raised if any paragraph is unrepresented. \\

Orphan Embeddings &
Embedding IDs must correspond to a known paragraph ID. &
orphan\_embeddings — Raised if extra embeddings are found. \\

Status Assignment &
"fail" if corpus ID mismatch exists; "warn" for other issues; "pass" if all checks succeed. &
Not a flag — Controlled by overall validation logic. \\
\hline
\end{longtable}

\renewcommand{\arraystretch}{0.8}
\begin{table}[H]
    \centering
    \footnotesize  
    \caption{Example consistency validation outputs: failing (left), warning (center), and passing (right).}
    \label{tab:consistency-validation-examples}
    \begin{tabular}{p{5cm} p{5cm} p{5cm}}
    \toprule
    \textbf{Failing Record} & \textbf{Warning Record} & \textbf{Passing Record} \\
    \midrule
    \begin{minipage}[t]{\linewidth}
    \raggedright\ttfamily
    "consistency\_validation": \{\\
    \quad "status": "fail",\\
    \quad "flags": \{\\
    \quad\quad "missing\_metadata\_id": false,\\
    \quad\quad "corpusid\_mismatch": true,\\
    \quad\quad "doi\_mismatch": false,\\
    \quad\quad "missing\_embeddings": false,\\
    \quad\quad "orphan\_embeddings": false,\\
    \quad\quad "doi\_missing\_sources": false\\
    \quad \},\\
    \quad "metrics": \{\\
    \quad\quad "corpus\_ids": \{\\
    \quad\quad\quad "filename\_id": 123456,\\
    \quad\quad\quad "corpus\_id": 123456,\\
    \quad\quad\quad "metadata\_id": 789012\\
    \quad\quad \},\\
    \quad\quad "doi\_set": [],\\
    \quad\quad "para\_count": 10,\\
    \quad\quad "embed\_count": 10,\\
    \quad\quad "missing\_para\_ids": [],\\
    \quad\quad "extra\_embed\_ids": [],\\
    \quad\quad "doi\_sources\_present": \{\\
    \quad\quad\quad "metadata\_doi": false,\\
    \quad\quad\quad "unpaywall\_doi": false,\\
    \quad\quad\quad "openalex\_doi": false\\
    \quad\quad \}\\
    \quad \},\\
    \quad "checked\_at": "2025-07-30T19:08:15.015264"\\
    \}
    \end{minipage}
    &
    \begin{minipage}[t]{\linewidth}
    \raggedright\ttfamily
    "consistency\_validation": \{\\
    \quad "status": "warn",\\
    \quad "flags": \{\\
    \quad\quad "missing\_metadata\_id": false,\\
    \quad\quad "corpusid\_mismatch": false,\\
    \quad\quad "doi\_mismatch": true,\\
    \quad\quad "missing\_embeddings": true,\\
    \quad\quad "orphan\_embeddings": false,\\
    \quad\quad "doi\_missing\_sources": false\\
    \quad \},\\
    \quad "metrics": \{\\
    \quad\quad "corpus\_ids": \{\\
    \quad\quad\quad "filename\_id": 101010,\\
    \quad\quad\quad "corpus\_id": 101010,\\
    \quad\quad\quad "metadata\_id": 101010\\
    \quad\quad \},\\
    \quad\quad "doi\_set": [\\
    \quad\quad\quad "10.1000/example",\\
    \quad\quad\quad "10.1000/other"\\
    \quad\quad ],\\
    \quad\quad "para\_count": 5,\\
    \quad\quad "embed\_count": 3,\\
    \quad\quad "missing\_para\_ids": [\\
    \quad\quad\quad "1P3", "1P4"\\
    \quad\quad ],\\
    \quad\quad "extra\_embed\_ids": [],\\
    \quad\quad "doi\_sources\_present": \{\\
    \quad\quad\quad "metadata\_doi": true,\\
    \quad\quad\quad "unpaywall\_doi": true,\\
    \quad\quad\quad "openalex\_doi": false\\
    \quad\quad \}\\
    \quad \},\\
    \quad "checked\_at": "2025-07-30T19:35:42.554110"\\
    \}
    \end{minipage}
    &
    \begin{minipage}[t]{\linewidth}
    \raggedright\ttfamily
    "consistency\_validation": \{\\
    \quad "status": "pass",\\
    \quad "flags": \{\\
    \quad\quad "missing\_metadata\_id": false,\\
    \quad\quad "corpusid\_mismatch": false,\\
    \quad\quad "doi\_mismatch": false,\\
    \quad\quad "missing\_embeddings": false,\\
    \quad\quad "orphan\_embeddings": false,\\
    \quad\quad "doi\_missing\_sources": false\\
    \quad \},\\
    \quad "metrics": \{\\
    \quad\quad "corpus\_ids": \{\\
    \quad\quad\quad "filename\_id": 202020,\\
    \quad\quad\quad "corpus\_id": 202020,\\
    \quad\quad\quad "metadata\_id": 202020\\
    \quad\quad \},\\
    \quad\quad "doi\_set": [\\
    \quad\quad\quad "10.1000/example"\\
    \quad\quad ],\\
    \quad\quad "para\_count": 8,\\
    \quad\quad "embed\_count": 8,\\
    \quad\quad "missing\_para\_ids": [],\\
    \quad\quad "extra\_embed\_ids": [],\\
    \quad\quad "doi\_sources\_present": \{\\
    \quad\quad\quad "metadata\_doi": true,\\
    \quad\quad\quad "unpaywall\_doi": true,\\
    \quad\quad\quad "openalex\_doi": true\\
    \quad\quad \}\\
    \quad \},\\
    \quad "checked\_at": "2025-07-30T20:05:08.123456"\\
    \}
    \end{minipage}
    \\
    \bottomrule
    \end{tabular}
    \end{table}

\end{document}